\documentclass[3p,times]{elsarticle}
\pdfoutput=1
\usepackage{ecrc}
\volume{00}
\firstpage{1}
\runauth{C. Ager et al.}
\usepackage{amssymb}
\usepackage{amsthm}
\newcounter{remcount}
\newtheorem{remark}[remcount]{Remark}
\usepackage{amsmath}
\usepackage{array}	
\usepackage{arydshln}
\usepackage[tight]{subfigure}		
\usepackage{graphicx}
\usepackage{color}
\usepackage{upgreek}
\usepackage[section]{placeins}
\usepackage{pict2e}
\biboptions{sort&compress,numbers}
\usepackage[figuresright]{rotating}
\definecolor{darkgreen}{rgb}{0,0.5,0}
\definecolor{green}{rgb}{0,1,0}
\definecolor{orange}{rgb}{1.0,0.5,0.0}
\definecolor{darkorange}{rgb}{0.6,0.3,0.0}

\newcommand{\tns}[1]{\underline{\boldsymbol{#1}}}

\newcommand{\mat}[1]{\underline{\boldsymbol{#1}}}
\newcommand{\partiald}[2]{\dfrac{ \partial #1}{\partial #2}}
\newcommand{\partialdt}[2]{\dfrac{ \partial^2 #1}{\partial {#2}^2}}

\newcommand{\derivn}[2]{\partial^{#1}_{\normal}#2}

\newcommand{\innerp}[3]{\left(#1, #2\right)_{#3}}
\newcommand{\innerpb}[3]{\left\langle #1, #2 \right\rangle_{#3}}
\newcommand{\jump}[1]{\ensuremath{\left[\!\left[#1\right]\!\right]} }
\newcommand{\grad}   { \boldsymbol{\nabla}    }
\newcommand{\gradRef}   { \boldsymbol{\nabla}_0    }

\renewcommand{\div} { \grad \! \cdot \!} 
\newcommand{\divRef} { \gradRef \! \cdot \!} 
\newcommand{\tr}   {\rm{tr}\,}

\newcommand{\zerovec}{\tns{0}}

\newcommand{\unity}{\tns{I}}

\newcommand{\norm}[2]{\ensuremath{\left|\left|#1\ensuremath\right|\right|_{#2}}}

\newcommand{\stress}{\tns{\sigma}}
\newcommand{\normal}{\tns{n}}
\newcommand{\tangent}{\tns{t}}

\newcommand{\btime}{t_0}
\newcommand{\etime}{t_E}
\newcommand{\timesfulltime}{\times [\btime, \etime]}
\newcommand{\Pnormal}{\left(\normal \otimes \normal\right)}
\newcommand{\Ptangent}{\left(\unity - \normal \otimes \normal\right)}

\newcommand{\strainenergy}[1]{\psi^{#1}}

\newcommand{\domain}{\Omega}
\newcommand{\fullbound}{\partial \domain}

\newcommand{\domainf}{\Omega^F}
\newcommand{\velocityf}{v}

\newcommand{\pressuref}{p}
\newcommand{\densityf}{\rho^F}
\newcommand{\velf}[1][]{\tns{\velocityf}^{F #1}}

\newcommand{\velfD}{\tns{\hat{\velocityf}}^F}
\newcommand{\velfB}{\tns{\mathring{\velocityf}}^F}
\newcommand{\pf}{\pressuref^F}

\newcommand{\viscf}{\mu^F}
\newcommand{\epsf}{\tns{\epsilon}^F}
\newcommand{\stressf}{\tns{\sigma}^F}

\newcommand{\bodyff}{\hat{\tns b}^F}

\newcommand{\normalf}{\tns n^F}

\newcommand{\tractionfN}{\tns{\hat{h}}^{F,N}}

\newcommand{\timef}{t}
\newcommand{\testvelf}[1][]{\delta \tns{\velocityf}^{F #1}}
\newcommand{\testpf}{\delta \pressuref^F}

\newcommand{\nboundf}{\interface^{F,N}}
\newcommand{\dboundf}{\interface^{F,D}}
\newcommand{\fullboundf}{\partial \domainf}

\newcommand{\timep}{t}
\newcommand{\domainp}{\Omega^P}
\newcommand{\refdomainp}{\Omega^P_0}
\newcommand{\displacementp}{u}
\newcommand{\velocityp}{v}
\newcommand{\pressurep}{p}
\newcommand{\porosity}{\phi}
\newcommand{\porosityB}{\mathring{\porosity}}
\newcommand{\velp}{\tns{\velocityp}^{P}}
\newcommand{\dispp}{\tns{\displacementp}^{P}}
\newcommand{\disppB}{\tns{\mathring{\displacementp}}^P}
\newcommand{\velpsB}{\tns{\mathring{\velocityp}}^{P^S}}
\newcommand{\velpB}{\tns{\mathring{\velocityp}}^{P}}
\newcommand{\velps}{ \partiald{\dispp}{\timep}}
\newcommand{\velpssmall}{\partial \dispp / \partial \timep}

\newcommand{\accps}{ \partialdt{\dispp}{\timep}}
\newcommand{\pp}{\pressurep^{P}}
\newcommand{\densitypf}{\densityf}
\newcommand{\refdensityps}{\rho^{P^S}_0}
\newcommand{\refavdensityps}{\tilde{\rho}^{P^S}_0}
\newcommand{\stresspkp}{\tns{S}^P}
\newcommand{\stressp}{\tns{\sigma}^P}
\newcommand{\bodyfpf}{\hat{\tns b}^{P^F}}
\newcommand{\bodyfp}{\hat{\tns b}^{P}}
\newcommand{\refbodyfp}{\bodyfp_0}

\newcommand{\tractionpfN}{\hat{h}^{P^F,N}}
\newcommand{\tractionpN}{\tns{\hat{h}}^{P,N}}
\newcommand{\reftractionpN}{\tractionpN_0}
\newcommand{\viscp}{\viscf}
\newcommand{\permeabpscalar}{k}
\newcommand{\permeabp}{\tns{\permeabpscalar}}
\newcommand{\matpermeabp}{\tns{\matpermeabpscalar}}
\newcommand{\matpermeabpscalar}{K}
\newcommand{\initmatpermeabpscalar}{\mathring{\matpermeabpscalar}}
\newcommand{\initmatpermeabp}{\mathring{\matpermeabp}}
\newcommand{\Jp}{J}
\newcommand{\strainenergyp}{\strainenergy{P}}

\newcommand{\strainglp}{\tns{E}}

\newcommand{\defgradp}{\tns{F}}
\newcommand{\testvelp}{\delta \tns{\velocityp}^P}
\newcommand{\testdispp}{\delta \tns{\displacementp}^P}

\newcommand{\testpp}{\delta \pressurep^P}
\newcommand{\dboundpf}{\interface^{P^F,D}}
\newcommand{\nboundpf}{\interface^{P^F,N}}
\newcommand{\dboundp}{\interface^{P,D}}
\newcommand{\nboundp}{\interface^{P,N}}
\newcommand{\refdboundp}{\dboundp_0}
\newcommand{\refnboundp}{\nboundp_0}

\newcommand{\fullboundp}{\partial \domainp}

\newcommand{\normalp}{\tns n^P}

\newcommand{\refnormalp}{\normalp_0}
\newcommand{\velpnD}{\hat{\velocityp}^P_n}
\newcommand{\disppD}{\tns{\hat{\displacementp}}^P}
\newcommand{\coord}{\tns x}
\newcommand{\refcoord}{\tns X}
\newcommand{\coordp}{\coord^P}
\newcommand{\refcoordp}{\refcoord^P}

\newcommand{\refdensitys}{\rho^{S}_0}

\newcommand{\timeso}{t}

\newcommand{\BJfac}{\beta_{BJ}}

\newcommand{\interface}{\Gamma}

\newcommand{\fpiinterface}{\interface^{FP}}

\newcommand{\sliplengh}{\kappa}

\newcommand{\adjointsign}{\zeta}

\newcommand{\identity}{\mat{I}}

\newcommand{\mycos}[1]{\text{cos}\left(#1\right)}
\newcommand{\mysin}[1]{\text{sin}\left(#1\right)}

\newcommand{\bjcoeff}{\alpha_{BJ}}

\newcommand{\jumpvaln}{\tns{\hat{g}}_n}
\newcommand{\jumpvalt}{\tns{\hat{g}}_t}
\newcommand{\jumpvals}{\tns{\hat{g}}_{\sigma}}
\newcommand{\jumpvalsn}{{\hat{g}}_{\sigma^n}}

\newcommand{\pfa}{\pf_{\mathcal{A}}}
\newcommand{\velfa}{\velf_{\mathcal{A}}}
\newcommand{\ppa}{\pp_{\mathcal{A}}}
\newcommand{\velpa}{\velp_{\mathcal{A}}}
\newcommand{\disppa}{\dispp_{\mathcal{A}}}
\newcommand{\velpsa}{ \partiald{\disppa}{\timep}}
\newcommand{\accpsa}{ \partialdt{\disppa}{\timep}}
\newcommand{\permeabpa}{ \permeabp_{\mathcal{A}}}
\newcommand{\stresspkpa}{ \stresspkp_{\mathcal{A}}}
\newcommand{\stresspa}{ \stressp_{\mathcal{A}}}
\newcommand{\stressfa}{ \stressf_{\mathcal{A}}}
\newcommand{\Jpa}{ \Jp_{\mathcal{A}}}
\newcommand{\defgradpa}{ \defgradp_{\mathcal{A}}}

\DeclareGraphicsExtensions{.pdf}
\begin{document}
\begin{frontmatter}
\dochead{}
\title{A Nitsche-based cut finite element method for the coupling of incompressible fluid flow with poroelasticity}
\author{C. Ager \corref{cor1}}
\ead{ager@lnm.mw.tum.de}
\cortext[cor1]{corresponding author}
\ead[url]{https://www.lnm.mw.tum.de/staff/christoph-ager}
\author{B. Schott}
\author{M. Winter}
\author{W.A. Wall}
\address{Institute for Computational Mechanics ,
				Technische Universit\"a{}t M\"u{}nchen,
				Boltzmannstr. 15, 85747 Garching b. M\"u{}nchen}
\begin{abstract}
The focus of this contribution is the numerical treatment of interface coupled problems concerning the interaction of incompressible fluid flow and permeable, elastic structures.
The main emphasis is on extending the range of applicability of available formulations on especially three aspects.
These aspects are the incorporation of a more general poroelasticity formulation, 
the use of the cut finite element method (CutFEM) to allow for large interface motion and topological changes of the fluid domain, 
and the application of a novel Nitsche-based approach to incorporate the Beavers-Joseph (-Saffmann) tangential interface condition. 
This last aspect allows one to extend the practicable range of applicability of the proposed formulation down to very low porosities and permeabilities
which is important in several examples in application.
Different aspects of the presented formulation are analyzed in a numerical example including spatial convergence, 
the sensitivity of the solution to the Nitsche penalty parameters,
varying porosities and permeabilities, and a varying Beavers-Joseph interface model constant.
Finally, a numerical example analyzing the fluid induced bending of a poroelastic beam provides evidence of the general applicability of the presented approach.

\end{abstract}
\end{frontmatter}
\vspace*{1cm}
\section{Introduction}
\label{sec:intro}
The interaction of an incompressible fluid with a permeable, elastic, and fluid-saturated structure is of great interest for various fields.
In geomechanics, applications such as analyzing groundwater flow including aquifers or the oil and gas flow in a permeable reservoir containing cracks are of interest.
In biomechanics, poroelastic structures appear for example in the modeling of the interaction of blood flow and permeable tissues.
Another important area of application are biofilms that are often modeled as simple solids but in many cases would be better represented as fluid filled poroelastic media
\cite{coroneo2014}. 
Lastly, the application, which is the incentive for developing the subsequent numerical method, 
is rough surface modeling in the context of fluid-structure-contact interaction \cite{ager2018}.

In recent years, several formulations to solve the interface coupled problem of incompressible flow and poroelasticity were presented 
\cite{showalter2005,badia2009,bukavc2015,zakerzadeh2016,ambartsumyan2017,luo2018} and novel approaches are still being developed to meet the arising challenges.
Therein, the governing equations inside of the poroelastic domain are usually based on the Biot-system \cite{biot1941}, where the fluid flow through the poroelastic matrix 
is modeled by a Darcy-like flow equation that is
volume-coupled to a linear solid mechanics model valid for ``small'' deformation.
In the fluid domain, the Stokes equations \cite{bukavc2015,ambartsumyan2017,luo2018,showalter2005} or, including the effect of 
convection, the Navier-Stokes equations \cite{badia2009,zakerzadeh2016} are applied. 
On the fluid-poroelastic interface, either the Beavers-Joseph-Saffmann \cite{bukavc2015,showalter2005,ambartsumyan2017,badia2009} or 
a ``no-slip'' condition \cite{bukavc2015,zakerzadeh2016,luo2018} in tangential interface direction are consulted.
Details on these interface conditions for the coupling of fluid and (rigid-)porous flow can be found in \cite{beavers1967,saffman1971,discacciati2009,cao2010,dangelo2011} and the references therein.

Motivated by the specific requirements of a fluid-poroelastic interaction (FPI) formulation to model rough surface contact in fluid-structure interaction (as in \cite{ager2018}), 
we extend the range of applicability of these formulations by several aspects.\\
First, a more general formulation for the poroelastic domain, which is based on the same fundamental physical equations as the classical Biot-system, is applied. 
This formulation allows one to take into account
large deformation and the motion of the poroelastic domain, a wide variety of material models by arbitrary strain energy density functions, and varying deformation dependent porosity
(see \cite{Chapelle2010b,Vuong2015,vuong2016} for details on this formulation and \cite{schrefler1998} for fundamentals).\\
Second, in order to allow for large deformation and motion or even topological changes of the fluid domain, a cut finite element method (CutFEM) is applied.
Herein, a non-interface-fitted, fixed-grid Eulerian computational mesh for the fluid domain is combined with an interface fitted computational mesh 
of the poroelastic domain in Lagrangian description w.r.t. the displacements of the poroelastic solid phase.
Development of the CutFEM, as it will be applied to fluid equations, started by analyses on the Poisson equation \cite{burman2012}, the Stokes equation \cite{burman2014_2,massing2014}, 
and finally, including advection, the Oseen equation \cite{schott2014,massing2016}.
It has been successfully applied to various applications including two-phase flow \cite{gross2007,hansbo2014,schott2015} 
and fluid-structure interaction \cite{burman2014,massing2015,schott2017,schott2017b}.
Herein, a weak imposition of the interface constraint through Nitsche-based methods and the stabilization of ``cut'' elements by a so-called Ghost-Penalty stabilization \cite{burman2010} are the predominant approaches.\\
The third extension concerns the tangential interface constraint enforcement using a novel Nitsche method, based on the formulations to incorporate Robin boundary conditions presented in \cite{Juntunen2009} 
for the Poisson equation and in \cite{winter2017} for general Navier conditions for the Oseen equation.
In contrast to the presented FPI formulations in the previously mentioned literature, 
where the tangential fluid stress on the interface is substituted by the kinematic relation given by the Beavers-Joseph-Saffmann condition,
this formulation prevents the ill-conditioning of the system to solve when considering small permeabilities or, in general, conditions close to the ``no-slip'' limit case on the interface.
Especially for FPI problems where the poroelastic media potentially approaches the impermeability limit, this is an essential requirement of the formulation for the imposition of the interface constraint.\\
Due to the focus on strong interactions between the fluid domain and the poroelastic structure, the finally resulting nonlinear system of equations is solved using 
a monolithic approach, which was already successfully developed and used for fluid-structure interaction \cite{heil2004,Kuttler2010,Mayr2015} and general coupled $n$-field problems \cite{verdugo2016}.

We analyze the presented formulation through a numerical example with an a priori known solution, which is achieved through the method of manufactured solutions.
In addition to the spatial convergence analysis for all essential domain and interface error norms, 
we determine the behavior of the formulation for variations of the Nitsche penalty parameters 
and obtain a recommendation for the choice of these parameters.
Furthermore, a test for varying porosity, and consequently varying the permeability through the Kozeny-Carman formula, 
allows one to compare the presented novel Nitsche method and the predominant ``Substitution'' approach. 
This is performed for moderate, down to very small porosities/permeabilities and therefore allows for a fundamental comparison of both approaches.
Finally, a comparison of the Beavers-Joseph (BJ) and Beavers-Joseph-Saffmann (BJS) condition for a variation of the interface model constant ranging from the limit cases ``slip'' to ``no-slip'' is performed.

The outline of this paper is as follows. In Section \ref{sec:goveq}, the governing equations of the fluid domain, poroelastic domain and the conditions on their common interface are discussed.
This is followed by a presentation of the applied numerical method in Section \ref{sec:FE}, including the discrete weak form with all additional contributions arising from the stabilization and the interface handling.
Herein, all constants occurring in the discrete weak formulation are listed.
In Section \ref{sec:ex1}, various aspects of the formulation are analyzed numerically with an example based on the method of manufactured solutions.
Finally, in Section \ref{sec:ex2}, the computational results of a general problem configuration, the fluid induced bending of a poroelastic beam, are presented
and a conclusion is given in Section \ref{sec:conclusion}.

\section{Governing equations}
\label{sec:goveq}
\begin{figure}[tb]
\centering
\def\svgwidth{0.45\textwidth}
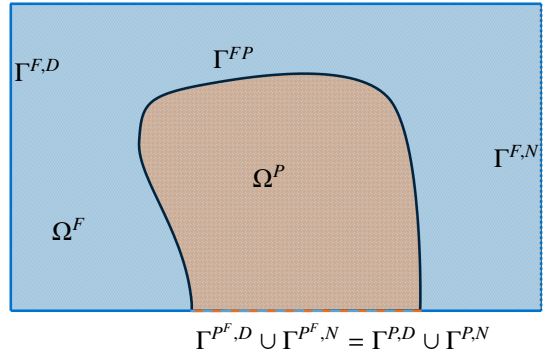
\caption{Fluid-Poroelastic interaction problem setup: fluid domain $\domainf$, poroelastic domain $\domainp$, common fluid-poroelastic interface $\fpiinterface$, and
boundaries $\Gamma^{F,D}$, $\Gamma^{F,N}$, $\Gamma^{P^F,D}$, $\Gamma^{P^F,N}$, $\Gamma^{P,D}$, and $\Gamma^{P,N}$ on the outer boundary.}
\label{fig:FPIsetup}
\end{figure}

In this section, the governing equations and an introduction to the notation for the FPI problem setup is given.
The principal configuration is visualized in Figure \ref{fig:FPIsetup}. It includes the fluid domain $\domainf$, the poroelastic domain $\domainp$, 
the common interface between both domains $\fpiinterface$, and the boundaries 
$\Gamma^{F,D}$, $\Gamma^{F,N}$, $\Gamma^{P^F,D}$, $\Gamma^{P^F,N}$, $\Gamma^{P,D}$, and $\Gamma^{P,N}$
where appropriate boundary conditions have to be applied.
The overall domain $\domain=\domainf \cup \domainp$ is given by the union of both individual domains 
and the outer boundary $\partial \domain = \Gamma^{F,D} \cup \Gamma^{F,N} \cup \Gamma^{P,D} \cup \Gamma^{P,N}
= \Gamma^{F,D} \cup \Gamma^{F,N} \cup\Gamma^{P^F,D} \cup \Gamma^{P^F,N}$ by the union of all boundaries.\\
This section starts with a presentation of the Navier-Stokes equations for the fluid domain and the system of equations for the poroelastic domain, including all boundary and initial conditions.
Finally, the conditions that have to be fulfilled on the common interface $\fpiinterface$ are introduced.

Starting from the initial point in time $\btime$, the physical response of the system until time $\etime$ is of interest.
The following indices and symbols of a scalar $*$ or vector-valued $\tns *$ quantity have a specific meaning.
The undeformed, reference/material configuration of a quantity is specified by the ``zero''-index: $*_0$ or $\tns *_0$, whereas a missing index refers to the current configuration 
(see \cite{Holzapfel2000} for details).
Time-dependent prescribed quantities at a boundary, an interface or in the domain are indicated by the ``hat''-symbol: $\hat{*}$ or $\hat{\tns *}$.
Quantities prescribed at $\btime$ are specified by the ``ring''-symbol: $\mathring{*}$ or $\mathring{\tns *}$.
The outer boundary of a domain $\domain^*$ is specified by $\partial \domain^*$.

\subsection{The fluid domain}
\label{sec:fluid}
In the fluid domain, an incompressible, viscous fluid should be considered for transient flows. 
Therefore, the Navier-Stokes equations, which include the balance of mass and momentum, make up the governing equations for the fluid domain:
\begin {align}
\densityf \partiald{\velf}{\timef} + \densityf \velf \cdot \grad \velf + \grad \pf - \div(2 \viscf \epsf (\velf)) - \densityf\bodyff  &= \zerovec   \quad \text{in} \;  \domainf \times [\btime,\etime],\label{eq:fluidm_eq}\\
\div \velf &= 0   \quad \text{in} \;  \domainf \times [\btime,\etime].\label{eq:fluidc_eq}
\end{align}
The fluid velocity and pressure are denoted by $\velf$ and $\pf$, the constant fluid density by $\densityf$, the dynamic viscosity by $\viscf$,
 the strain-rate tensor by $\epsf (\velf) = \frac{1}{2}\left[\grad \velf + \left(\grad \velf\right)^T\right]$, and  the body force per unit mass by $\bodyff$.
An adequate initial velocity field $\velfB$ is prescribed as:
\begin{align}
\velf = \velfB \quad \text{in} \;  \domainf \times \left\lbrace t_0 \right\rbrace.
\label{eq:fluid_init}
\end{align}
Boundary conditions on the subset of the outer boundary $\dboundf\cup\nboundf=\fullbound\cap\fullboundf$ have to be considered, to finalize the description of the fluid problem.
Hereby, the fluid velocity $\velfD$ on Dirichlet boundaries $\dboundf$ 
and the fluid traction $\tractionfN$ on Neumann boundaries $\nboundf$ are prescribed as:
\begin{align}
\velf = \velfD \quad \text{on} \;  \dboundf \times [\btime,\etime], \qquad
\stressf \cdot \normalf = \tractionfN \quad \text{on} \;  \nboundf\times [\btime,\etime].
\label{eq:fluid_bcs}
\end{align}
Herein, the Cauchy fluid stress defined by $\stressf = -\pf \identity + 2 \viscf \epsf(\velf)$ and the outward-pointing unit normal vector of the fluid domain $\normalf$ are applied.
The interface conditions on the common interface $\fpiinterface = \fullboundf\setminus\left(\dboundf\cup\nboundf\right)$,
which is not part of the outer boundary $\fullbound~\cap~\fpiinterface=\emptyset$,  are discussed in Section \ref{sec:fpinterface}.

\subsection{The poroelastic domain}
\label{sec:poro}
As already stated in the introduction, the governing equations of the poroelastic domain model the homogenized incompressible flow in a fluid saturated elastic solid matrix.
Here, a formulation which is more flexible and has a wider range of validity than the classical Biot-system \cite{biot1941} is applied.
This homogenized, poroelastic model was successfully developed and applied in \cite{Chapelle2010b, Vuong2015,vuong2016, vuong2017}.
Therein, the flow is also modeled by a Darcy flow based equation.
Large deformations, finite strain, and an arbitrary choice of the modeled macroscopic material are included in the balance of the momentum of the poroelastic mixture.
Additionally, deformation and fluid load induced variations of the porosity are considered in the model.
The governing equations are given as:
\begin{align}
\left.
\partiald{\porosity}{\timep}
\right|_{\refcoordp}
+ \porosity \div \velps + \div \left[ \porosity \left(\velp - \velps \right) \right] &= 0    \quad \text{in} \;  \domainp \times [\btime,\etime],
\label{eq:poroc_eq}\\
\left.
\densitypf \partiald{\velp}{\timep}
\right|_{\refcoordp}
- \densityf\velps\cdot\grad\velp +\grad \pp - \densitypf \bodyfpf +\viscp  \porosity \permeabp^{-1} \cdot \left(\velp-\velps\right) &= \zerovec    \quad \text{in} \;  \domainp \times [\btime,\etime],
\label{eq:porom_eq}\\
\refavdensityps \accps - \divRef \, \left( \defgradp \cdot \stresspkp \right) - \refavdensityps \refbodyfp
- \Jp \porosity \left(\defgradp \right)^{-T} \cdot \gradRef \pp 
- \viscp \Jp \porosity^2 \permeabp^{-1} \cdot \left( \velp - \velps \right) &= \zerovec   \quad \text{in} \;  \refdomainp \times [\btime,\etime],
\label{eq:mixturem_eq}
\end{align}
and include the balance of mass \eqref{eq:poroc_eq} and momentum  \eqref{eq:porom_eq} of the fluid phase in current configuration, 
and the balance of momentum \eqref{eq:mixturem_eq} for the whole poroelastic mixture (consisting of fluid and solid) in material configuration on a macroscopic scale.
The balance of mass of the solid phase is already included implicitly (see \cite{Chapelle2010b,Vuong2015} for details).
In the equations, the average microscopic state in the poroelastic domain is represented, which can be observed from a macroscopic view.
No fluctuations due to the porous microstructure are represented.
Quantities occurring in equations \eqref{eq:poroc_eq}-\eqref{eq:mixturem_eq} therefore represent an average state in the poroelastic media.

The displacement vector is denoted by $\dispp = \coordp - \refcoordp$, which describes the motion of a material point of the solid phase 
(with position $\refcoordp$ at initial time $\timeso = \btime$) 
due to the deformation of the elastic matrix to the current position $\coordp$.
Furthermore, the macroscopic averaged initial density of the solid phase is specified as $\refavdensityps = (1-\porosity) \refdensityps$, 
with $\refdensityps$ being the associated averaged initial density of the solid phase.
The body force acting on the poroelastic mixture per unit macroscopic averaged solid matrix mass is denoted by $\refbodyfp$.
The macroscopic deformation gradient is specified as $\defgradp$, 
the material divergence operator as $\divRef \tns *$, 
the determinant of the macroscopic deformation gradient as $\Jp=\det{\defgradp}$,
and the homogenized second Piola-Kirchhoff stress tensor as $\stresspkp$.
The Cauchy stress is given by $\stressp = \frac{1}{\Jp} \defgradp \cdot \stresspkp \cdot \left( \defgradp \right)^T$ 
and will be required when considering the interaction on the interface $\fpiinterface$ in the current configuration.
The porosity, which is the fluid volume fraction inside the poroelastic domain, is denoted by $\porosity$,
the velocity and the pressure of the fluid phase by $\velp$ and $\pp$,
the body force acting on the fluid per unit mass by $\bodyfpf$,
and the spatial permeability of the poroelastic matrix by $\permeabp=\left(\Jp\right)^{-1}\defgradp \cdot \matpermeabp \cdot \left(\defgradp\right)^T$.
The latter is dependent on the corresponding prescribed material permeability $\matpermeabp$.
A macroscopic strain energy function represents the material behavior of the poroelastic solid:
\begin{align}
\strainenergyp(\strainglp, \Jp (1-\porosity)) = \strainenergy{P,skel}(\strainglp)+\strainenergy{P,vol}(\Jp (1-\porosity))+\strainenergy{P,pen}(\strainglp, \Jp (1-\porosity)).
\end{align}
The three contributions therein account for the strain energy due to macroscopic deformation of the solid phase $\strainenergy{P,skel}$, 
the volume change of the solid phase due to changing fluid pressure $\strainenergy{P,vol}$,
and to guarantee positive porosity $\strainenergy{P,pen}$ of the poroelastic model (see \cite{Chapelle2010b,Vuong2015}).
Two constitutive relations complete the system of equations:
\begin{align}
\stresspkp &= 
\partiald{\strainenergyp(\strainglp, \Jp (1-\porosity)=\text{const.})}{\strainglp}
- \pp\Jp\left(\defgradp\right)^{-1}\cdot\left(\defgradp\right)^{-T}, 
\quad \strainglp = \frac{1}{2}\left[\left(\defgradp\right)^T\cdot\defgradp - \identity\right], 
\quad \defgradp = \identity + \partiald{\dispp}{\refcoordp},
\label{eq:poro_constitutive1}\\
\pp &= 
-\partiald{\strainenergyp(\strainglp=\text{const.}, \Jp(1-\porosity))}{(\Jp(1-\porosity))}
\label{eq:poro_constitutive2}.
\end{align}
There the Green-Lagrange strain tensor is denoted by $\strainglp$.
The following conditions at the initial point in time $t=t_0$ are necessary for the poroelastic problem:
\begin{align}
\dispp= \disppB \quad \text{in} \;  \refdomainp \times \left\lbrace t_0 \right\rbrace, \qquad
\velps = \velpsB \quad \text{in} \;  \refdomainp \times \left\lbrace t_0 \right\rbrace, \qquad
\porosity = \porosityB \quad \text{in} \;  \domainp \times \left\lbrace t_0 \right\rbrace, \qquad
\velp = \velpB \quad \text{in} \;  \domainp \times \left\lbrace t_0 \right\rbrace.
\label{eq:poro_init}
\end{align}
Herein, the initial displacement, initial solid phase velocity, initial porosity, and initial fluid velocity field are denoted by $\disppB$, $\velpsB$, $\porosityB$, and $\velpB$, respectively.
Adequate boundary conditions on the subset of the outer boundary  $\dboundp\cup\nboundp= \dboundpf\cup\nboundpf = \fullbound\cap\fullboundp$  have to be prescribed as:
\begin{alignat}{2}
\velp \cdot \normalp &= \velpnD\quad \text{on} \;  \dboundpf\times [\btime,\etime],\qquad
-\pp \normalp &&= \tractionpfN \normalp \quad \text{on} \;  \nboundpf\times [\btime,\etime],\label{eq:porof_bcs}\\
\dispp &= \disppD \quad \text{on} \;  \refdboundp\times [\btime,\etime],\qquad
\left( \defgradp \cdot \stresspkp \right) \cdot \refnormalp &&= \reftractionpN \quad \text{on} \;  \refnboundp\times [\btime,\etime].\label{eq:poro_bcs}
\end{alignat}
Therein, the scalar normal fluid velocity of the Darcy like flow on Dirichlet boundaries $\dboundpf$ is specified as $\velpnD$, 
the traction in normal direction on Neumann boundaries $\nboundpf$ as $\tractionpfN$,
the displacement of the poroelastic domain on Dirichlet boundaries $\refdboundp$ as $\disppD$,  and 
the traction acting onto the poroelastic mixture on Neumann boundaries $\refnboundp$ as $\reftractionpN$ , with $\normalp$ being the outward-pointing unit normal vector of domain $\domainp$ on the boundary.
Again, the interface conditions on the common interface $\fpiinterface = \fullboundp \setminus \left( \dboundpf \cup \nboundpf \right) = \fullboundp \setminus \left( \dboundp \cup \nboundp \right)$ 
are discussed in the following Section \ref{sec:fpinterface}.

\subsection{The common interface between the fluid and poroelastic domains}
\label{sec:fpinterface}
The conditions on the common interface between a viscous fluid and the poroelastic domain are specified in analogy to the frequently analyzed coupling of viscous flow and porous flow.
Hereby, the following conditions need to be fulfilled on the interface $\fpiinterface = \fullboundp \cap \fullboundf$:
\begin{align}
\zerovec &= \stressf \cdot \normalf - \stressp \cdot \normalf - \jumpvals \quad &&\text{on} \; \fpiinterface \timesfulltime,
\label{eq:fpi_dyneq}\\
0 &= \normalf \cdot  \stressf \cdot \normalf + \pp - \jumpvalsn \quad &&\text{on} \; \fpiinterface \timesfulltime,
\label{eq:fpi_dyneqp}\\
0 &= \left[\velf - \velps - \porosity \left( \velp - \velps \right) - \jumpvaln \right]\cdot \normalf\quad &&\text{on} \; \fpiinterface \timesfulltime,
\label{eq:fpi_massb}\\
0 &= \left[\velf - \velps - \BJfac \porosity \left( \velp - \velps \right) + \sliplengh \normalf\cdot \stressf - \jumpvalt \right] \cdot \tangent_i \quad i=1,2 \quad &&\text{on} \; \fpiinterface \timesfulltime
\label{eq:fpi_bj}.
\end{align}
The dynamic stress balance in the current configuration between the Cauchy stresses of fluid and the entire poroelastic mixture is represented by \eqref{eq:fpi_dyneq}.
Furthermore, a dynamic stress balance between the fluid pressure inside of the poroelastic domain and the stress components of the viscous fluid, in normal direction, are enforced by \eqref{eq:fpi_dyneqp}.
The kinematic constraint in interface normal direction \eqref{eq:fpi_massb} guarantees mass-balance on the interface.

A constraint in tangential direction on the viscous fluid is still missing.
To include effects arising from the boundary layer in the porous flow, which cannot be represented by the Darcy equation, the so-called Beavers-Joseph (BJ) condition \cite{beavers1967} ($\BJfac = 1$) is considered in \eqref{eq:fpi_bj}.
Herein, a proportionality of the viscous shear stress and the relative velocity slip in tangential direction between the adjacent fluids on both sides of the interface is proposed.
The tangential vectors $\tangent_i$, which are orthogonal to the normal vector $\normalf$, define the tangential plane of the interface.
The proportionality constant $\sliplengh$ depends on the permeability $\permeabp$ of the porous structure, the fluid viscosity $\viscf$, and the positive model parameter $\bjcoeff$ 
which has to be verified experimentally. Then, $\sliplengh$ can be computed as:
 \begin{align}
 \kappa  = \left(\alpha_{BJ} \viscf \sqrt{3}\right)^{-1}\sqrt{\text{tr}(\permeabp)}.
\label{eq: bjcoeff}
\end{align}
In cases with small permeability, a simplified condition, which does not include the seepage velocity, the Beavers-Joseph-Saffmann (BJS) condition \cite{saffman1971,jones1973} ($\BJfac = 0$), can be applied.
Due to the nonexistent tangential stress contribution on the boundaries/interfaces of the Darcy equation,
this condition - which does not include the tangential porous fluid velocity - is mostly analyzed and applied.
For the Beavers-Joseph condition it is shown in \cite{cao2010} that well-posedness is established for small $\bjcoeff$. 
In \cite{gartling1996, burman2007}, numerical computations for a ``no-slip'' condition between viscous and porous fluid velocity (Beavers-Joseph condition with $\bjcoeff^{-1} = 0$) show oscillations of the 
porous velocity close to the interface. To analyze the behavior of the two different approaches ($\BJfac = 1$ and $\BJfac = 0$), a comparison of both methods when varying the parameter $\bjcoeff$ 
is presented in Section \ref{sec:varbj}.

\begin{remark}[Occurring jump contributions in conditions \eqref{eq:fpi_dyneq} - \eqref{eq:fpi_bj}]
\label{rem:fpi_jump_terms}
In all conditions \eqref{eq:fpi_dyneq} - \eqref{eq:fpi_bj}, additional jump contributions $\jumpvals$, $\jumpvalsn$, $\jumpvaln$, and $\jumpvalt$ are incorporated.
In general, these contributions have to vanish when considering the physical correct conditions, but they will be employed in Section \ref{sec:ex1} for the application of the method of manufactured solutions.
This simplifies the choice of possible prescribed analytic solutions, as non-vanishing physical conditions can be considered.
\end{remark}

\section{Finite element discretization}
\label{sec:FE}
In this section, all essential aspects to solve the FPI problem based on the finite element method numerically, are presented.
First, the semi-discrete weak form is built up from the governing equations presented in the previous section. It is extended by essential discrete stabilization contributions
and completed by the Nitsche-based interface contributions.
Finally, the time discretization by the one-step-$\theta$ scheme is applied to the semi-discrete form, where $\theta$ is the implicit time integration factor of the one-step-$\theta$ scheme and $\Delta t$ the size of each time step.

Starting from here, all quantities in the following are spatially discretized, including the primary unknowns, the test functions in the weak form as well as the domains and interfaces.
However, no index $h$ is added to indicate these discrete quantities, and the double meaning of the continuous quantities is accepted for the sake of simplicity of the presentation.
Below, the expressions $\innerp{*}{*}{\Omega^*}$, $\innerpb{*}{*}{\partial \Omega^*}$, and $\innerpb{*}{*}{\mathcal{F}_{*}} = \sum_{\mathcal{F} \in \mathcal{F}_{*}}\innerpb{*}{*}{\mathcal{F}}$
denote the $\mathcal{L}^2$-inner product integrated in the domain $\Omega^*$, on the boundary or interface $\partial \Omega^*$,
and on all faces of face set $\mathcal{F}_{*}$, respectively.
The jump operator is given by $\jump{*}_\mathcal{F} = \jump{*} = \left( *^+_\mathcal{F}-*^-_\mathcal{F}\right)$, where the ``$+$'' and ``$-$'' sign specifies the evaluation of quantity ``$*$'' in
positive or negative face normal direction.

\subsection{Semi-discrete weak forms of the fluid domain and the poroelastic domain}
The time continuous, semi-discrete weak forms valid in the fluid domain and poroelastic domain are derived from equations 
\eqref{eq:fluidm_eq} - \eqref{eq:fluidc_eq} and \eqref{eq:poroc_eq} - \eqref{eq:mixturem_eq}, respectively.
Herein, required contributions on interface $\fpiinterface$  to account for the interface constraints are omitted and will be presented in Section \ref{sec:num_fpinterface}:
\begin{align}
&\mathcal{W}^F\left[\left(\testvelf, \testpf\right),\left(\velf, \pf\right)\right] = \innerp{\testvelf}{\densityf \partiald{\velf}{t}}{\domainf} + \innerp{\testvelf}{\densityf \velf \cdot \grad \velf}{\domainf} 
-\innerp{\div \testvelf}{\pf}{\domainf}\nonumber\\
&\qquad+\innerp{\epsf (\testvelf)}{2 \viscf \epsf (\velf)}{\domainf}
-\innerp{\testvelf}{\densityf\bodyff}{\domainf} 
-\innerpb{\testvelf}{\tractionfN}{\nboundf}
+\innerp{\testpf}{\div \velf}{\domainf},\label{eq:w_fluid}\\
&\mathcal{W}^P\left[\left(\testvelp, \testdispp, \testpp\right),\left(\velp,\dispp,\pp\right)\right] = \innerp{\testpp}{\partiald{\porosity}{\timep}}{\domainp} + \innerp{\testpp}{\porosity \div \velps}{\domainp}
-\innerp{\grad \testpp}{\porosity \left( \velp - \velps \right) }{\domainp}\nonumber\\
&\qquad+ \innerpb{\testpp}{\porosity \normalp \cdot \left( \velp - \velps \right) }{\fullboundp}
+\innerp{\testvelp}{\densitypf \partiald{\velp}{\timep}}{\domainp} - \innerp{\div \testvelp}{\pp}{\domainp} - \innerp{\testvelp}{\densitypf \partiald{\dispp}{\timep}\grad \velp}{\domainp} \nonumber\\
&\qquad+\innerp{\testvelp}{\viscp  \porosity \permeabp^{-1} \cdot  \velp}{\domainp}
-\innerp{\testvelp}{\viscp \porosity \permeabp^{-1} \cdot \velps}{\domainp} - \innerp{\testvelp}{\densitypf \bodyfpf}{\domainp}
- \innerpb{\testvelp}{\tractionpfN \normalp}{\nboundpf}\nonumber\\
&\qquad+\innerp{\testdispp}{\refavdensityps \accps}{\refdomainp} 
+\innerp{\gradRef \testdispp}{\defgradp\cdot\stresspkp}{\refdomainp}
+\innerp{\testdispp}{\viscp \Jp \porosity^2 \permeabp^{-1} \cdot \velps}{\refdomainp}
-\innerp{\testdispp}{\viscp \Jp  \porosity^2 \permeabp^{-1} \cdot \velp}{\refdomainp}\nonumber\\
&\qquad-\innerp{\testdispp}{\Jp \porosity \left(\defgradp\right)^{-T} \gradRef \pp}{\refdomainp}
-\innerp{\testdispp}{\refavdensityps \refbodyfp}{\refdomainp} 
-\innerpb{\testdispp}{\reftractionpN}{\refnboundp}
.\label{eq:w_poro}
\end{align}
Herein, $\left(\testvelf, \testpf,\testvelp, \testdispp, \testpp\right)$ are the corresponding test functions of the primary unknowns $\left(\velf, \pf,\velp,\dispp,\pp\right)$.
The discrete approximation space is constructed using continuous piece-wise polynomials of order $k$ in each discrete element.
It equals the discrete solution space for pressures $\left(\pf,\pp\right)$ and the single components of the vector-valued solution space for velocities and displacements $\left(\velf ,\velp,\dispp \right)$.
The discrete test space of the functions $\left(\testvelf, \testpf,\testvelp, \testdispp, \testpp\right)$ is equal to their corresponding space for the primal unknowns.
As usual, for Dirichlet type boundary conditions on $\Gamma^{*,D}$, the velocity and displacement solutions $\velfD, \velpnD, \disppD$ are directly incorporated into the discrete solution space,
and the discrete test space is modified accordingly.

To guarantee inf-sup stability for the interpolation of the pressure and the velocity with equal polynomial order $k$, to control convective instabilities, and to ensure mass conservation,
discrete stabilization operators are added.
Consequently, the weakly consistent, symmetric contributions \eqref{eq:CIP_F} and \eqref{eq:CIP_P} of the continuous interior penalty (CIP) method
are added to the weak forms of the fluid equations \eqref{eq:w_fluid} and \eqref{eq:w_poro}.
This method, proposed by \cite{burman2006} for the Oseen problem, includes an additional temporal reactive scaling as presented in \cite{massing2016}, which arises from the 
temporal discretization of the semi-discrete weak form with the one-step-$\theta$ scheme.
\begin{align}
\mathcal{W}^F_{\mathcal{S}}\left[\left(\testvelf, \testpf\right),\left(\velf, \pf\right)\right] =\nonumber\\ \innerpb{\tau_u^F \jump{\grad \testvelf}}{\jump{\grad \velf}}{\mathcal{F}_{\domainf}}
+\innerpb{\tau_p^F \jump{\grad \testpf}}{\jump{\grad \pf}}{\mathcal{F}_{\domainf}}
+\innerpb{\tau_{div}^F \jump{\div \testvelf}}{\jump{\div \velf}}{\mathcal{F}_{\domainf}}\nonumber\\
\tau_u^F = \gamma_u \left(\densityf\norm{\velf}{\infty,\mathcal{F}}\right)^2  h_{\mathcal{F}}^3 \left(\Phi^F_{\mathcal{F}}\right)^{-1}, \quad
\tau_p^F= \gamma_p h_{\mathcal{F}}^3 \left(\Phi^F_{\mathcal{F}}\right)^{-1}, \quad
\tau_{div}^F= \gamma_{div} h_{\mathcal{F}} \Phi^F_{\mathcal{F}},\nonumber\\
\Phi^F_{\mathcal{F}}= \viscf + h_{\mathcal{F}} c_v \densityf\norm{\velf}{\infty,\mathcal{F}} +   h_{\mathcal{F}}^2 c_t \frac{\densityf}{\theta \Delta t}
\label{eq:CIP_F}
\end{align}

\begin{align}
\mathcal{W}^P_{\mathcal{S}}\left[\left(\testvelp, \testpp\right),\left(\velp, \pp\right)\right] = \innerpb{\tau_p^P \jump{\grad \testpp}}{\jump{\grad \pp}}{\mathcal{F}_{\domainp}}
+\innerpb{\tau_{div}^P \jump{\div \testvelp}}{\jump{\div \velp}}{\mathcal{F}_{\domainp}}\nonumber\\
\tau_p^P= \gamma_p h_{\mathcal{F}}^3 \left(  \Phi^P_{\mathcal{F}}\right)^{-1}, \quad
\tau_{div}^P= \gamma_{div} h_{\mathcal{F}} \Phi^P_{\mathcal{F}}, \quad
\Phi^P_{\mathcal{F}}= h_{\mathcal{F}}^2 \left(c_k \frac{\viscf \porosityB}{\initmatpermeabpscalar}  + c_t \frac{\densitypf}{\theta \Delta t} \right)
\label{eq:CIP_P}
\end{align}
Arising from the poroelastic equation \eqref{eq:w_poro}, the reactive contribution is extended by the physical reaction coefficient in \eqref{eq:CIP_P}.
The constants are set to $\gamma_p = 0.05$, $\gamma_u = 0.05$ , $\gamma_{div} = 10^{-3}\gamma_p$, $c_t = 1/12$, $c_k = 1$, $c_v = 1/6$.
The face sets $\mathcal{F}_{\domainf}$ and $\mathcal{F}_{\domainp}$ include all inner faces between elements associated to the discretization of the corresponding domain
(see Figure \ref{fig:discret} (left) ).
The mesh size parameter $h_{\mathcal{F}}$ characterizes the maximum diameter of both elements connected to the face $\mathcal{F}$.
The velocity norm $\norm{\velf}{\infty,\mathcal{F}}$ equals the maximal velocity component within both elements connected to $\mathcal{F}$.
Therefore, all stabilization parameters $\tau_p^F,\tau_u^F,\tau_{div}^F,\tau_p^P,$ and $\tau_{div}^P$ are constant on each face $\mathcal{F}$.

\begin{remark}[Definition of the reactive contribution in the poroelastic stabilization \eqref{eq:CIP_P}]
\label{rem:coerc}
Here and in the presented numerical examples the constant, initial, scalar, reactive contribution ($\porosityB, \initmatpermeabpscalar$) for the reactive stabilization in the poroelastic domain are considered for simplicity.
In the case of large variations in porosity or permeability, the actual computed quantities ($\porosity, \permeabp$) should be considered, however leading to an additional non-linearity in the system.
Also in case of a relevant anisotropy of the permeability tensor $\permeabp$, further adaptions to $\Phi^P_{\mathcal{F}}$ have to be considered 
(for residual-based stabilization, see e.g. \cite{barrios2015}).
\end{remark}
\begin{remark}[Other fluid stabilization techniques]
\label{rem:other_stab}
In general other methods to stabilize these discrete weak forms could be applied, a comparison of different stabilization methods for incompressible flow problems is given in \cite{braack2007}.
\end{remark}

\subsubsection{The CutFEM applied to the fluid domain}
\begin{figure}[t]
\centering
\def\svgwidth{0.49\textwidth}
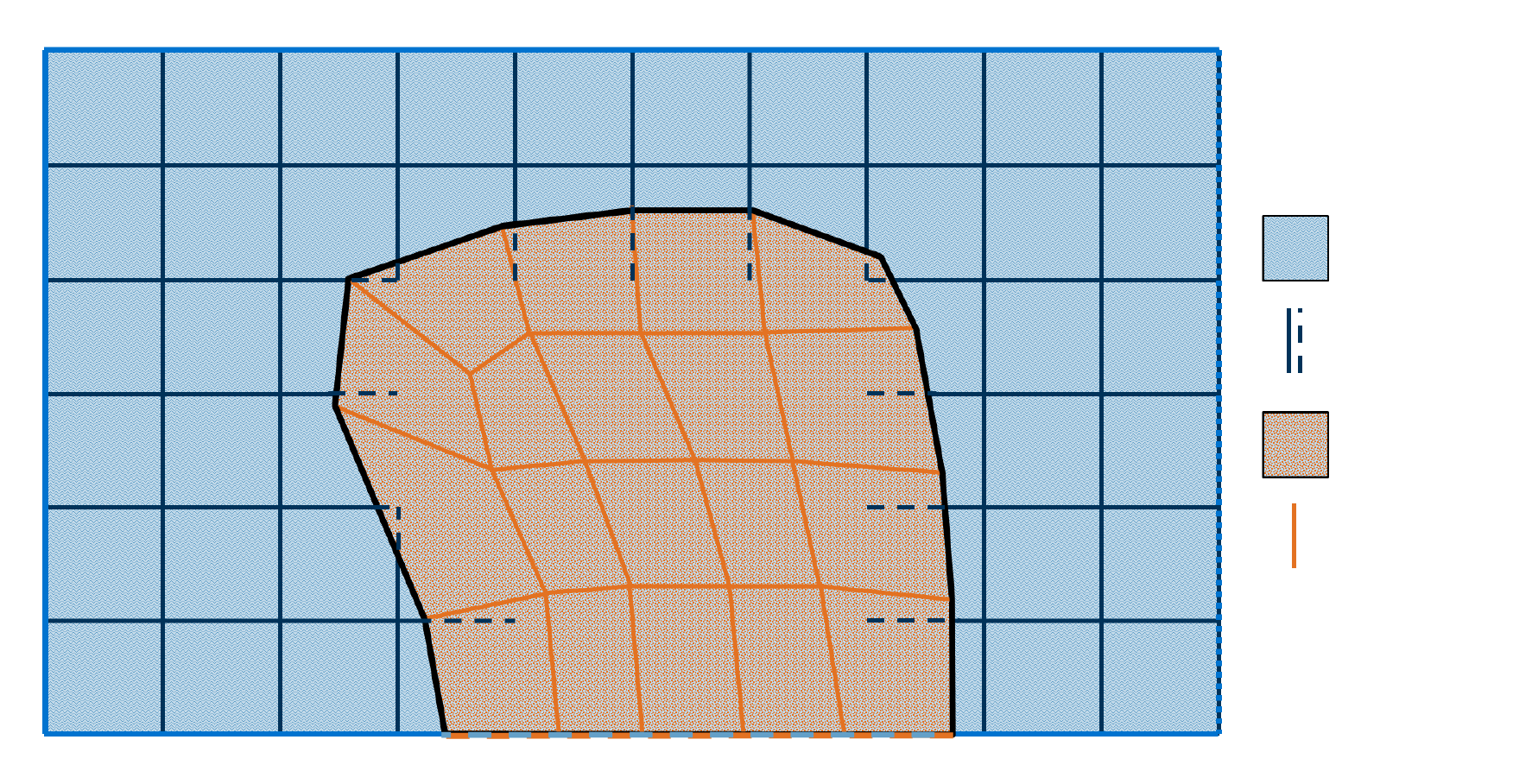
\def\svgwidth{0.49\textwidth}
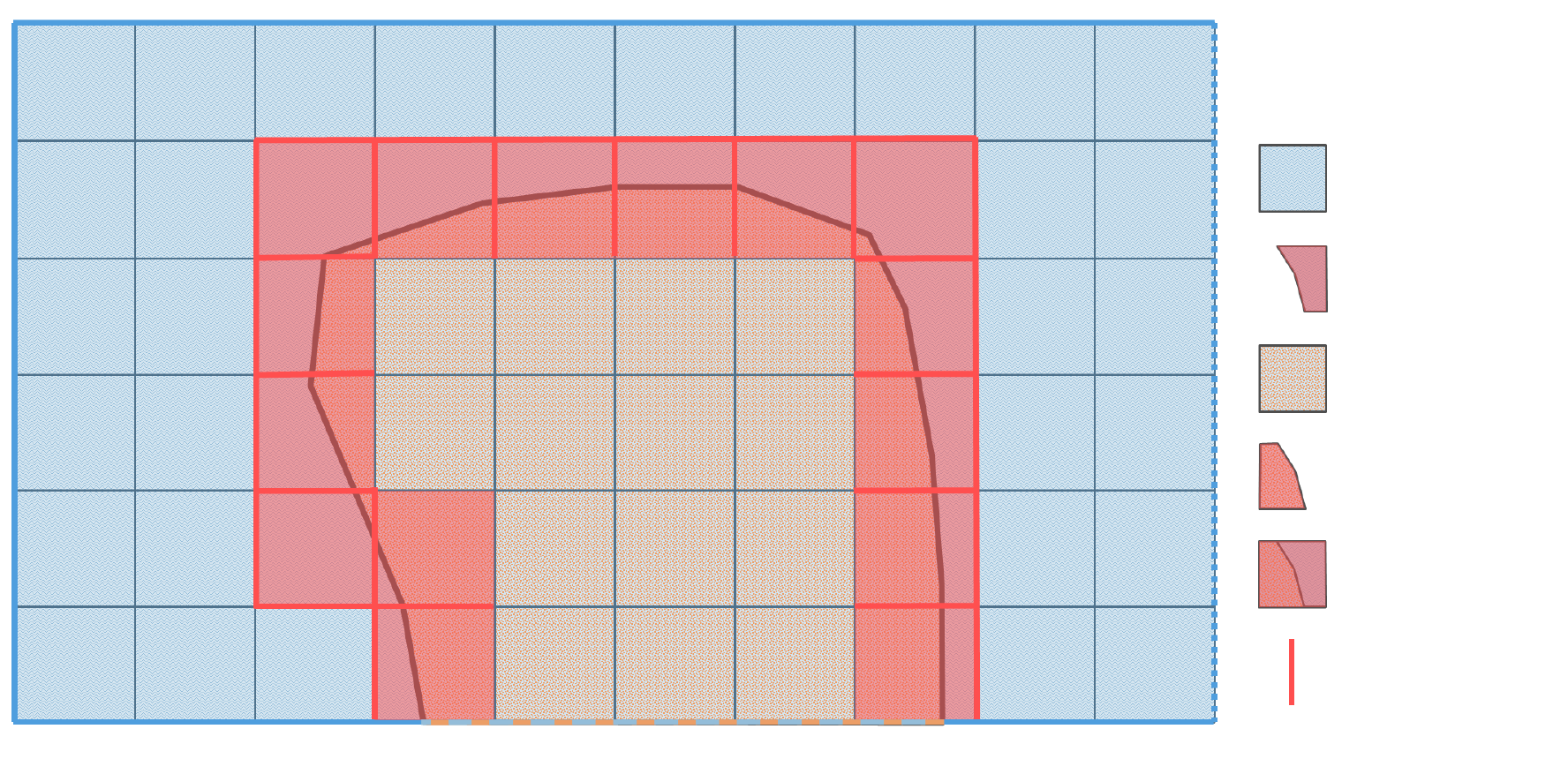
\caption{Boundary and interface fitted discretization $\mathcal{T}^P$ of the poroelastic domain $\domainp$ and 
visualization of the inner face sets ($\mathcal{F}_{\domainf}$ and $\mathcal{F}_{\domainp}$), where the CIP-stabilization is evaluated (left).
A non-fitted discretization $\mathcal{T} = \mathcal{T}^F \cup \mathcal{T}_{\fpiinterface} \cup \mathcal{T}^0$
represents the fluid domain $\domainf$ through a set of elements in $\mathcal{T}^F$ and the physical subdomain $\mathcal{T}^F_{\fpiinterface}$ of the elements in $\mathcal{T}_{\fpiinterface}$.
The nonphysical domain, which equals the poroelastic domain $\domainp$, consists of a set of elements in $\mathcal{T}^0$ 
and the nonphysical subdomain $\mathcal{T}^0_{\fpiinterface}$ of the elements in $\mathcal{T}_{\fpiinterface}$.
For all inner faces $\mathcal{F}_{\fpiinterface}$ of $\mathcal{T}^F \cup \mathcal{T}_{\fpiinterface}$, which are connected to one element in $\mathcal{T}_{\fpiinterface}$, the ``ghost penalty'' stabilization is applied (right).
}
\label{fig:discret}
\end{figure}
As announced in the introduction, to allow for large motion and deformation of the poroelastic domain and to enable topological changes of the fluid domain, 
a non-interface-fitted discretization of the fluid domain is enabled by the CutFEM.
The discretization concepts for the poroelastic and fluid domain are visualized in Figure \ref{fig:discret}.
While for the poroelastic domain $\domainp$ a boundary and interface fitted computation mesh $\mathcal{T}^P$ is applied, 
the fluid domain $\domainf$ is only a subset of the computational mesh $\mathcal{T} = \left(\mathcal{T}^F \cup \mathcal{T}_{\fpiinterface} \cup \mathcal{T}^0\right)$.
With this configuration, the discrete interface location $\fpiinterface$ is given by the outer boundary of the computational mesh of the poroelastic domain $\fpiinterface \subseteq \partial \domainp$.
Making use of this interface geometry, the fluid domain can be recovered as the elements in the set $\mathcal{T}^F$ and the physical subdomain of elements in $\mathcal{F}_{\fpiinterface}$
split by the interface $\fpiinterface$.
By making use of the poroelastic outward unit normal vector $\normalp$, the physical fluid subdomain can be determined.
For this part of the computational mesh, the weak form \eqref{eq:w_fluid} with the $\mathcal{L}^2$-inner products have to be integrated.
This is performed with standard Gaussian quadrature rule for all uncut elements in $\mathcal{T}^F$,
and using the method described in \cite{sudhakar2014}, which utilizes the divergence theorem, for the physical part of the cut elements in $\mathcal{T}_{\fpiinterface}$.
For the remaining, nonphysical subdomain of the computational mesh, no volume integration has to be performed.

Computational nodes and their potential degrees of freedom, which have vanishing shape functions in the physical fluid domain and therefore do not contribute to the weak form, 
are not added to the discrete solution and test function space for the fluid velocity and the pressure.

Due to the independent deformation of the poroelastic domain $\domainp$, its boundary $\fullboundp$, and the independent motion of the interface $\fpiinterface$
relative to the computation fluid mesh, the numerical method has to be robust with respect to any intersection configuration.
In particular, interface positions which lead to a very small contribution of single discrete degrees of freedom to the weak form \eqref{eq:w_fluid}, have
to be treated properly.
If not done so, such configurations can lead to an ill-conditioned system of equations due to the decreased influence of single degrees of freedom.
Furthermore, a loss of discrete stability, due to the Nitsche-based approach for the imposition of interface conditions (see Section \ref{sec:num_fpinterface}), 
which relies on sufficient control of the discrete fluid solution in the fluid domain $\domainf$ extended by the 
``ghost'' domain $\mathcal{T}^0_{\fpiinterface}$, could occur.

Additional ``ghost penalty'' stabilization allows for the extension of the physical solution of the fluid domain into the ``ghost'' domain.
This can be achieved by adding an additional symmetric and weakly consistent ``ghost'' penalty operator, similar to the CIP operators in \eqref{eq:CIP_F}, to the weak form \eqref{eq:w_fluid}.
Such a stabilization technique was first presented in \cite{burman2010} and analyzed in \cite{burman2012} for the Poisson's problem, extended to the Stokes problem in \cite{massing2014},
and to the Oseen problem in \cite{massing2016}.
The additional ``ghost penalty'' operator applied in the numerical examples is:
\begin{align}
\mathcal{W}^F_{\mathcal{G}}\left[\left(\testvelf, \testpf\right),\left(\velf, \pf\right)\right] =
\sum_{1\leq j \leq k}\innerpb{\tau_p^{GP,j}\jump{\derivn{j}{\testpf}}}{\jump{\derivn{j}{\pf}}}{\mathcal{F}_{\fpiinterface}} \nonumber\\
+\innerpb{\tau_{div}^{GP,1} \jump{\div \testvelf}}{\jump{\div \velf}}{\mathcal{F}_{\fpiinterface}}
+\sum_{1\leq j \leq k}\innerpb{\tau_u^{GP,j}\jump{\derivn{j}{\testvelf}}}{\jump{\derivn{j}{\velf}}}{\mathcal{F}_{\fpiinterface}},\nonumber\\
\tau_p^{GP,1} = \tau_p^F,\quad 
\tau_p^{GP,2} = 0.05 h_{\mathcal{F}}^2 \tau_p^{GP,1},\nonumber\\
\tau_u^{GP,1} = \tau_u^{F} + \gamma_{\nu}^{GP} h_{\mathcal{F}} \viscf + \gamma_{t}^{GP} h_{\mathcal{F}}^3 \frac{\densityf}{\theta \Delta t},\quad
\tau_{div}^{GP,1}= \tau_{div}^{F}, \quad
\tau_u^{GP,2} = 0.05 h_{\mathcal{F}}^2 \left(\tau_u^{GP,1} + \tau_{div}^{GP,1} \right)
\label{eq:GP}.
\end{align}
Herein, $\derivn{j}{*}$ denotes the $j^{\text{th}}$ derivative of $*$ in face ${\mathcal{F}}$ normal direction.
In principle, these operators penalize deviations from an extended smooth solution of an extension of the discrete fluid solution in $\domainf$ into the ``ghost'' domain $\mathcal{T}^0_{\fpiinterface}$.
Contrary to the CIP operators \eqref{eq:CIP_F}, also jumps of higher order derivatives of the pressure and the velocity are penalized.
Furthermore, additional viscous and reactive parts compared to the CIP stabilization are required, which are included in the parameter $\tau_u^{GP,j}$.
As the numerical examples are discretized by bi-linear quadrilateral elements, and therefore derivatives for $k>2$ vanish, stabilization parameters for $k>2$ are not considered.
The parameters for the derivatives of first order $\tau_p^{GP,1},\tau_u^{GP,1}, \tau_{div}^{GP,1}$ are chosen equal to the CIP parameter, whereas the 
additional parameters in the viscous and reactive scaling are set to $\gamma_{\nu}^{GP} = 0.1$ and $\gamma_{t}^{GP} = 0.001$.
Stabilization parameters for $2^{\text{nd}}$ order derivatives are based on the first order parameters, but include an additional weighting and scaling with the mesh size parameter $ h_{\mathcal{F}}$.
\begin{remark}[Choice of ``ghost penalty'' operators]
\label{rem:ghostp}
In order to reduce the variants of different ``ghost penalty'' operators, the velocity-based contribution related to the convective, viscous and reactive part are considered with the same operator
$\innerpb{\jump{\derivn{j}{\testvelf}}}{\jump{\derivn{j}{\velf}}}{\mathcal{F}_{\fpiinterface}}$ in stabilization parameter $\tau_u^{GP,1}$.
For derivatives of order $j>1$, also the incompressibility contribution is combined into this operator, as can be seen in $\tau_u^{GP,2}$ \cite{massing2016}.
\end{remark}

\subsection{The Nitsche-based method on the common interface between fluid domain and poroelastic domain}
\label{sec:num_fpinterface}
Up until now, the interface conditions between the fluid and the poroelastic domain on the interface $\fpiinterface$ were not incorporated in the weak forms \eqref{eq:w_fluid}, \eqref{eq:w_poro}
and the additional stabilization contributions \eqref{eq:CIP_F}, \eqref{eq:CIP_P}, and \eqref{eq:GP}.
Due to the successful application of the CutFEM and the weak imposition of boundary/interface conditions by Nitsche-based methods 
(see e.g. \cite{burman2014, schott2015, burman2015cutfem, schott2016, schott2017}), this combination is also selected here.
Therefore, an additional integration on the discrete interface $\fpiinterface$, which is given by the boundary of the deformed poroelastic domain $\fullboundp$, has to be performed.
Due to the different constraints in normal \eqref{eq:fpi_massb} and tangential \eqref{eq:fpi_bj} direction on the interface $\fpiinterface$,
separate treatment of both directions can be applied.
\subsubsection{Normal direction}
In normal direction, the contributions to the weak form for Nitsche's method, similar to the methods in \cite{dangelo2011,bukavc2015}, including the consistency, an adjoint-consistency and a penalty term, 
reads as follows:
\begin{align}
\mathcal{W}^{FP,n}\left[\left(\testvelf, \testpf,\testvelp, \testdispp, \testpp\right),\left(\velf, \pf,\velp, \dispp, \pp\right)\right]=\nonumber\\
\innerpb{\testvelp+\testdispp-\testvelf}{\stressf  \cdot \normalf \cdot \Pnormal}{\fpiinterface}
- \innerpb{\testvelp}{\jumpvalsn \normal \cdot \Pnormal}{\fpiinterface} - \innerpb{\testdispp}{\jumpvals \cdot \Pnormal}{\fpiinterface}  \nonumber\\
-\innerpb{\testpf \cdot \normalf + \adjointsign 2 \viscf \epsf (\testvelf) \cdot \normalf }{\left[{\velf} {- \velps} - \porosity \left( {\velp} {- \velps} \right)  - \jumpvaln \right]\cdot \Pnormal}{\fpiinterface}\nonumber\\
 +\frac{\phi^F_{\Gamma}}{\gamma_n h_{\Gamma}}\innerpb{\testvelf - \testvelp - \testdispp}{\left[{\velf} - {\velps} - \porosity \left( {\velp} - {\velps} \right) - \jumpvaln \right]\cdot \Pnormal}{\fpiinterface}\nonumber,\\
 \phi^F_{\Gamma}  = \viscf + h_{\Gamma} c_{v,\Gamma} \densityf \norm{\velf}{\infty,\Gamma} + h_{\Gamma}^2  c_{t,\Gamma} \frac{\densityf}{\theta \Delta t}.
 \label{eq:w_fpi_n}
\end{align}
This leads solely to contributions in normal direction of the interface $\fpiinterface$, as all terms are projected in the normal direction by the projection matrix $\Pnormal$.
The terms in second line originate from the integration by parts in the derivation of the weak form \eqref{eq:w_fluid} 
and are therefore called consistency terms. Herein, the interface stress is chosen to be represented by the fluid stress.
The last two terms in this line vanish for zero stress jumps in \eqref{eq:fpi_dyneq} and \eqref{eq:fpi_dyneqp}, which, in general, is the physical relevant case.
All terms in the following lines are added in order to obtain a stable and convergent discrete numerical scheme and to enforce the kinematic constrain \eqref{eq:fpi_massb}. 
Consistency is guaranteed due to the included kinematic constraint \eqref{eq:fpi_massb}, wherefore these contributions vanish in the case of an exact constraint fulfillment.
An additional adjoint-inconsistent pressure term in line three balances the pressure contribution of fluid stress in the consistency boundary integrals of line two.
For the viscous adjoint-consistency term, an adjoint-consistent $\adjointsign = 1$ or adjoint-inconsistent $\adjointsign = -1$ variant can be chosen.
Finally, in the last line, a penalty term guarantees the stability of the numerical method, if a sufficiently small constant $\gamma_n$ is chosen.
The dependence of the resulting error norms of the numerical scheme on the penalty parameter $\gamma_n$ is analyzed in the numerical example presented in Section \ref{sec:varpen} for both choices of $\adjointsign$.
The additive scaling of $\phi^F_{\Gamma}$ aims for a stable numerical scheme in all regimes of the fluid equations, as can be found in \cite{massing2016}.
Herein, $\norm{\velf}{\infty,\Gamma}$ is the maximal fluid velocity component at the current point in space on the interface $\fpiinterface$. 
The convective and reactive constants are specified to: $c_{v,\Gamma}=1/6, c_{t,\Gamma} = 1/12$. The mesh size parameter $h_{\Gamma}$ is computed by 
the ratio of element volume and the part of the area of interface $\fpiinterface$ at the local intersected element in $\mathcal{T}_{\fpiinterface}$.

\begin{remark}[Alternative scaling of the penalty terms]
\label{rem:coerc}
In case the partial integration of the porosity gradient is not performed in the derivation of the weak form of the balance of linear momentum of the poroelastic mixture \eqref{eq:mixturem_eq} 
(weak form presented in \cite{Vuong2015}, basically $\grad \porosity \longrightarrow \grad \pp$ is not performed),
a statement on the scaling of the penalty terms can be acquired.
For this formulation, the consistency terms of the weak form \eqref{eq:mixturem_eq} include only the averaged structural stress $\stress^{P^S}=\stressp + \porosity \pp \identity$, 
which results in the normal dynamic equilibrium (replacing constraint \eqref{eq:fpi_dyneq} in normal direction):
\begin{align}
 0 = \left(1-\porosity\right)\normal \cdot \stressf \cdot \normal - \normal \cdot \stress^{P^S} \cdot \normal \quad &&\text{on} \; \fpiinterface \timesfulltime.
\end{align}
With an additional multiplication of the weak poroelastic fluid equation with the porosity $\porosity$
(to result in a symmetric physical reactive contribution in domain $\domainp$ of the porous fluid phase and the porous solid phase)
, the resulting consistency terms in normal direction for an interface stress representation by the fluid stress are: 
\begin{align}
+\innerpb{\porosity\testvelp+\left(1-\porosity\right)\testdispp-\testvelf}{\stressf  \cdot \normalf \cdot \Pnormal}{\fpiinterface}.
\label{eq:const_alt}
\end{align}
Analogous to coercivity analyses for the Nitsche's method for the weak imposition of boundary conditions (e.g. \cite{burman2012}), it can be stated that the interface semi-norm specified by the
left test-function part $\left(\porosity\testvelp+\left(1-\porosity\right)\testdispp-\testvelf\right)$ of \eqref{eq:const_alt}, should be balanced
by an equal symmetric contribution of the penalty terms.
Due to the same structure of this consistency test-function part and the kinematic constraint \eqref{eq:fpi_massb}, an additional, 
multiplicative scaling of the penalty term including test-function $\testdispp$ with $\left(1-\porosity\right)$ fulfills these requirements. 
A significant influence of this modification is expected for a vanishing fluid phase close to the limit case $\porosity = 1$, which would result in a vanishing penalty contribution.
Nevertheless, due to the slightly different weak form applied here, this additional scaling is not applied.
\end{remark}

\subsubsection{Tangential direction}
For the weak imposition of the tangential constraint \eqref{eq:fpi_bj}, two different methods are presented and compared in Sections \ref{sec:varperm} and \ref{sec:varbj}. 
\paragraph{Substitution approach}
The first method is presented for the validation and comparison of the subsequent Nitsche-based approach. 
Herein, the tangential interface traction is substituted, making use of relation \eqref{eq:fpi_bj}, by a kinematic relation. 
Similar ``Substitution'' methods were applied in \cite{badia2009, bukavc2015,ambartsumyan2017} to incorporate the Beavers-Joseph (-Saffmann) condition.
To enforce the tangential constraint by the ``Substitution'' method, the following contribution is added to the weak form:
\begin{align}
\mathcal{W}^{FP,t,Sub}\left[\left(\testvelf, \testpf,\testvelp, \testdispp, \testpp\right),\left(\velf, \pf,\velp, \dispp, \pp\right)\right]=\nonumber\\
\innerpb{\testvelf-\testdispp}{\frac{1}{\sliplengh}\left[{\velf} {- \velps} - \BJfac\porosity \left( {\velp} {- \velps} \right) - \jumpvalt \right]\cdot \Ptangent}{\fpiinterface}
- \innerpb{\testdispp}{\jumpvals \cdot \Ptangent}{\fpiinterface}.
 \label{eq:w_fpi_t_sub}
\end{align}
Herein, the tangential boundary integrals, arising from integration by parts when deriving \eqref{eq:w_fluid} and \eqref{eq:w_poro}, are substituted by the terms in \eqref{eq:w_fpi_t_sub}.
The principal structure and sign of this term equals a penalty term and therefore can be categorized as a positive contribution in a coercivity analysis.
Due to the division by $\sliplengh$ in the prefactor, for decreasing $\sliplengh$ these terms starts to dominate the overall problem formulation. 
This worsens the conditioning of the discrete system of equations to solve and leads to increasing error norms, as analyzed in Section \ref{sec:varperm}. 
The second term arises from the representation of the fluid interface stress by the kinematic constraint in \eqref{eq:fpi_bj} and due to the non vanishing stress jump $\jumpvals$
in the dynamic equilibrium \eqref{eq:fpi_dyneq}.

\paragraph{Nitsche-based approach}
The second presented method, which does not suffer from this conditioning problem for small $\kappa$, is based on the Nitsche's method for general boundary conditions first presented in \cite{Juntunen2009}
for the Poisson problem. The extension to the Oseen problem is given in \cite{winter2017}.
The tangential interface terms when applying this approach are: 
\begin{align}
\mathcal{W}^{FP,t,Nit}\left[\left(\testvelf, \testpf,\testvelp, \testdispp, \testpp\right),\left(\velf, \pf,\velp, \dispp, \pp\right)\right]=\nonumber\\
 \innerpb{\testdispp-\testvelf}{ \stressf  \cdot \normalf \cdot \Ptangent}{\fpiinterface} - \innerpb{\testdispp}{\jumpvals \cdot \Ptangent}{\fpiinterface}\nonumber\\
+\adjointsign\frac{\gamma_t h_{\Gamma}}{\kappa\viscf+\gamma_t h_{\Gamma}}\innerpb{-2 \viscf \epsf (\testvelf) \cdot \normalf }
 {\left[{\velf} {- \velps} - \BJfac\porosity \left( {\velp} {- \velps} \right) + {\kappa \stressf \cdot \normalf} - \jumpvalt \right]\cdot \Ptangent}{\fpiinterface}\nonumber\\
 +\frac{\viscf}{\kappa\viscf+\gamma_t h_{\Gamma}}\innerpb{\testvelf - \testdispp}
 {\left[{\velf} - {\velps} - \BJfac\porosity \left( {\velp} - {\velps} \right) + {\kappa \stressf \cdot \normalf}  - \jumpvalt \right]\cdot \Ptangent}{\fpiinterface}.
 \label{eq:w_fpi_t_nit}
\end{align}
Similar to the presented Nitsche method in normal direction, the interface stress is represented by the fluid stress.
Therefore, the consistency integrals in tangential direction, which arise from the partial integration in the derivation of the weak forms \eqref{eq:w_fluid} and \eqref{eq:w_poro}, result in the contributions of line two.
Again, a nonphysical contribution in case $\jumpvals \neq \zerovec$ arises from the interface stress representation as fluid stress.
The following terms in the last two lines are a consistent addition due to the inclusion of the kinematic constraint \eqref{eq:fpi_bj}, for what reason these contributions vanish in the case of the exact solution.
In line three, an  adjoint-consistent ($\adjointsign = 1$) or an adjoint-inconsistent ($\adjointsign = -1$) term is added.
As can be seen from the occurring prefactor, they balance the consistency integrals in line two and the consistency like contribution 
$\left(\viscf\left(\kappa\viscf+\gamma_t h_{\Gamma}\right)^{-1}\innerpb{\testvelf - \testdispp}
 {{\kappa \stressf \cdot \normalf}\cdot \Ptangent}{\fpiinterface}\right)$
in the following penalty terms.
Finally, the penalty term in the last line aims for the stability of the numerical scheme for a sufficiently small penalty parameter $\gamma_t$.
The prefactor of the penalty term also results from the two contributions of the consistency boundary integrals and the consistency like contribution in the penalty terms itself.
In Section \ref{sec:varpen}, the required penalty parameter $\gamma_t$ for the adjoint-consistent and adjoint-inconsistent variant is analyzed.

\begin{remark}[Nonexistent tangential poroelastic fluid penalty contribution]
\label{rem:tang_pen}
By comparison of the Nitsche contributions in normal $\mathcal{W}^{FP,n}$ and tangential $\mathcal{W}^{FP,t,Nit}$ direction, it can be observed that the tangential penalty contribution of the poroelastic fluid equation 
is nonexistent. 
Considering the analogy to the coercivity analyses for general boundary conditions (see \cite{Juntunen2009}),
the interface semi-norm specified by the test-functions in the consistency terms $\left(\testdispp-\testvelf\right)$ of \eqref{eq:w_fpi_t_nit} has to be balanced by a
symmetric penalty contribution.
By analyzing the Beaver-Joseph-Saffmann condition case ($\BJfac=0$), the structure of the kinematic constraint \eqref{eq:fpi_bj} equals this contribution and therefore
applying the same jump of test-functions $\left(\testdispp-\testvelf\right)$ for the penalty terms results in the desired result.
Also, from a modeling point of view, this ``missing'' tangential penalty is reasonable, as a Darcy fluid cannot compensate a tangential boundary stress.
For the Beavers-Joseph condition, this argumentation does not directly hold anymore.
Still, it can be clearly seen from computed results (not presented here) that a tangential penalty contribution tested on the Darcy-based equation should not be added to the weak form $\mathcal{W}^{FP,t,Nit}$.
For example, one recognizes oscillations of the velocity $\velp$  in the porous domain close to the interface $\fpiinterface$ even for a simple configuration such as the parallel flow of viscous and porous fluid
when adding this penalty contribution.
\end{remark}

\begin{remark}[Combination of projected consistency terms]
\label{rem:cond_terms}
In the case when contributions $\mathcal{W}^{FP,n}$ and $\mathcal{W}^{FP,t,Nit}$ are combined, no projection of the fluid- and poroelastic consistency terms is required as they can be directly combined,
and thus the implementation can be simplified:
\begin{align}
\innerpb{\testdispp-\testvelf}{ \stressf  \cdot \normalf}{\fpiinterface} - \innerpb{\testdispp}{\jumpvals}{\fpiinterface} = 
\innerpb{\testdispp-\testvelf}{\stressf  \cdot \normalf \cdot \Pnormal}{\fpiinterface}
- \innerpb{\testdispp}{\jumpvals \cdot \Pnormal}{\fpiinterface} \nonumber\\
+\innerpb{\testdispp-\testvelf}{ \stressf  \cdot \normalf \cdot \Ptangent}{\fpiinterface} - \innerpb{\testdispp}{\jumpvals \cdot \Ptangent}{\fpiinterface}.
\label{eq:fpi_simpl}
\end{align}
Due to the nonexistent viscosity and therefore tangential contribution of the Darcy-based fluid model in domain $\domainp$, 
the projection of the consistency contributions in the poroelastic fluid equation 
$\innerpb{\testvelp}{\stressf  \cdot \normalf \cdot \Pnormal}{\fpiinterface}- \innerpb{\testvelp}{\jumpvalsn \normal \cdot \Pnormal}{\fpiinterface} $
has to remain untouched in $\mathcal{W}^{FP,n}$.
\end{remark}

\subsection{Time discretization and final system}
\label{sec:sec_fin}
By summing up all contributions discussed beforehand, the final semi-discrete weak form for the coupled fluid-poroelastic interaction problem is set up as follows.
Find $\left(\velf, \pf ,\velp,\dispp,\pp\right)$, such that for all $\left(\testvelf, \testpf,\testvelp, \testdispp, \testpp\right)$:
\begin{align}
\mathcal{W}^F\left[\left(\testvelf, \testpf\right),\left(\velf, \pf\right)\right] + \mathcal{W}^F_{\mathcal{S}}\left[\left(\testvelf, \testpf\right),\left(\velf, \pf\right)\right]
+\mathcal{W}^F_{\mathcal{G}}\left[\left(\testvelf, \testpf\right),\left(\velf, \pf\right)\right]\nonumber\\
+\mathcal{W}^P\left[\left(\testvelp, \testdispp, \testpp\right),\left(\velp,\dispp,\pp\right)\right] + \mathcal{W}^P_{\mathcal{S}}\left[\left(\testvelp, \testpp\right),\left(\velp, \pp\right)\right]\nonumber\\
+\mathcal{W}^{FP,n}\left[\left(\testvelf, \testpf,\testvelp, \testdispp, \testpp\right),\left(\velf, \pf,\velp, \dispp, \pp\right)\right]\nonumber\\
+\mathcal{W}^{FP,t,*}\left[\left(\testvelf, \testpf,\testvelp, \testdispp, \testpp\right),\left(\velf, \pf,\velp, \dispp, \pp\right)\right]= 0.
\end{align}
Therein the ``$*$''-symbol, specifies either ``$Sub$'' for the ``Substitution'' method or ``$Nit$'' for the Nitsche-based variant to enforce the tangential interface condition.

As a next step, time discretization by the one-step-$\theta$ scheme is applied and the final nonlinear system is solved with a Newton-Raphson like method.
As this is already done in a similar manner for the CutFEM fluid-structure interaction, the reader is referred to \cite{schott2017}.
Details are shown there and the references mentioned therein.
One aspect that should still be highlighted is that, due to the motion of interface $\fpiinterface$, 
different fluid velocity and pressure solution spaces may occur during the solution procedure.
To enable the time integration and the nonlinear solution procedure, a reconstruction of the solution of a previous step has to be performed.
In this context, the previous step indicates either the previous timestep or the previous iteration step in the Newton-Raphson like nonlinear solution procedure.
This solution reconstruction and the algorithmic details on the nonlinear solution strategy can be found in \cite{schott2017} and are therefore not presented here.

\section{Numerical example: Computational analysis of the formulation}
\label{sec:ex1}
In this section, the presented formulation will be analyzed numerically.
The aim is to verify that the expected properties, already analyzed in literature for the simplified variants of this problems,
namely for the Stokes/Darcy coupling, the Stokes/Biot-system coupling, the CutFEM applied on the Oseen equations,
the Nitsche's method for general Navier boundary condition, and the fluid-structure interaction are still valid for this problem configuration.
Therefore, a problem setup that results in a known analytic solution is constructed by the method of manufactured solutions.
Independent of the choice of the boundary and interface locations, this known solution should be valid in the respective domain.
Comparison of the computed solutions with the analytic solution allows one to analyze different aspects of the formulation.
\subsection{Analytic solution and problem setup}
The analytic solution is chosen as follows:
\label{ssec:analytsol}
\begin{align}
 \velfa(\coord,t)&= 
 \begin{bmatrix} - A^F\,\mycos{B\, \pi\, x} \mysin{B \, \pi\, y} g_u(t) \\
 A^F\,\mysin{B\, \pi\, x} \mycos{B \, \pi\, y} g_u(t) \end{bmatrix} ,
  \label{eq:sol_velf}\\
 \pfa(\coord,t)&= 
  -\frac{1}{4}\,\left(\mycos{2\,B\, \pi\, x} + \mycos{2\,B\, \pi\, y}\right) \densityf g_p(t),
  \label{eq:sol_p}\\
 \velpa(\coord,t)&= 
 \begin{bmatrix} - A^P\,\mycos{B\, \pi\, x} \mysin{B \, \pi\, y} g_u(t) \\
 A^P\,\mysin{B\, \pi\, x} \mycos{B \, \pi\, y} g_u(t)\end{bmatrix} ,
 \label{eq:sol_velp}\\
 \ppa(\coord,t)&= 
  -\frac{1}{4}\,\left(\mycos{2\,B\, \pi\, x} + \mycos{2\,B\, \pi\, y}\right) \densityf g_p(t),
 \label{eq:sol_pp}\\
 \disppa(\refcoord,t)&= 
 \begin{bmatrix} - A^{P^S}\,\mycos{B\, \pi\, X} \mysin{B \, \pi\, Y} \frac{\left( 1 - g_u(t)\right)}{-2\, C^2 \,\pi^2 \,\viscf\left(\densityf\right)^{-1}} \\
A^{P^S}\,\mysin{B\, \pi\, X} \mycos{B \, \pi\, Y} \frac{\left( 1 - g_u(t)\right)}{-2\, C^2 \,\pi^2 \,\viscf\left(\densityf\right)^{-1}} \end{bmatrix} ,
\label{eq:sol_up}
\end{align}
$\text{with the time dependent functions:} \quad
g_u(t) = e^{-2\, C^2 \,\pi^2 \,\viscf \left(\densityf\right)^{-1}\, t}
\quad \text{and}\quad
g_p(t) = e^{-4\, C^2 \,\pi^2 \,\viscf\left(\densityf\right)^{-1}\, t}.$

Herein, the analytic velocity and pressure solution in the fluid domain is denoted by $\velfa$ and $\pfa$.
The analytic velocity, pressure, and displacement solution  in the poroelastic domain is denoted by $\velpa$, $\ppa$, and $\disppa$, respectively.
The components of the two dimensional position vector $\coord = \left[x,y\right]^T$ in current configuration are specified as $x$ and $y$ and
the components of the material position vector $\refcoord = \left[X,Y\right]^T$ as $X$ and $Y$.
The solution is chosen in a way that fulfills the balance of fluid mass in \eqref{eq:fluidc_eq} and \eqref{eq:poroc_eq} in the case of assuming constant porosity $\porosity$.
Additionally, by specifying the solution amplitudes to $A^F=0.1$, $A^P=0.21$, and $A^{P^S}=-0.01$, the mass-balance \eqref{eq:fpi_massb} 
and the kinematic contribution part of the BJ or BJS condition \eqref{eq:fpi_bj} on the interface $\fpiinterface$ are fulfilled without additional contributions
$(\jumpvaln, \jumpvalt)$
at the initial point in time due to the vanishing deformation.
The space constant $B$, and the time constant $C$ influence the spatial and temporal gradients of the given solution. 
As the focus in the following should be on the spatial errors, $B=1.0$ is chosen to be larger than $C=0.01$.

The dynamic viscosity of the fluid is specified to $\viscf = 1.0$. 
To characterize the porous flow resistance, an isotropic material permeability $\matpermeabp = \matpermeabpscalar \cdot \identity$, 
with scalar value $\matpermeabpscalar = 0.1$, is prescribed.
The porosity is specified to be constant in space and time and set to $\porosity = 0.5$ to fulfill equation \eqref{eq:poroc_eq} by the analytic solution.
Therefore, equation \eqref{eq:poro_constitutive2} is not evaluated, as the strain energy function $\strainenergy{P,vol}(\Jp (1-\porosity))$ is defined implicitly.
No penalty contribution of the strain energy function is considered, i.e. $\strainenergy{P,pen}(\strainglp, \Jp (1-\porosity)) = 0$.
The macroscopic deformation of the solid phase is given by a Neo-Hookean material model with the hyperelastic strain energy function with Young's modulus $E = 1000$ and Poisson ratio $\nu = 0.3$:
\begin{align}
\label{ex1:strainenergysp}
\strainenergy{P,skel} = c \left[\text{tr}\left(\left(\defgradp\right)^T\cdot\defgradp\right)-3\right]+\frac{c}{\beta}\left(\left(\Jp\right)^{-2 \beta}-1\right), \quad
c = \frac{E}{4(1+\nu)}, \quad \beta = \frac{\nu}{1-2 \nu}.
\end{align}
The averaged initial density of the solid phase is chosen to be equal to the fluid density $\refdensitys = \densityf = 1.0$.
If not denoted otherwise, the Beavers-Joseph coefficient on the interface $\fpiinterface$ is set to $\bjcoeff = 1.0$.
In order to fulfill the balance of momentum, defined by equations \eqref{eq:fluidm_eq}, \eqref{eq:porom_eq}, and \eqref{eq:mixturem_eq},
with the analytic solutions \eqref{eq:sol_velf}--\eqref{eq:sol_up}, the following body force is applied:
\begin{align}
\densityf \bodyff &= \densityf \partiald{\velfa}{\timef} +  \densityf \velfa \cdot \grad \velfa - \div\stressfa,\quad 
\stressfa = -\pfa \identity + 2 \viscf \epsf(\velfa),
\label{eq:ex1_bf_f}\\
\densitypf \bodyfpf &= \left.\densitypf \partiald{\velpa}{\timep}\right|_{\refcoordp} 
- \densityf\velpsa\cdot\grad\velpa +\grad \ppa +\viscp  \porosityB \permeabpa^{-1} \cdot \left(\velpa-\velpsa\right),
\label{eq:ex1_bf_pf}\\
\refavdensityps \refbodyfp &= \refavdensityps \accpsa 
- \divRef \, \left( \defgradpa \cdot 
\stresspkpa
\right)
+ \left(1-\porosityB  \right) \Jpa \grad \ppa 
- \viscp \Jpa \porosityB^2  \permeabpa^{-1} \cdot \left( \velpa - \velpsa \right),
\label{eq:ex1_bf_ps}
\end{align}
where the analytic deformation gradient is defined as $ \defgradpa = \identity + \partiald{\disppa}{\refcoordp}$, the determinate as $\Jpa=\det{\defgradpa}$, 
the analytic homogenized second Piola-Kirchhoff stress tensor as $\stresspkpa = 2 c \identity - 2 c \Jpa^{-2 \beta} \defgradpa^{-1}\cdot\defgradpa^{-T}$, 
the analytic Cauchy stress tensor as $\stresspa = \Jpa^{-1} \defgradpa \cdot \stresspkpa \cdot \defgradpa^T$, 
and the analytic inverse spatial permeability as $\permeabpa^{-1}=\Jp \defgradpa^{-T} \cdot \matpermeabp^{-1} \cdot \left(\defgradpa\right)^{-1}$.
The analytic solution fulfills the interface conditions \eqref{eq:fpi_dyneq} - \eqref{eq:fpi_bj}, if the constraint-jumps and traction-jumps $\jumpvals, \jumpvalsn, \jumpvaln, \jumpvalt$ are defined by the
analytic solution as:
\begin{align}
\jumpvals&=\stressfa \cdot \normalf - \stresspa \cdot \normalf,
\label{eq:ex1_tj_s}\\
\jumpvalsn&=\normalf \cdot  \stressfa \cdot \normalf + \ppa,
\label{eq:ex1_tj_sn}\\
\jumpvaln&=\velfa - \velpsa - \porosity \left( \velpa - \velpsa \right),
\label{eq:ex1_cj_n}\\
\jumpvalt&=\velfa - \velpsa - \BJfac \porosity \left( \velpa - \velpsa \right) + \sliplengh \normalf\cdot \stressfa.
\label{eq:ex1_cj_t}
\end{align}

\begin{remark}[Evaluation of the body force \eqref{eq:ex1_bf_f}-\eqref{eq:ex1_bf_ps}, the traction-jump \eqref{eq:ex1_tj_s}-\eqref{eq:ex1_tj_sn}, and the constraint-jump contributions \eqref{eq:ex1_cj_n}-\eqref{eq:ex1_cj_t}]
\label{rem:evaluation_bf}
All body force contributions \eqref{eq:ex1_bf_f}-\eqref{eq:ex1_bf_ps} are evaluated at the computational nodes and re-interpolated with the standard shape functions of the corresponding finite elements.
Compared to a direct volume integration, the additional error does not deteriorate the error convergence order of the subsequent computation.
Contrary to this, the traction-jump and constraint-jump contributions are considered directly for the integration on the interface $\fpiinterface$ at every point in space.
Hereby, the material position vector $\refcoord$ is computed by re-interpolation of the initial nodal coordinates.
All volume contributions \eqref{eq:ex1_bf_f}-\eqref{eq:ex1_bf_ps} depend solely on the point in space and time for the given analytic solution, 
whereas the interface contributions \eqref{eq:ex1_tj_s}-\eqref{eq:ex1_cj_t} additionally depend on the normal vector $\normalf$. 
Therefore, contributions \eqref{eq:ex1_bf_f}-\eqref{eq:ex1_cj_t} were computed and simplified symbolically by Maple as time- and space-dependent functions in advance as far as possible.
The evaluation of these functions and the multiplication of the components by the discrete normal vector in each numerical integration point for the interface contributions is performed during the computation.
\end{remark}

\begin{figure}[tb]
\centering
\def\svgwidth{0.3\textwidth}
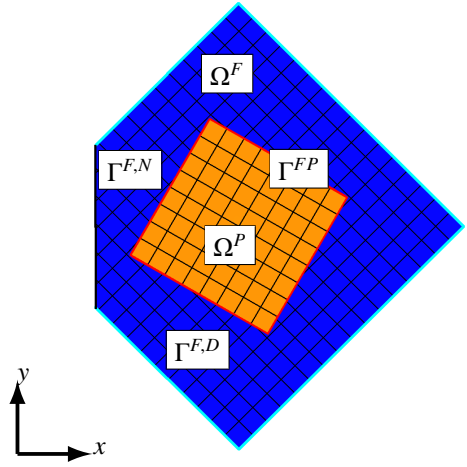
\caption{Geometric setup and computational meshes. The fluid domain $\domainf$, the poroelastic domain $\domainp$, the common interface $\fpiinterface$,
and boundaries with Neumann boundary condition $\Gamma^{F,N}$ or Dirichlet boundary condition $\Gamma^{F,D}$ are visualized here.
Black lines indicate the computational mesh for a mesh size of $h=0.0625$, corresponding to $16\times 16$ bi-linear elements to discretize the fluid domain and $8 \times 8$ bi-linear elements
to discretize the poroelastic domain.}
\label{fig:ex1_discretization}
\end{figure}

These additional body force and jump contributions allow fulfilling the analytic solution, 
and therefore the application of the method of manufactured solution,
independent of the alignment of the interface, the boundaries, and the domains.
The setup, as shown in Figure \ref{fig:ex1_discretization}, includes a wide variety of different intersections of single fluid elements by the interface $\fpiinterface$ for different mesh resolutions, 
while still allowing for the use of structured discretizations.
The poroelastic domain $\domainp$ is a square of size $0.5 \times 0.5$, which is rotated by an angle of $\alpha = 30^\circ{}$ around its center.
The fluid domain $\domainf$ is described by a square with size $1.0 \times 1.0$, rotated by an angle of $\beta = 45^\circ{}$ around its center,
where a part is removed by a vertical cut (horizontal distance from the square center $\Delta x = -0.45$), and the part occupied by the poroelastic domain $\domainp$ is excluded. 
On $\Gamma^{F,D}$, where the fluid discretization is matching, the analytic velocity \eqref{eq:sol_velf} is prescribed as a Dirichlet boundary condition.
At $\Gamma^{F,N}$, which is non-matching to the fluid discretization, the analytic fluid traction $\stressfa \cdot \normalf$ is prescribed as a Neumann boundary condition.

\begin{figure}[tb]
\centering
\def\svgwidth{0.3\textwidth}
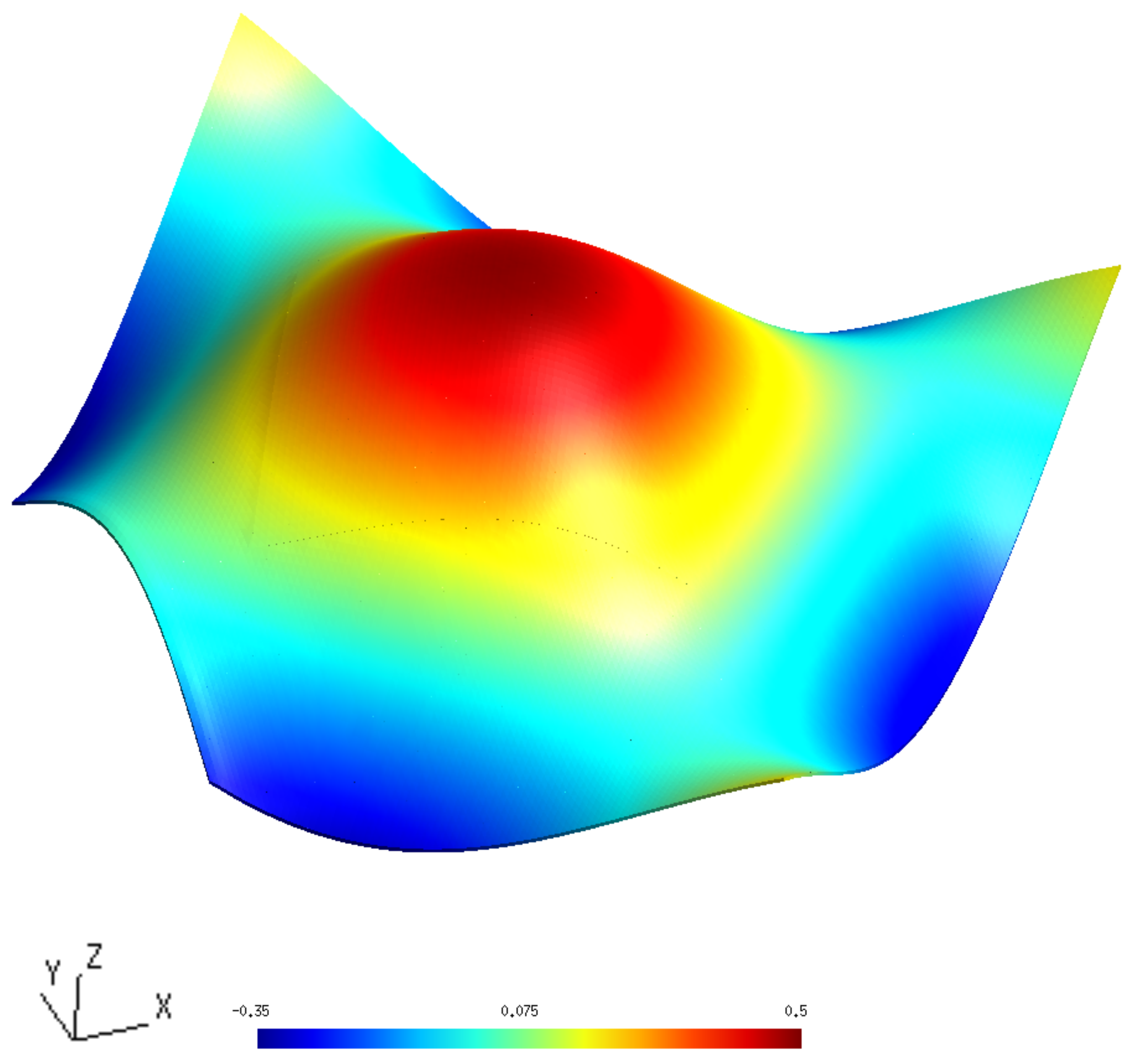
\def\svgwidth{0.3\textwidth}
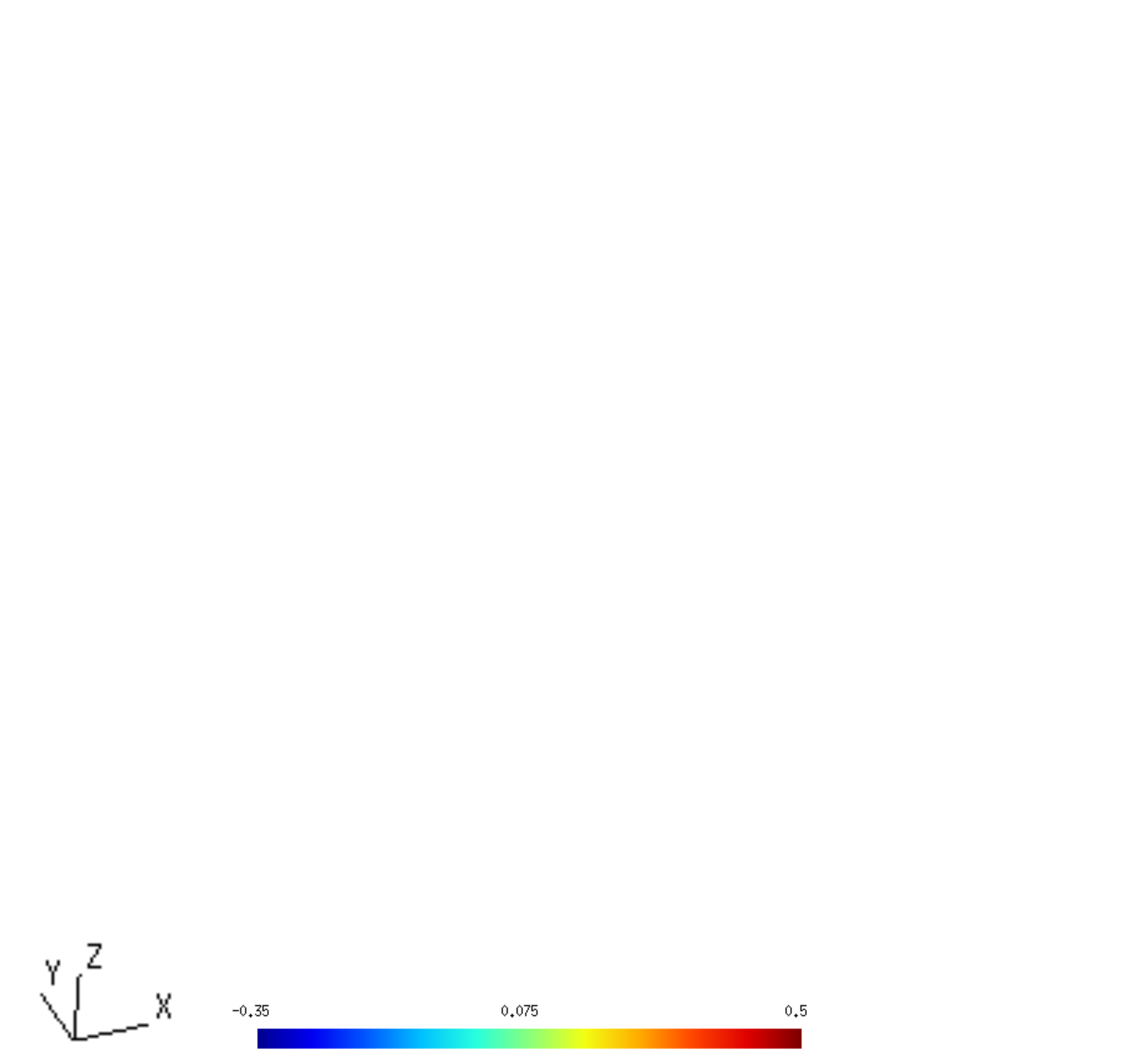
\def\svgwidth{0.3\textwidth}
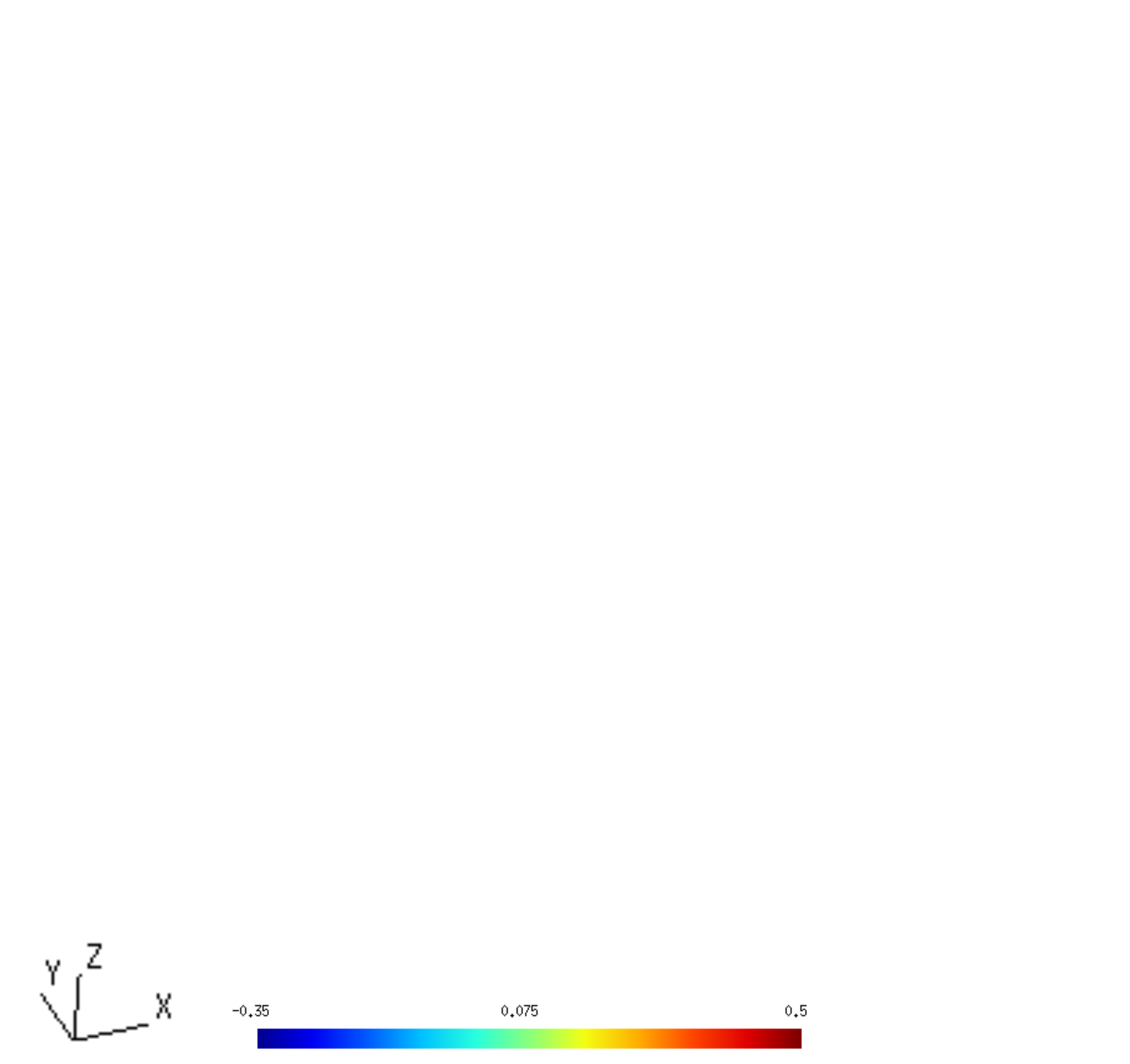
\caption{Computed pressure, velocity magnitude and displacement magnitude solution at $t = 0.1$ for $h=0.0078125$, $\Delta t = 0.05$, $\xi=-1$, $\gamma_n^{-1} = 45.0$, $\gamma_t^{-1} = 45.0$, $\bjcoeff = 1.0$,
Beavers-Joseph condition $\BJfac=1$ incorporated by \eqref{eq:w_fpi_t_nit}. Visualized by color-code and scalar warp in positive $z$-direction.}
\label{fig:ex1_solution}
\end{figure}

For the spatial discretization, four-noded, bi-linear, quadrilateral elements are considered.
(This is accomplished through a discretization with one layer of eight-noded, tri-linear hexahedreal elements.)
Discretization in time is performed by the backward Euler scheme $\theta = 1$, with a time step length of $\Delta t = 0.05$ if not denoted otherwise.
The final point in time of interest is set to $\etime = 0.1$. The initial state is given by the analytic solution:
\begin{align}
\velfB(\coord) = \velfa(\coord,0), \quad
\velpB(\coord) = \velpa(\coord,0),\quad
\disppB(\refcoord) = \disppa(\refcoord,0),\quad
\velpsB(\refcoord) = \velpsa(\refcoord,0).
\label{eq:ex1_initsol}
\end{align}
In Figure \ref{fig:ex1_solution}, the computed solution for a specific set of parameters is visualized for all computed unknowns, 
namely the pressure $(\pf,\pp)$, the velocity $(\velf,\velp)$ and the displacement $\dispp$.

To quantify the performance of the proposed formulation, the following $L_2$ error norms integrated in the domains $\domainf$ and $\domainp$, as well as on the interface $\fpiinterface$ are consulted:
\begin{align}
&\norm{\velf-\velfa}{\domainf},
\norm{\pf-\pfa}{\domainf},
\norm{\grad \velf- \grad \velfa}{\domainf},
\norm{\velp-\velpa}{\domainp},
\norm{\pp-\ppa}{\domainp},
\norm{\dispp-\disppa}{\refdomainp},
\norm{\gradRef\dispp-\gradRef\disppa}{\refdomainp},\nonumber\\
&\norm{\pf-\pfa}{\fpiinterface},
\norm{\left(\grad \velf- \grad \velfa\right)\cdot\normal}{\fpiinterface},
\norm{\pp-\ppa}{\fpiinterface},
\norm{\left(\grad\dispp-\grad\disppa\right)\cdot\normal}{\fpiinterface},\nonumber\\
\mathcal{E}_n = &\left|\left|\,\left[
\left(\velf - \velps - \porosity\left(\velp-\velps\right)\right)
-\left(\velfa - \velpsa - \porosity\left(\velpa-\velpsa\right)\right)
\right]\Pnormal\,  \right|\right|_{\fpiinterface},\nonumber\\
\mathcal{E}_t = &\left|\left|\,\left[
\left(\velf - \velps - \porosity\BJfac\left(\velp-\velps\right)\right)
-\left(\velfa - \velpsa - \porosity\BJfac\left(\velpa-\velpsa\right)\right)
\right]\Ptangent\,  \right|\right|_{\fpiinterface}.
\end{align}
While the norms in the first line quantify the overall domain error, specific interface error norms in line two quantify the components of the interface traction error,
and, finally, two norms quantify the error of the kinematic components of constraints \eqref{eq:fpi_massb} and \eqref{eq:fpi_bj} in tangential and normal direction separately.
\begin{remark}[Missing data points]
\label{rem:mis_datapoints}
For some of the following computations, data points are not plotted in the diagrams.
These solutions were not computed, as a result of exceeding a maximum number of iterations in the Newton-Raphson like scheme or due to nonphysical intermediate displacement states leading to 
problems in the geometric intersection algorithm.
As neighboring points in the presented graphs already show an increase of the computed error and therefore this behavior is expected, this was not investigated further.
\end{remark}
\FloatBarrier
\subsection{Spatial convergence analysis}
\label{sec:conv}
\begin{figure}[hbtp]
\begin{minipage}[hbt]{0.32\textwidth}
\raggedleft
\def\figscaling{0.47}
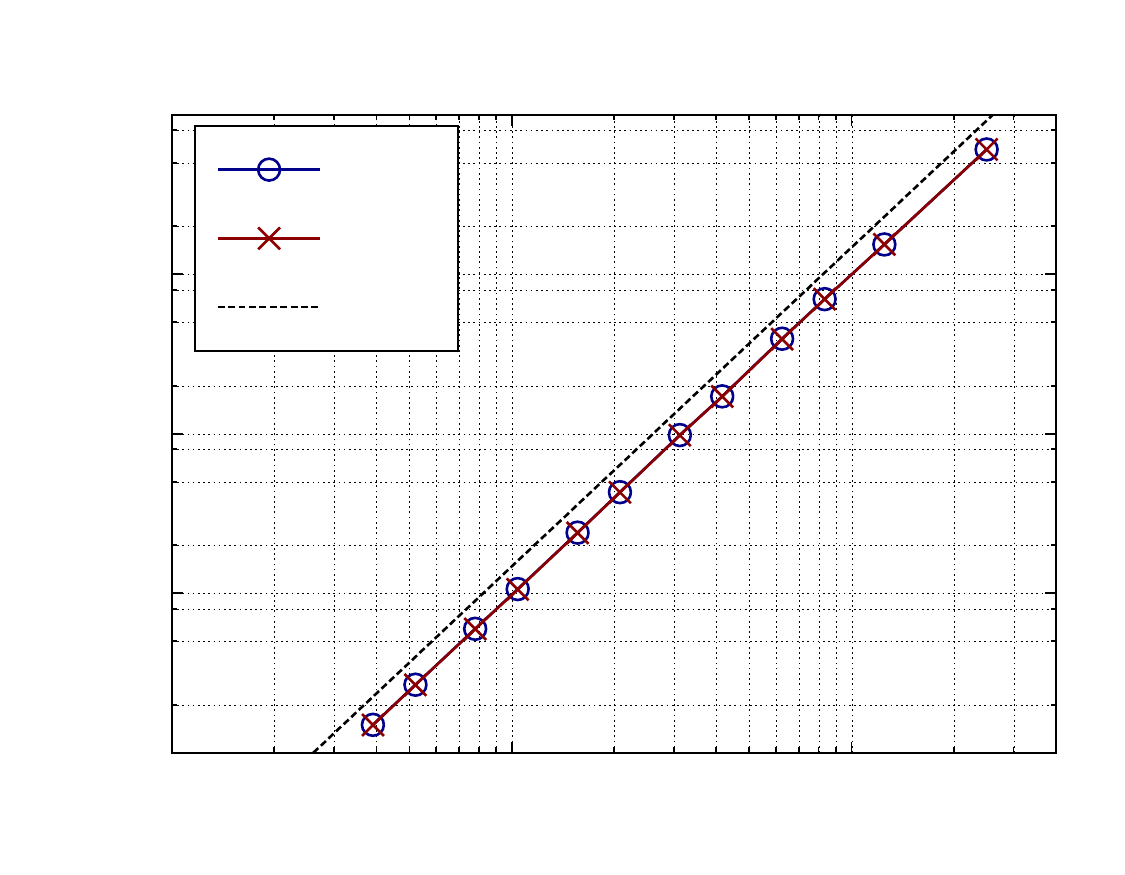
\end{minipage}
\begin{minipage}[hbt]{0.32\textwidth}
\centering
\def\figscaling{0.47}
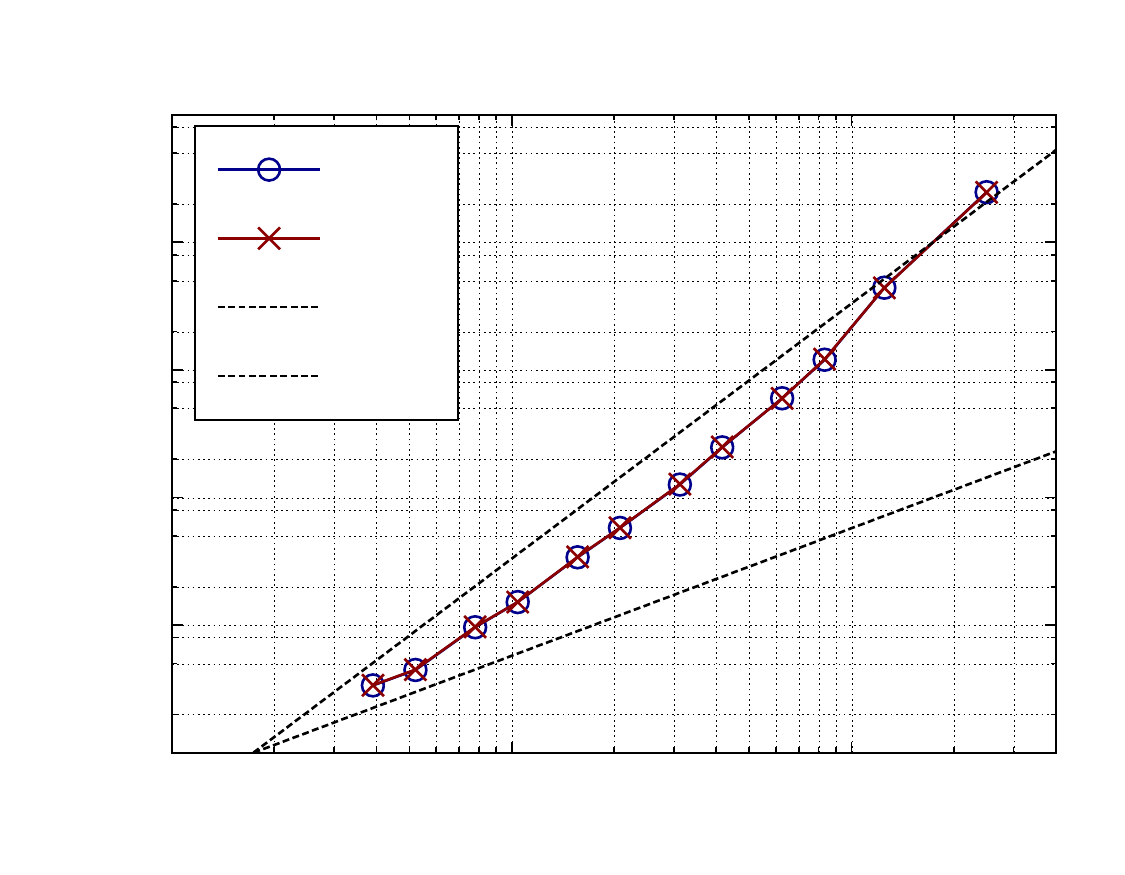
\end{minipage}
\begin{minipage}[hbt]{0.32\textwidth}
\centering
\def\figscaling{0.47}
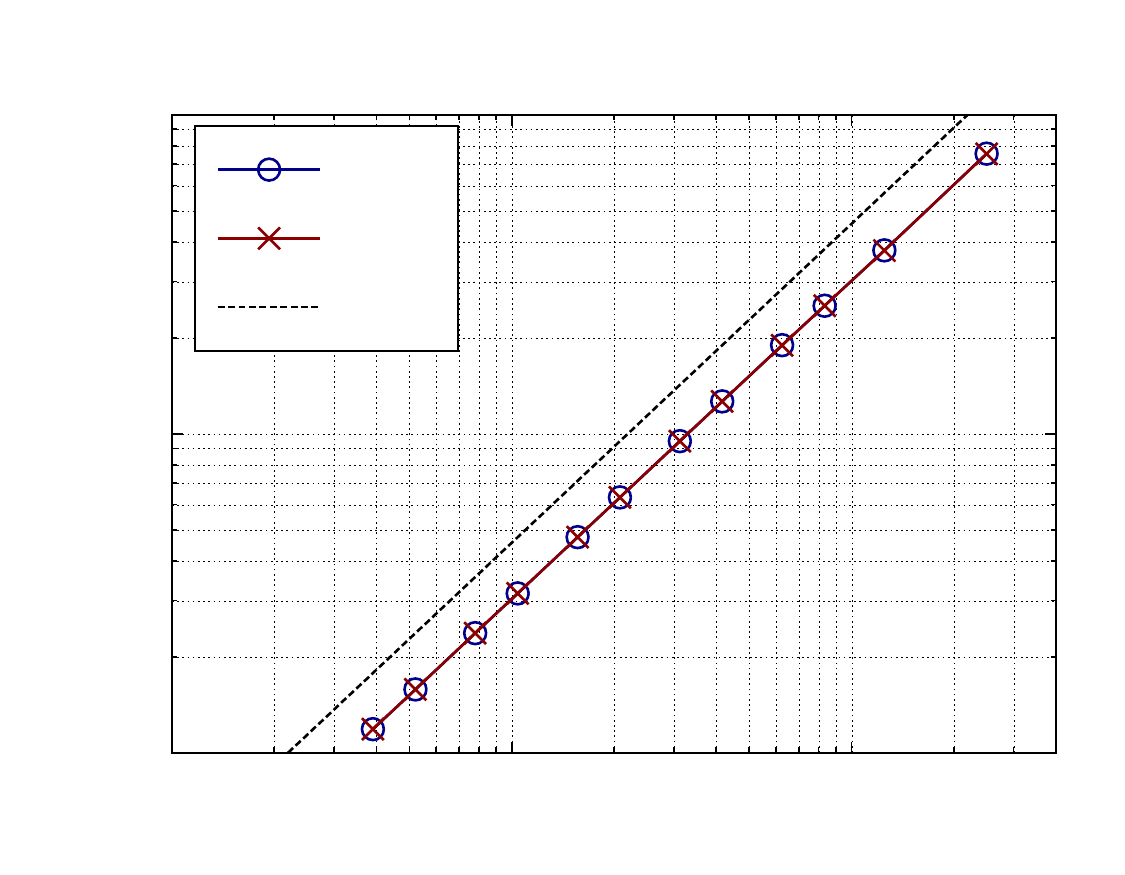
\end{minipage}

\begin{minipage}[hbt]{0.32\textwidth}
\raggedleft
\def\figscaling{0.47}
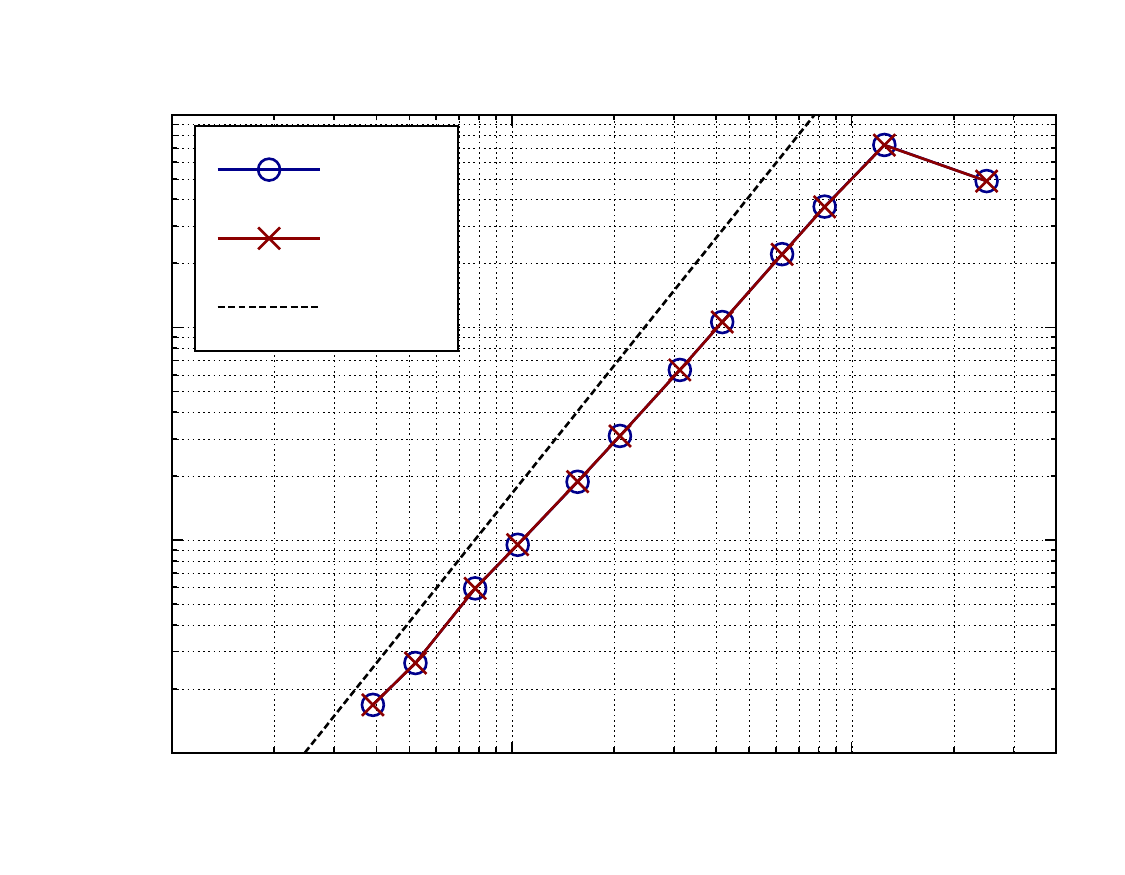
\end{minipage}
\begin{minipage}[hbt]{0.32\textwidth}
\centering
\def\figscaling{0.47}
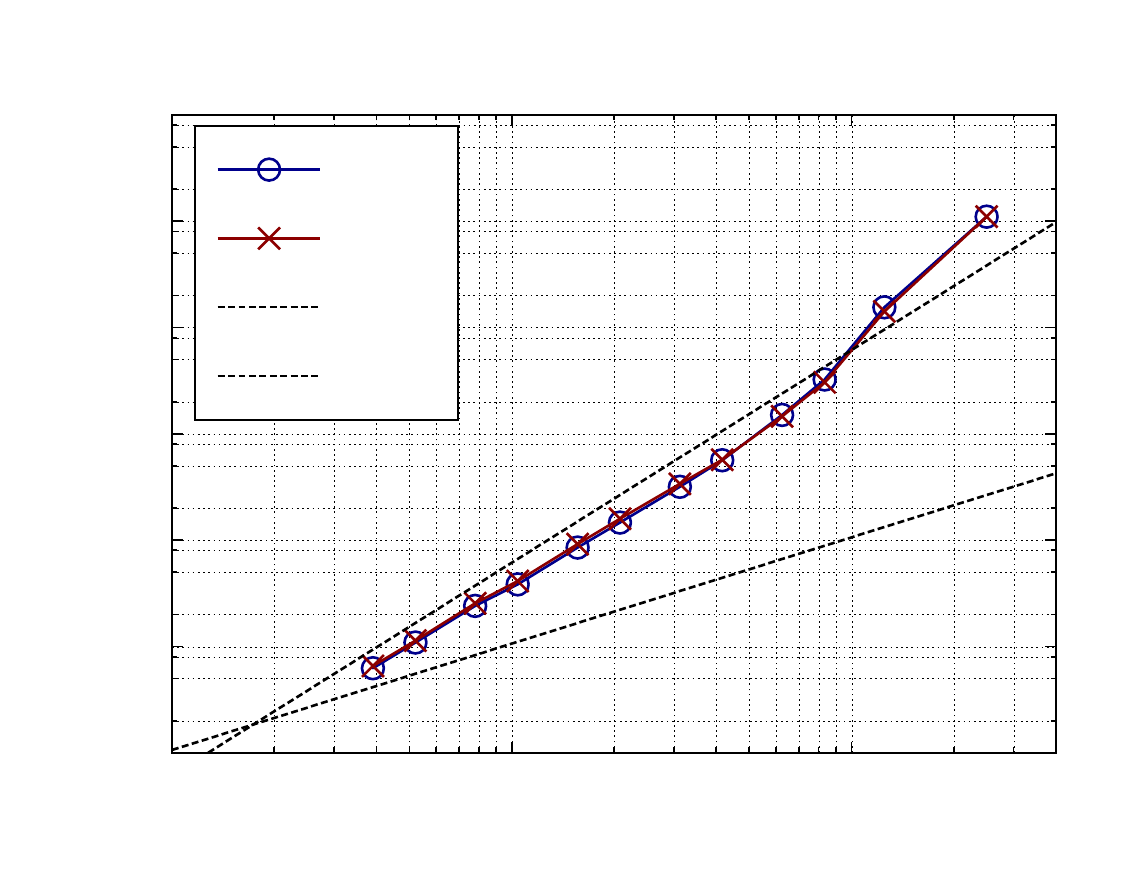
\end{minipage}
\begin{minipage}[hbt]{0.32\textwidth}
\centering
\def\figscaling{0.47}
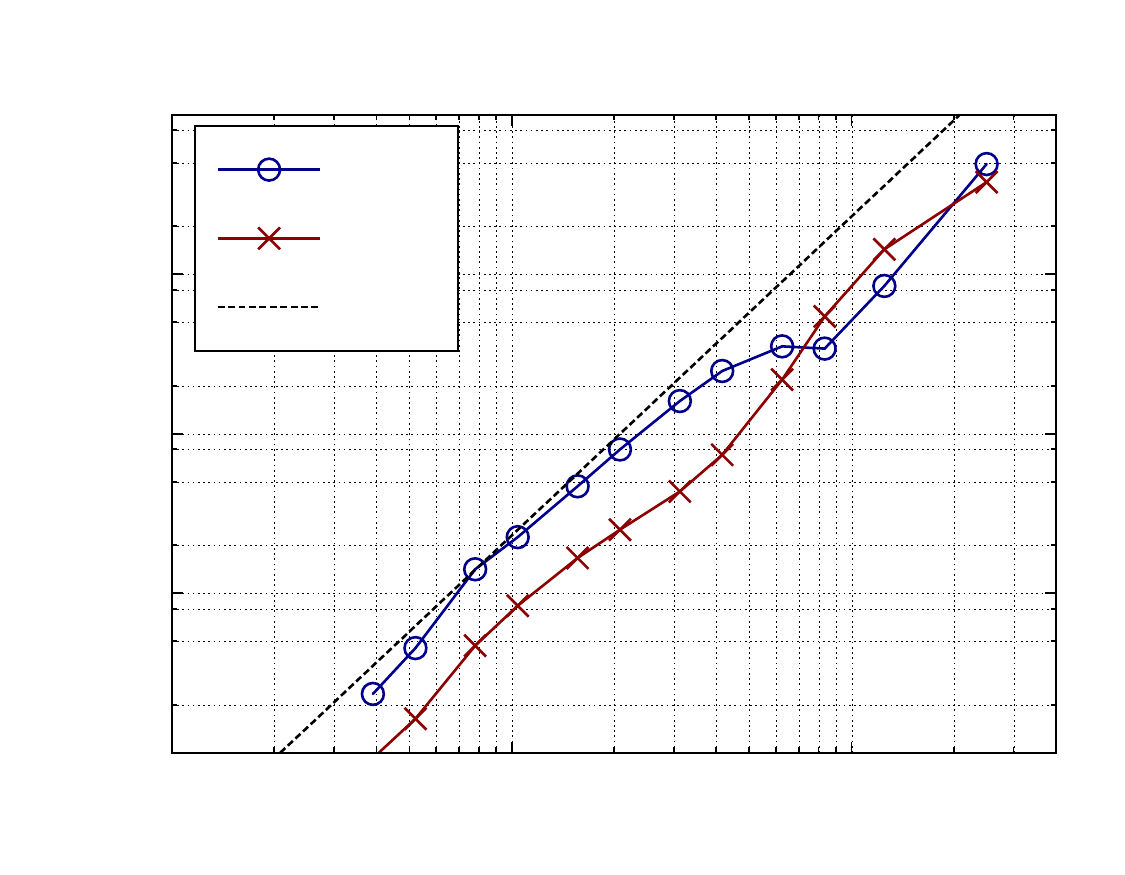
\end{minipage}

\begin{minipage}[hbt]{0.32\textwidth}
\raggedleft
\def\figscaling{0.47}
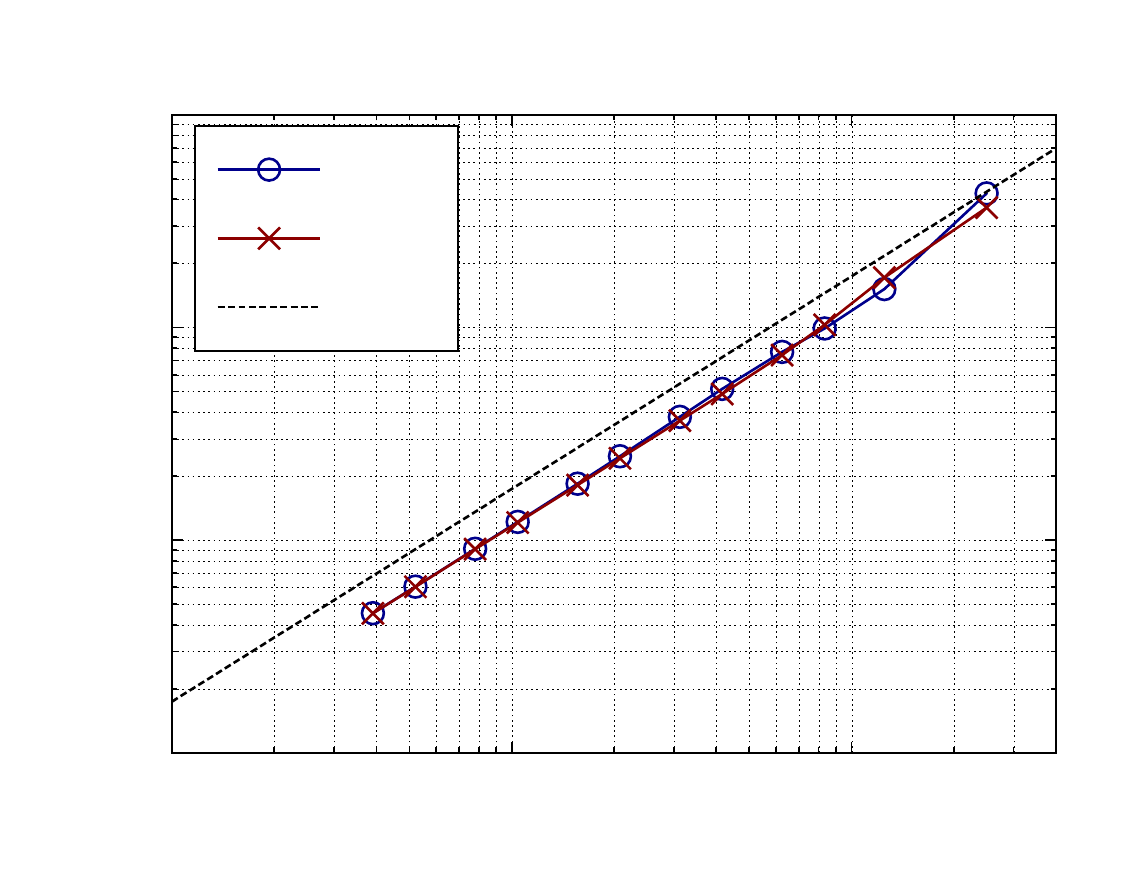
\end{minipage}
\begin{minipage}[hbt]{0.32\textwidth}
\raggedleft
\def\figscaling{0.47}
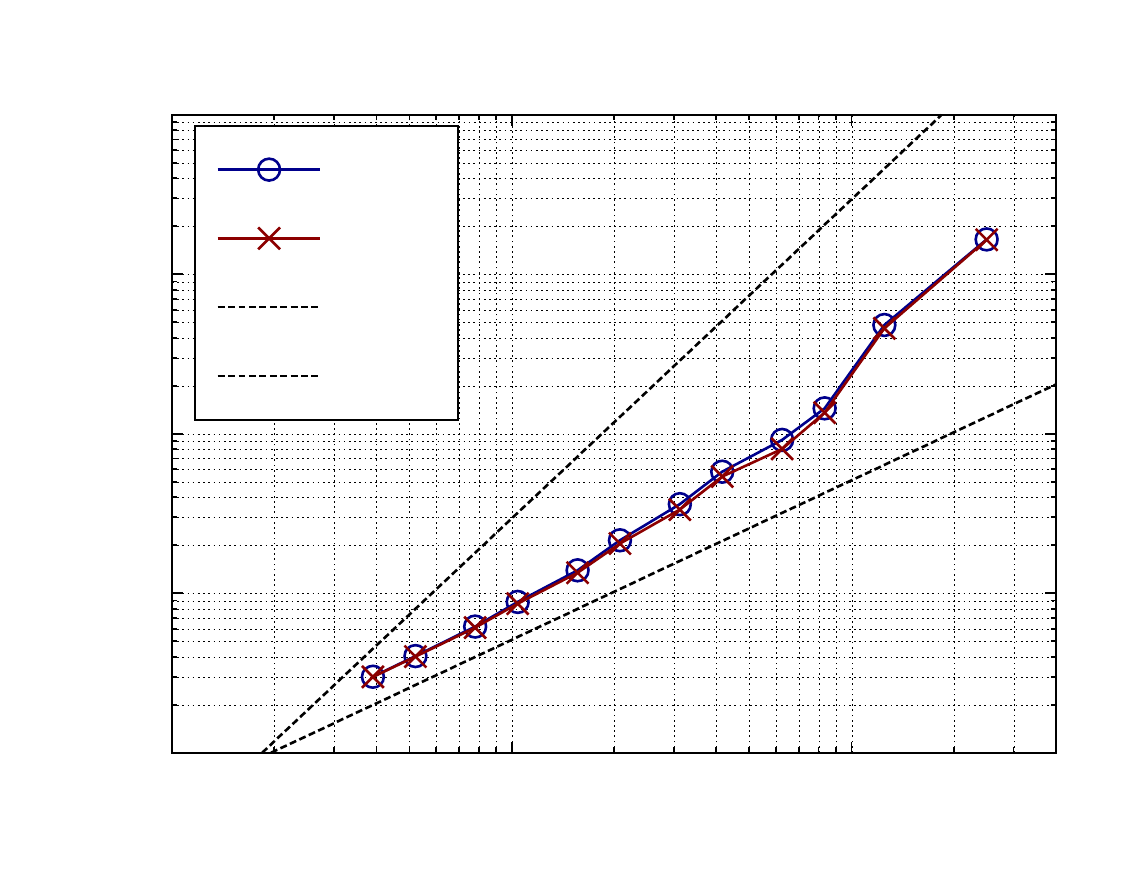
\end{minipage}
\begin{minipage}[hbt]{0.32\textwidth}
\centering
\def\figscaling{0.47}
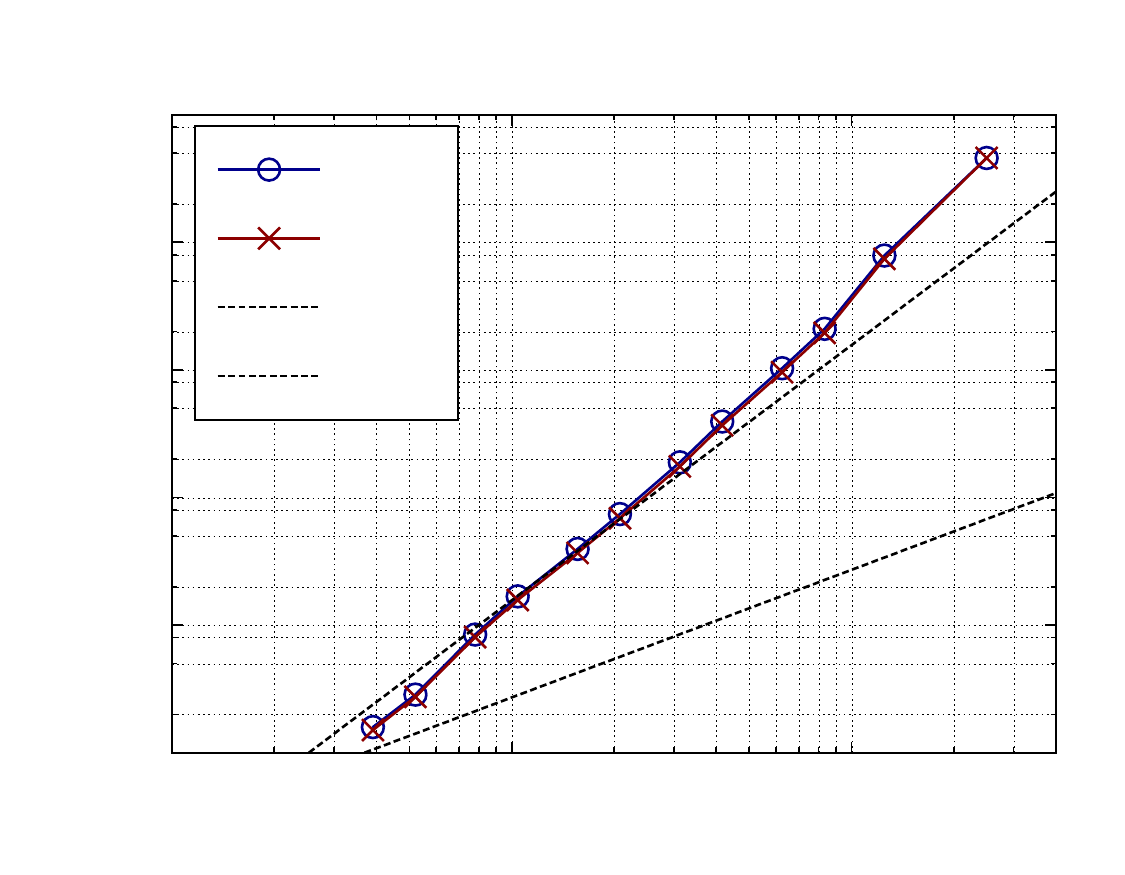
\end{minipage}

\centering
\begin{minipage}[hbt]{0.32\textwidth}
\centering
\def\figscaling{0.47}
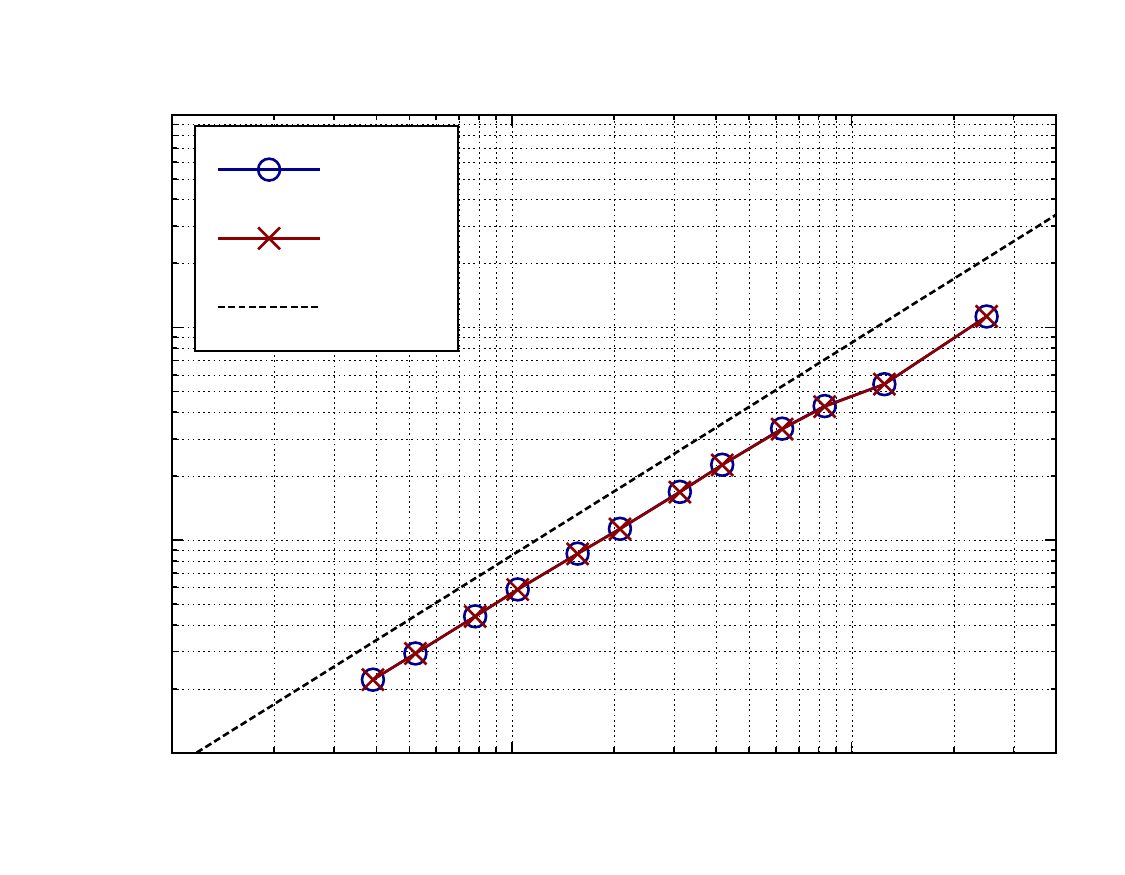
\end{minipage}
\begin{minipage}[hbt]{0.32\textwidth}
\centering
\def\figscaling{0.47}
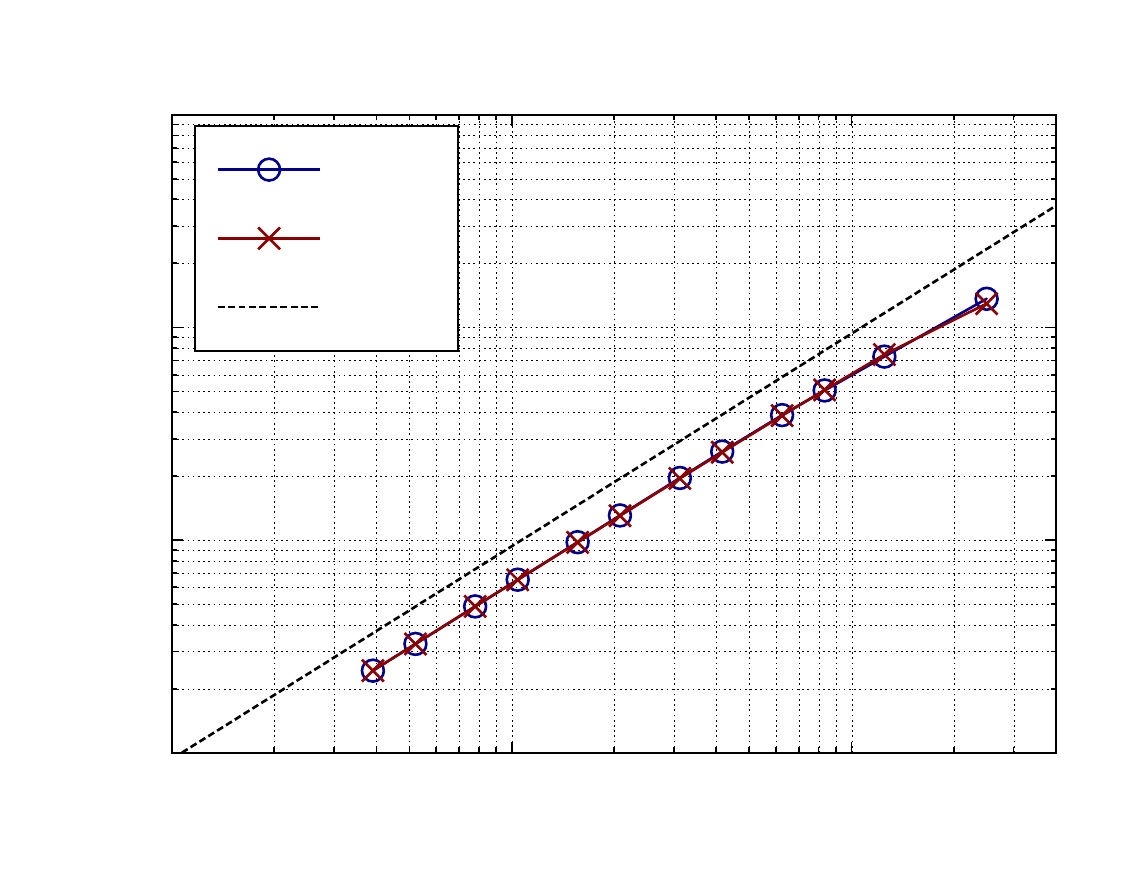
\end{minipage}

\centering
\begin{minipage}[hbt]{0.32\textwidth}
\centering
\def\figscaling{0.47}
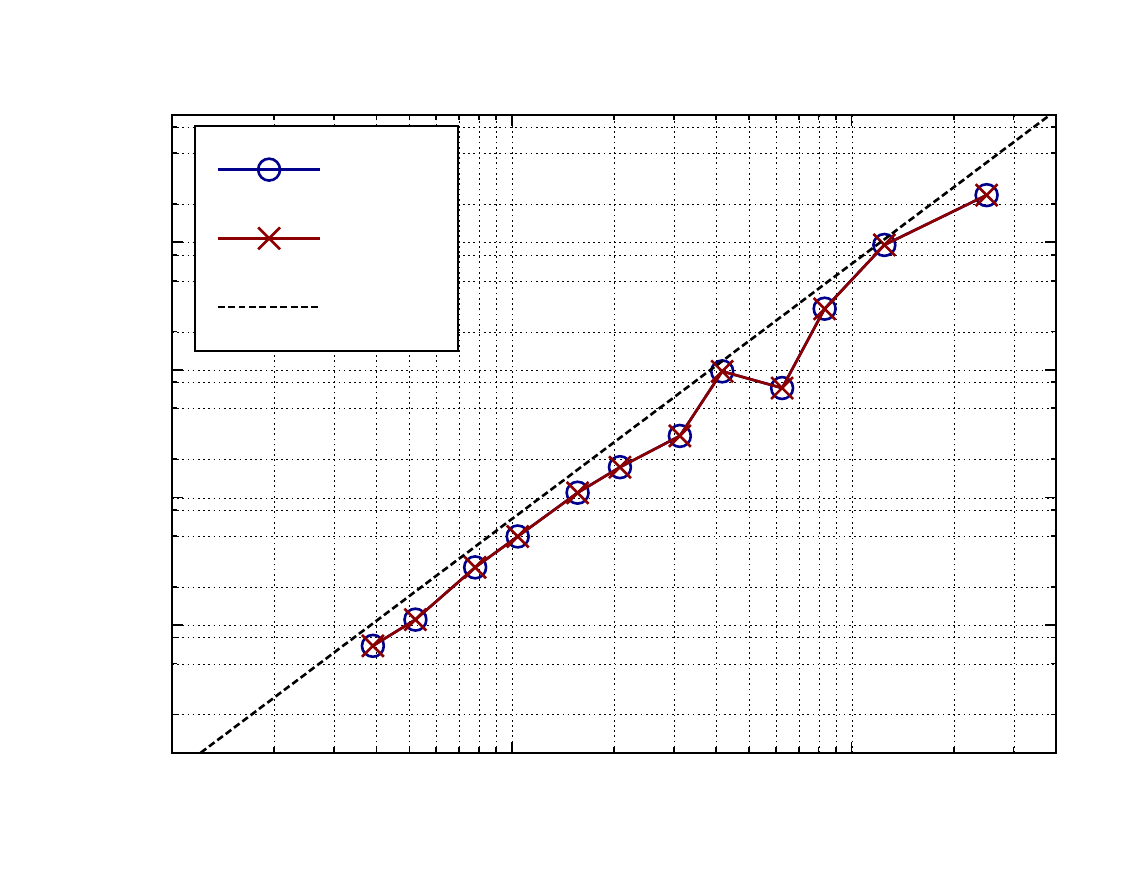
\end{minipage}
\begin{minipage}[hbt]{0.32\textwidth}
\centering
\def\figscaling{0.47}
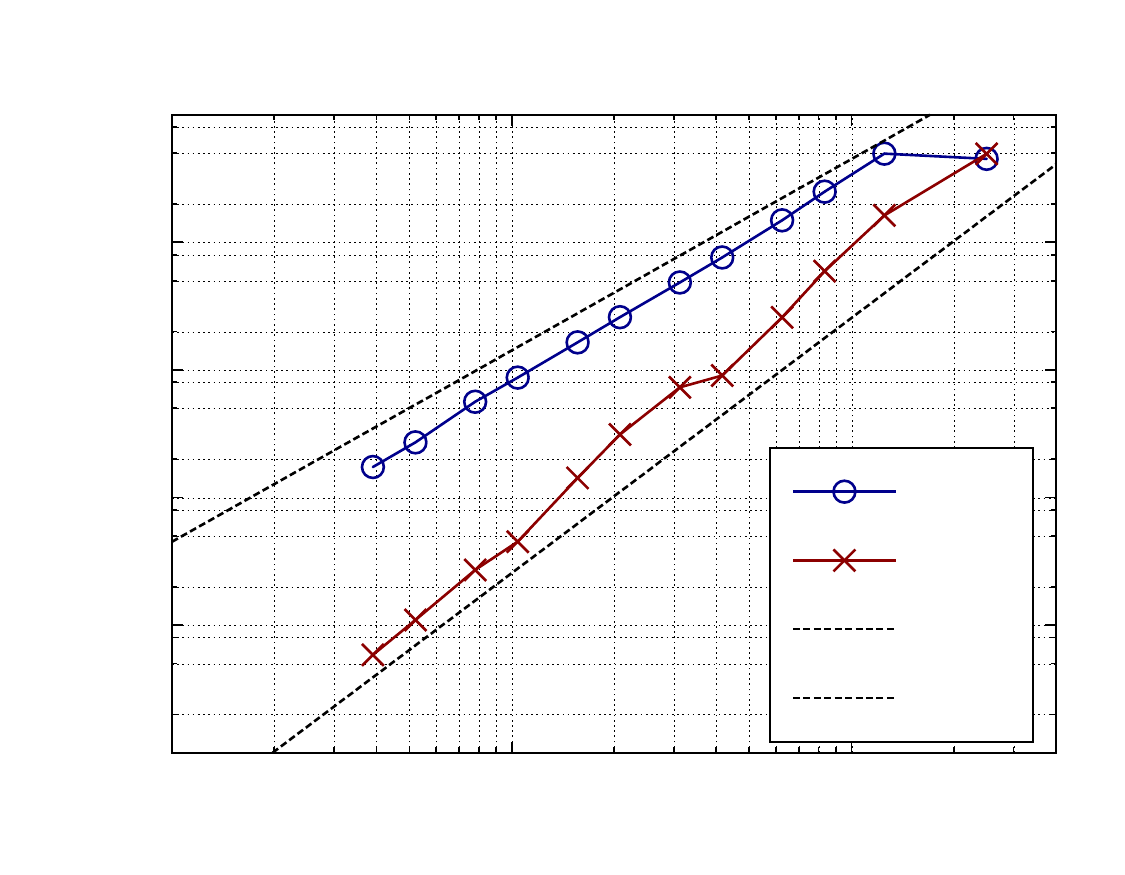
\end{minipage}
\caption{Computed domain and interface error norms for the spatial convergence study.
Considered is the adjoint-inconsistent ($\adjointsign = -1$) Nitsche-based approach ``Nit'' with contribution \eqref{eq:w_fpi_t_nit} to the weak form and the inverse penalty parameters $\gamma_n^{-1}=\gamma_t^{-1}=45$.
}
\label{fig:conv}
\end{figure}

First, a spatial convergence analysis for the coupled FPI setup is performed.
Indications for convergence rates to be expected for the formulation, can be found in the contributions 
for the Stokes/Darcy coupling \cite{burman2007, badia2009_2}, 
for the Stokes/Biot-system coupling \cite{badia2009}, 
for the applied poroelastic formulation \cite{vuong2016}, 
for the CutFEM applied on the Stokes equation \cite{burman2014} and applied on the Oseen equation \cite{massing2016} including the general Navier boundary condition \cite{winter2017},
and for the CutFEM applied on the fluid-structure interaction \cite{massing2015}.

The mesh size $h$, which equals the edge length of the quadrilateral elements of squared shape, is varied in $h\in\left[0.25,0.0039062\right]$.
For this analysis, the Nitsche penalty parameters are set to $\gamma_n^{-1}=\gamma_t^{-1}=45$ and the adjoint-inconsistent variant $\left(\xi=-1\right)$ is applied.
To prevent the temporal error from exceeding the spatial error, the two finest meshes $h=0.0052083$ and $h=0.0039062$ are discretized in time with half the time step length of $\Delta t = 0.025$.

The error norms, computed for the BJ and BJS interface condition, are presented in Figure \ref{fig:conv}.
In all domain error norms and almost all interface error norms, the expected convergence rates can be observed starting for smaller values of the mesh size than $h<0.1$.
No difference between both interface conditions can be observed.
The exception is the displacement domain error, where a difference between both methods can be observed potentially due to the small absolute error level.
For the normal error $\mathcal{E}_n$, a noticeable kink can be observed, which is most likely due to the varying fluid element intersection configurations.
Nevertheless, the overall convergence rate is not altered for both of these cases.
The only significant difference can be observed in the tangential kinematic error $\mathcal{E}_t$.
Here, a convergence rate of $h^2$ can only be observed for the BJS variant, while the BJ condition leads to $h^{3/2}$.
This is a consequence of the additionally occurring tangential velocity in the poroelastic fluid constraint \eqref{eq:fpi_bj} in the case of BJ ($\BJfac=1$).
Nevertheless, for large enough values of the coefficient $\sliplengh \viscf$, the full tangential constraint error will be
dominated by the first order convergence of the fluid gradient for BJ and BJS.

\subsection{Sensitivity of the formulation for variations of the Nitsche penalty parameters $\gamma_n$ and $\gamma_t$}
\label{sec:varpen}
\begin{figure}[hbtp]
\begin{minipage}[hbt]{0.32\textwidth}
\centering
\def\figscaling{0.5}
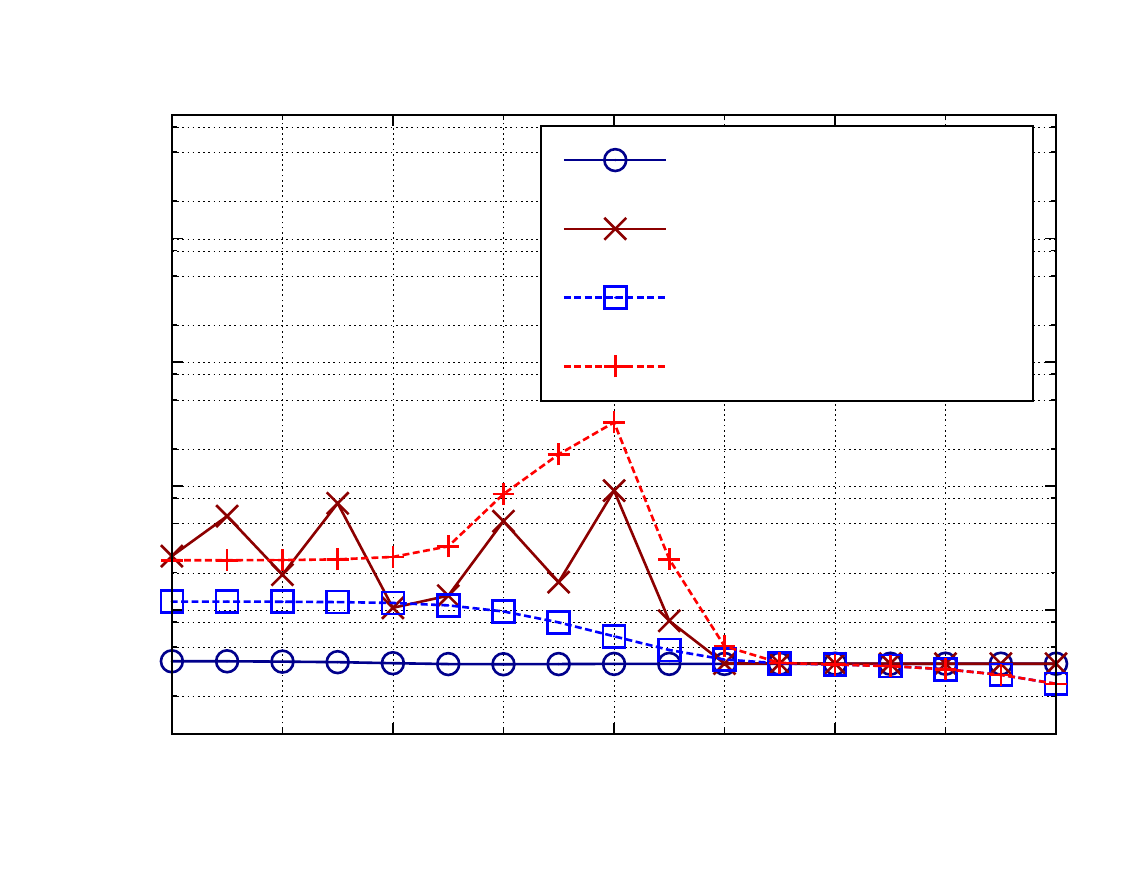
\end{minipage}
\begin{minipage}[hbt]{0.32\textwidth}
\centering
\def\figscaling{0.5}
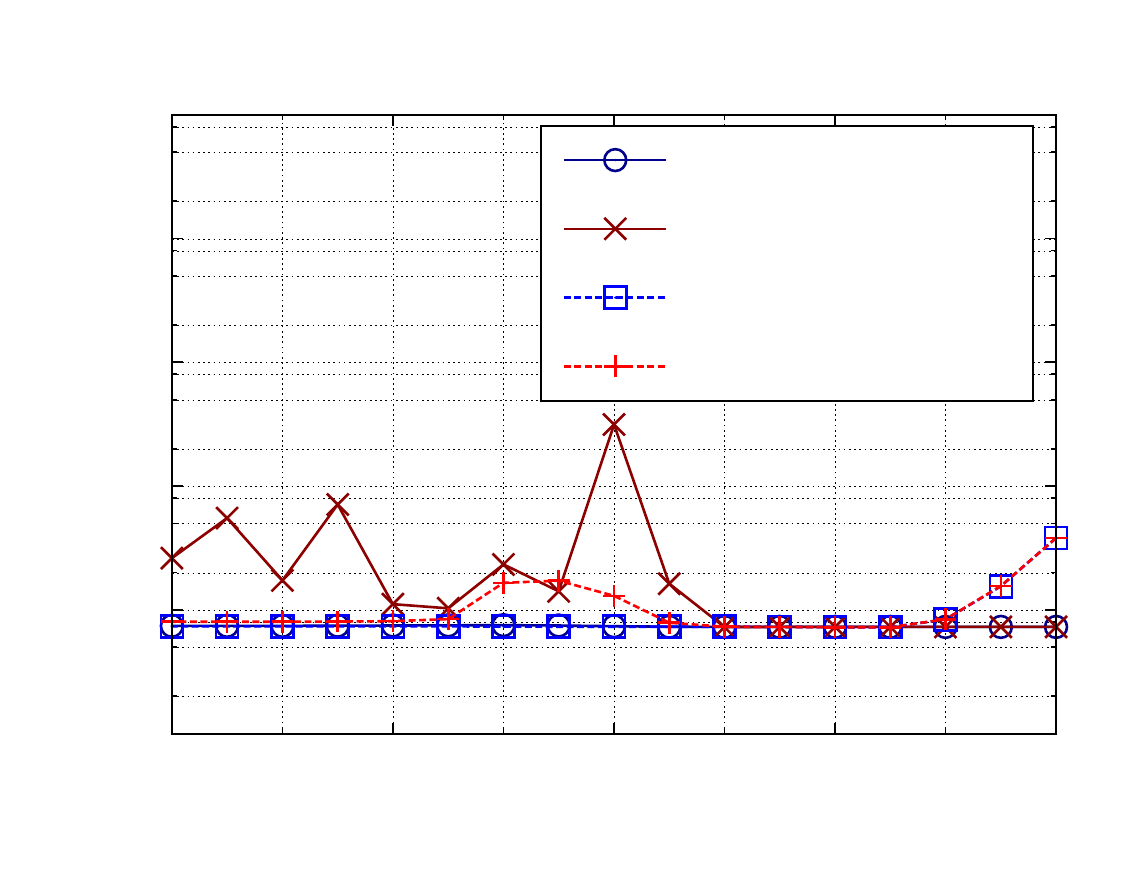
\end{minipage}
\begin{minipage}[hbt]{0.32\textwidth}
\raggedleft
\def\figscaling{0.5}
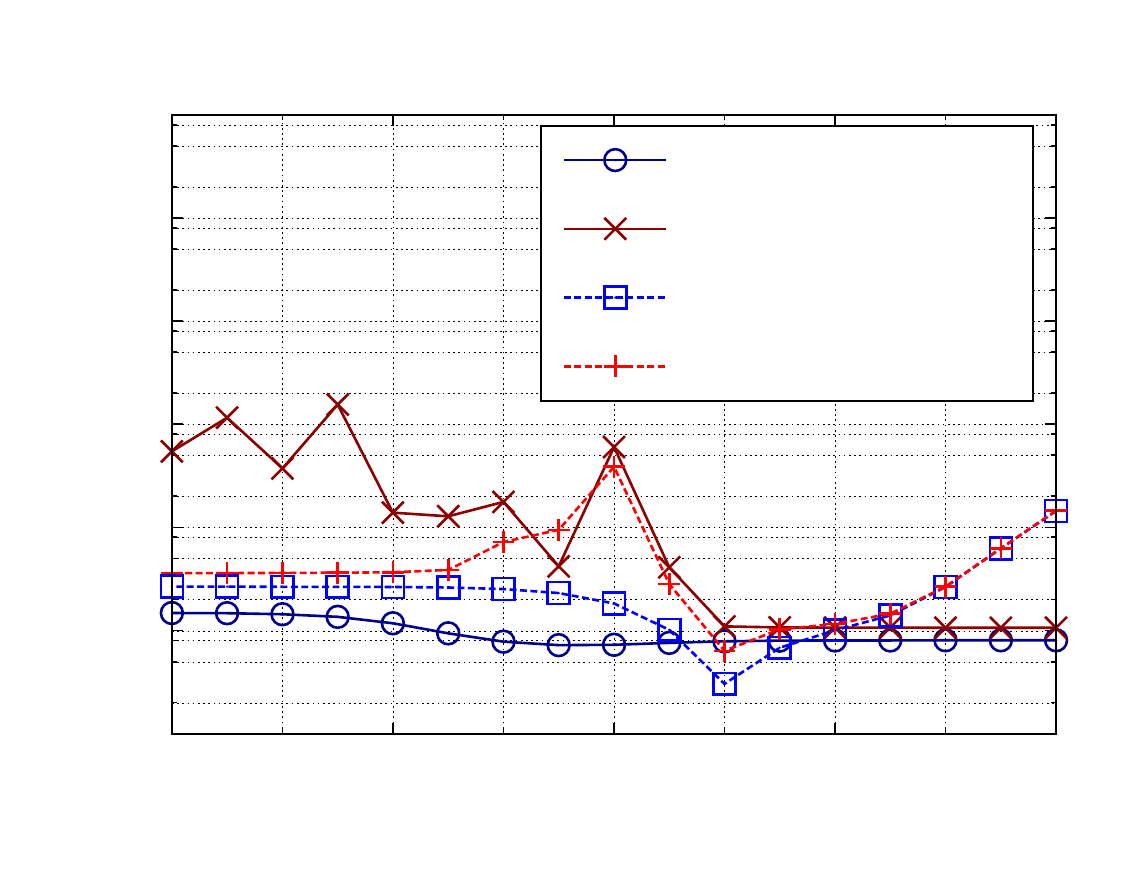
\end{minipage}

\begin{minipage}[hbt]{0.32\textwidth}
\centering
\def\figscaling{0.5}
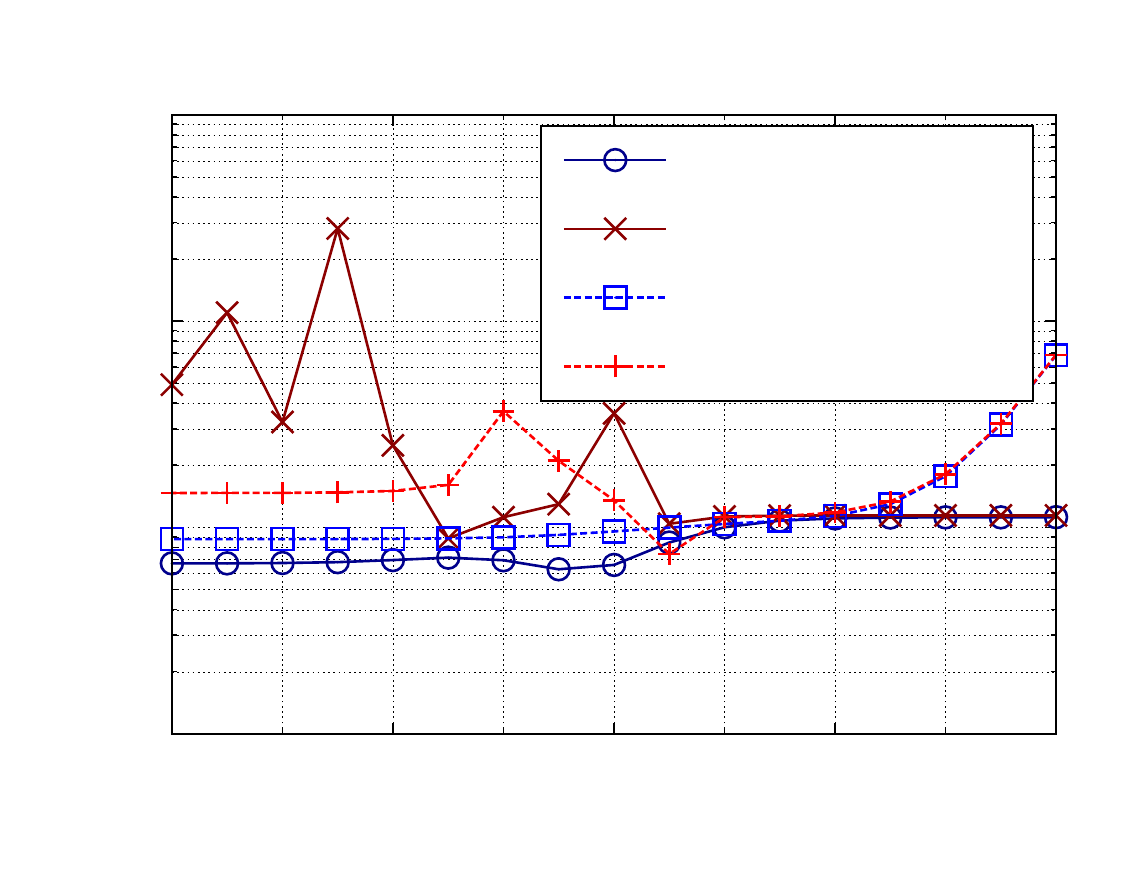
\end{minipage}
\begin{minipage}[hbt]{0.32\textwidth}
\centering
\def\figscaling{0.5}
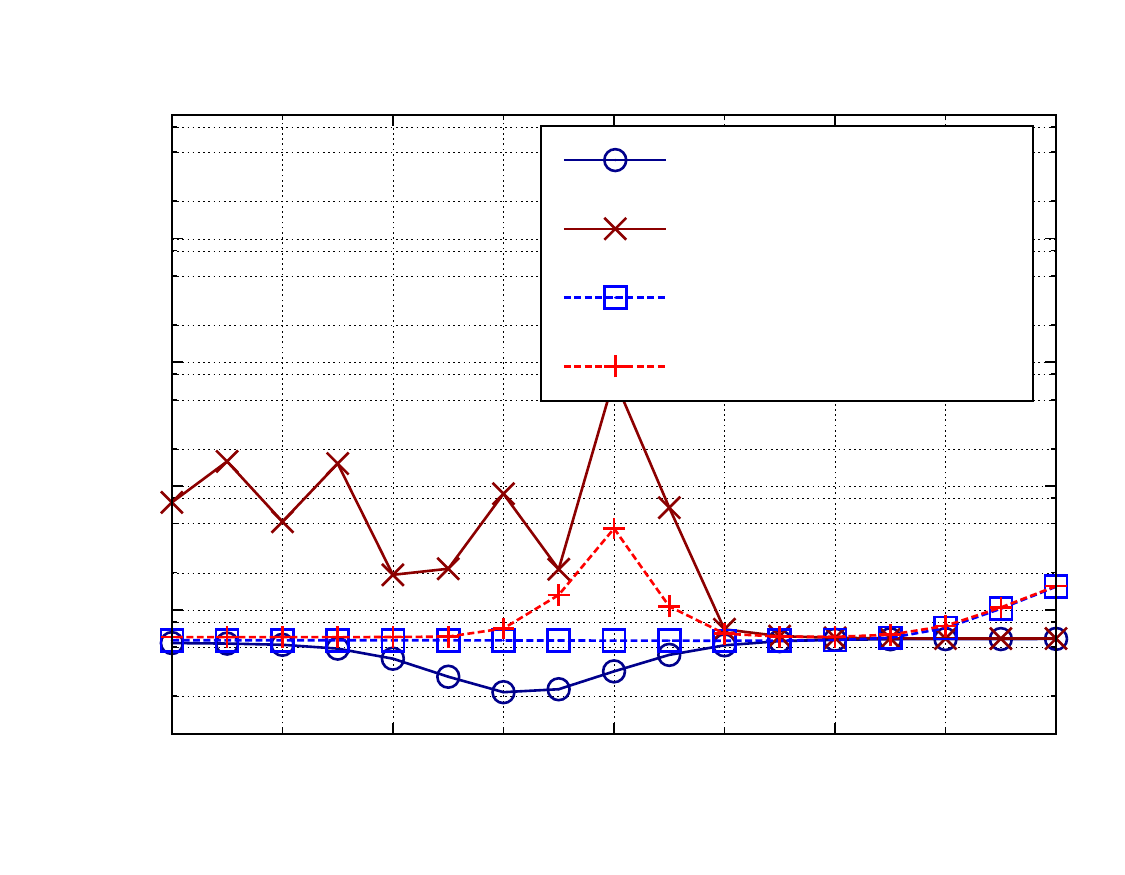
\end{minipage}
\begin{minipage}[hbt]{0.32\textwidth}
\centering
\def\figscaling{0.5}
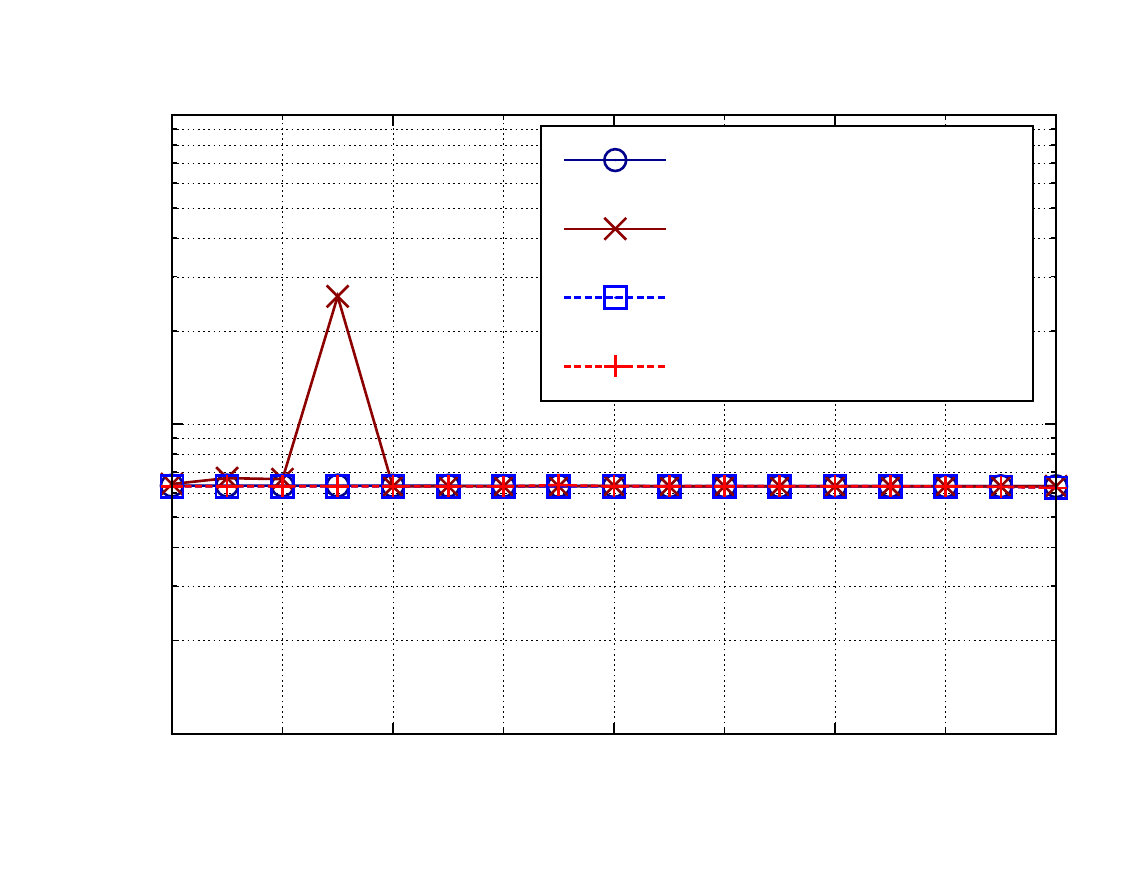
\end{minipage}
\caption{Computed interface error norms for varying the Nitsche penalty parameters $\gamma_t^{-1}$ and $\gamma_n^{-1}$. 
$\left(*=t,\xi=-1\right)$: adjoint-inconsistent variant, varying penalty parameter $\gamma_t^{-1}$, fixed penalty parameter $\gamma_n^{-1} = 45$.
$\left(*=t,\xi=1\right)$: adjoint-consistent variant, varying penalty parameter $\gamma_t^{-1}$, fixed penalty parameter $\gamma_n^{-1} = 45$.
$\left(*=n,\xi=-1\right)$: adjoint-inconsistent variant, varying penalty parameter $\gamma_n^{-1}$, fixed penalty parameter $\gamma_t^{-1} = 45$.
$\left(*=n,\xi=1\right)$: adjoint-consistent variant, varying penalty parameter $\gamma_n^{-1}$, fixed penalty parameter $\gamma_t^{-1} = 45$.
Computed for mesh size $h=0.0078125$.}
\label{fig:res_varpen}
\end{figure}

In \eqref{eq:w_fpi_n} and \eqref{eq:w_fpi_t_nit}, the two penalty parameters $\gamma_n$ and $\gamma_t$ were introduced.
For the weak imposition of boundary and interface conditions by the Nitsche-based method, it is expected that both parameters are required to be sufficiently small in the case of 
a formulation with the adjoint-consistent terms ($\xi = 1$).
Whereas for the variant with the adjoint-inconsistent terms ($\xi = -1$), no lower limit is expected.
Both parameters are varied independently to detect whether a different choice of both parameters is more beneficial.
In Figure \ref{fig:res_varpen}, a representative selection of the computed error norms dependent on the inverse of the chosen penalty parameters is shown.

For the adjoint-inconsistent variant $\left(\xi=-1\right)$, a smooth dependence of the computed error norms and the parameters $\gamma_t^{-1}$ and $\gamma_n^{-1}$ can be identified,
which indicates a discrete stable formulation for the whole parameter range.
Increasing error norms can be identified for the pressure and normal constraint for decreasing penalty parameters. 
For the adjoint-consistent variant $\left(\xi=1\right)$, it can be observed that a penalty parameter smaller than $\gamma_*^{-1}<10$ results in 
large variations of the error for small variations of the penalty parameter.
This agrees with the expectation of a loss in coercivity for this formulation.
Both variants show an increasing computed error for almost all error norms in the case of large normal penalty parameters.
The sole exception is the error of kinematic constraint in normal direction $\mathcal{E}_n$, which is directly enforced by this penalty contribution, where a decreasing error for an increasing penalty parameter can be observed.

As a result of this computed study, a value for both inverse penalty parameters $\gamma_t^{-1}$ and $\gamma_n^{-1}$, independent of the adjoint term symmetry, in the range of $(10,100)$ is recommended.
This ensures discrete stability and a good compromise between kinematic constraint enforcement and errors of the pressure and velocity/displacement gradients.
The parameter range and the dependence of the computed errors and penalty parameters is in good agreement with the results presented in \cite{schott2014,massing2016,winter2017}.

\begin{remark}[Adjoint-inconsistent method with $\sliplengh^{-1} = 0$]
\label{rem:nonzerogp}
From the results presented above, also a variant with $\gamma_t^{-1} = 0$ for skew-symmetric adjoint terms $\xi=-1$ seems possible. 
As already pointed out in \cite{winter2017}, this is not possible when approaching the ``full slip'' limit $\sliplengh^{-1} = 0$, as the system to be solved becomes ill-conditioned.
This can be directly seen in the second to last line of the tangential Nitsche terms in \eqref{eq:w_fpi_t_nit}, where the prefactor of the stress term 
${\left(\sliplengh\gamma_t h_{\Gamma}\right)}{\left(\sliplengh\viscf+\gamma_t h_{\Gamma}\right)}^{-1} $ dominates the overall weak form.
\end{remark}

\subsection{Comparison of tangential ``Substitution'' and ``Nitsche'' variant for varying porosity $\porosity$, permeability $\permeabp$}
\label{sec:varperm}
As presented in Section \ref{sec:num_fpinterface}, two different approaches for incorporating the BJ or BJS condition in tangential direction 
are analyzed.
First, the ``Substitution'' approach is analyzed, which is, to the best of the authors' knowledge, exclusively applied in literature for this type of condition, due to its simplicity and good numerical stability.
Nevertheless, when approaching the ``no-slip''-limit ($\sliplengh = 0$), the $\sliplengh^{-1}$ contribution leads to an ill-conditioned system for the ``Substitution'' approach.
Therefore, the second approach based on the Nitsche method is compared to the ``Substitution'' approach for a wide range of the coefficient $\sliplengh$ 
(specified by relation \eqref{eq: bjcoeff}).

To analyze the performance of both numerical approaches for varying physical parameters, a relation between the porosity $\porosity$ and the permeability $\permeabp$ for the poroelastic media is considered.
One possibility is the Kozeny-Carman formula (see e.g. \cite{Coussy:04}), which will be applied in the following by:
\begin{align}
\matpermeabp = \matpermeabpscalar_{\text{ref}} \frac{1-\porosity_{\text{ref}}^2}{\porosity_{\text{ref}}^3} \frac{\left(\Jp\porosity\right)^3}{1-\left(\Jp\porosity\right)^2}\unity,
\label{eq:kozcar}
\end{align}
with the reference porosity $\porosity_{\text{ref}} = 0.5$ and the scalar, material reference permeability $\matpermeabpscalar_{\text{ref}}=0.1$ being prescribed.
Then, by a reduction of the porosity $\porosity$ the trace of the material permeability $\text{tr}(\matpermeabp)$ and therefore the trace of the permeability $\text{tr}(\permeabp)$ decreases.
This finally results in a shrinking coefficient $\sliplengh$.
To analyze the difference in both formulations for a modification of solely physical parameters, the porosity is prescribed in the range of $\porosity = \left[10^{-6},5 \cdot 10^{-1} \right]$
and the corresponding permeability is computed leading to a modified coefficient $\sliplengh$.
To analyze the dependence of the Beavers-Joseph coefficient $\bjcoeff$, the computations are performed for $\bjcoeff = 1$ and $\bjcoeff= 10$.

Due to the considered low porosity and permeability, the relative velocity between the poroelastic fluid and poroelastic solid of the analytic solution is reduced.
For this comparison, the amplitudes in the analytic solution \eqref{eq:sol_velp} and \eqref{eq:sol_up} are chosen to an equal value of $A^P=A^{P^S}=-10^{-5}$.
As the solution \eqref{eq:sol_velp} is given in the current configuration and the solution \eqref{eq:sol_up} in material configuration, 
a relative velocity for the deformed configuration still occurs.

A representative selection of the computed error norms is presented in Figure \ref{fig:varperm}.
First, in analyzing the Nitsche-based approach and comparing the computed results for the different coefficients $\bjcoeff = 1$ and $\bjcoeff= 10$, no dependence of the computed error norms can be observed.
Moreover, the variation of the porosity does not lead to a significant change of the computed errors.
Solely for large porosities, close to $\porosity = 0.5$, an increasing error can be observed for all variants.
This is likely due to the prescribed analytic solution with small relative velocity in the poroelastic domain, 
whereas the pressure gradient term cannot be balanced by the reactive term due to the moderate coefficient $\viscf\porosity\permeabp^{-1}$.
The ``Substitution'' approach, on the other hand, leads to increasing error norms for a small porosity (and permeability) for almost all quantities.
This is the expected behavior, due to the scaling $\sliplengh^{-1}$ of the substitution term. 
The kinematic tangential error $\mathcal{E}_t$, however, reduces for this approach in the small porosity limit,
as the global system reduces to this constraint for $\sliplengh = 0$, which is the expected behavior.

Finally, one can conclude that both variants perform well for a wide range of the porosity $\porosity$ and the permeability $\permeabp$.
For the computed problem setup, the ``Substitution'' approach leads to a comparable error in comparison to the Nitsche-based approach for porosities down to $\porosity = 10^{-3}$.
Nevertheless, when it is essential to approach the impermeability limit $\porosity = 0$ and $\permeabp = \zerovec$
(e.g. considering contacting rough surfaces by a poroelastic model \cite{ager2018}), the substitution approach fails due to ill-conditioning.

\subsection{Comparison of Beavers-Joseph and Beavers-Joseph-Saffmann interface condition}
\label{sec:varbj}
\begin{figure}[p]
\begin{minipage}[hbt]{0.32\textwidth}
\centering
\def\figscaling{0.5}
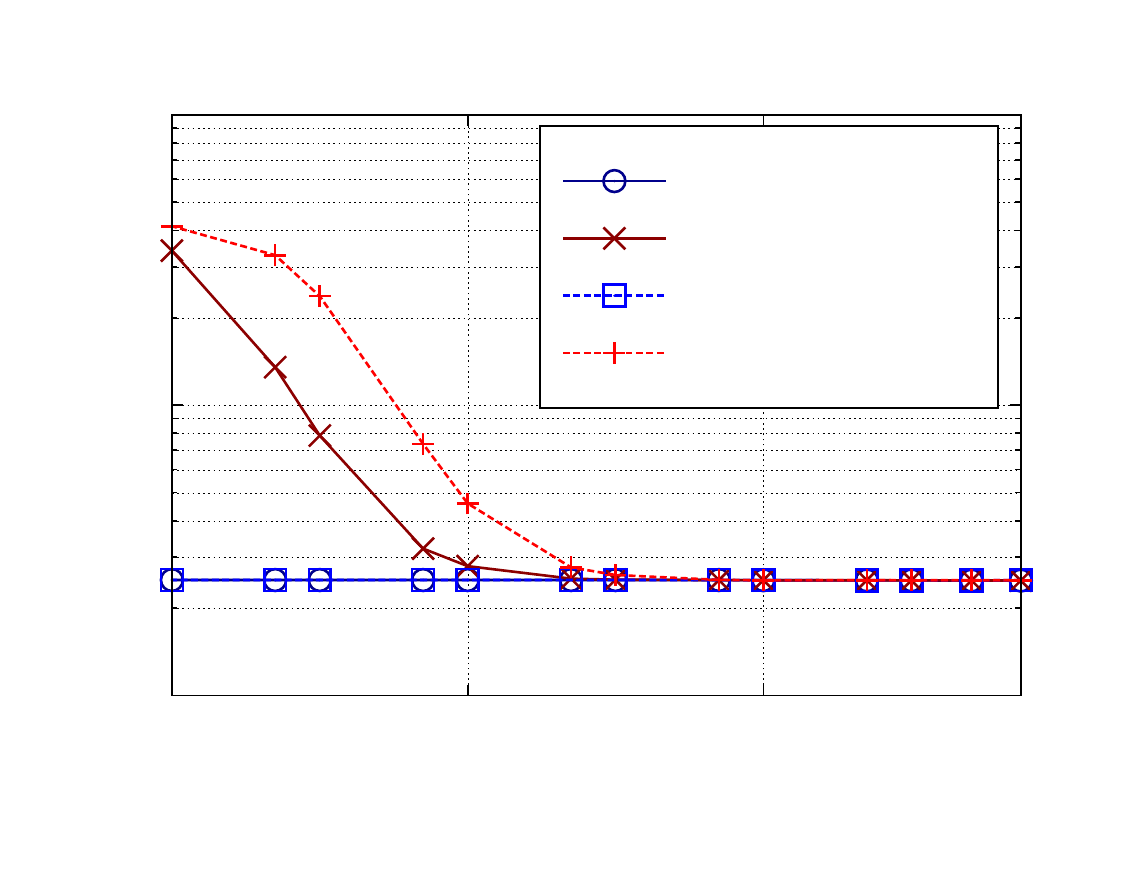
\end{minipage}
\begin{minipage}[hbt]{0.32\textwidth}
\centering
\def\figscaling{0.5}
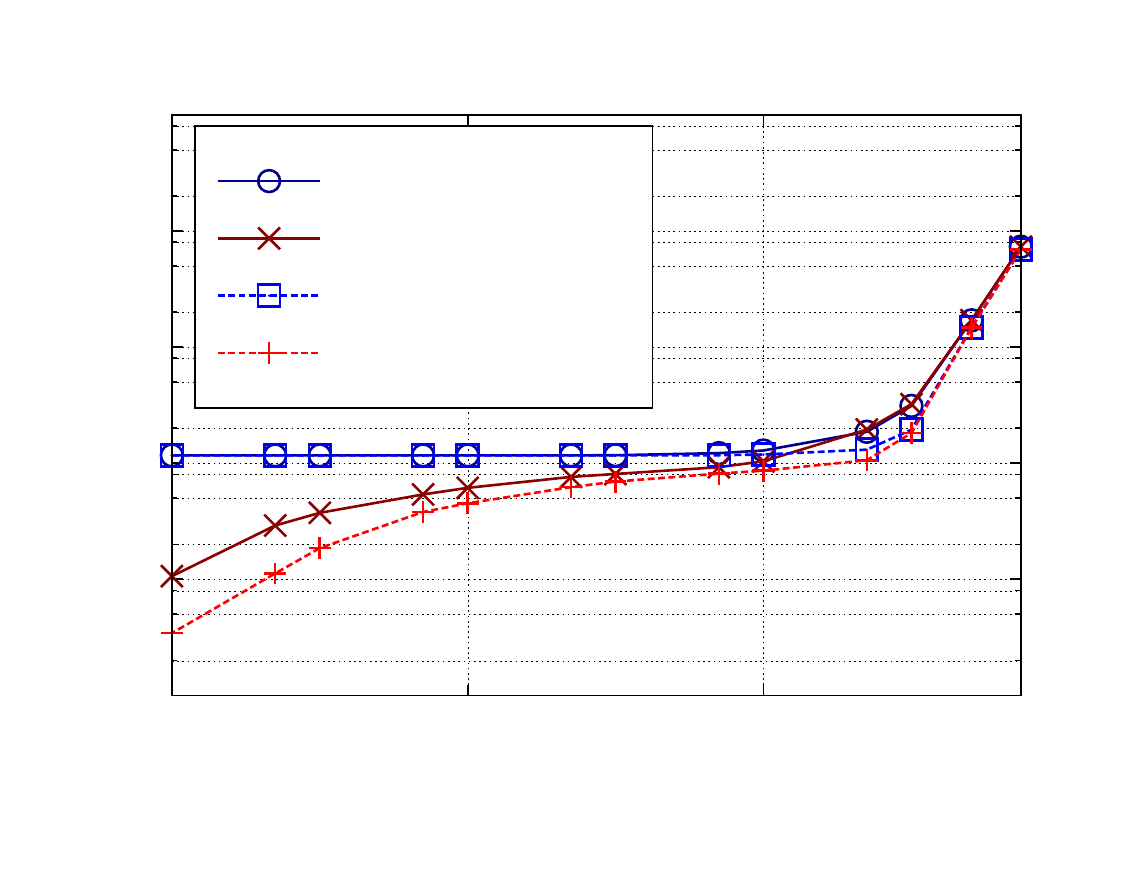
\end{minipage}
\begin{minipage}[hbt]{0.32\textwidth}
\raggedleft
\def\figscaling{0.5}
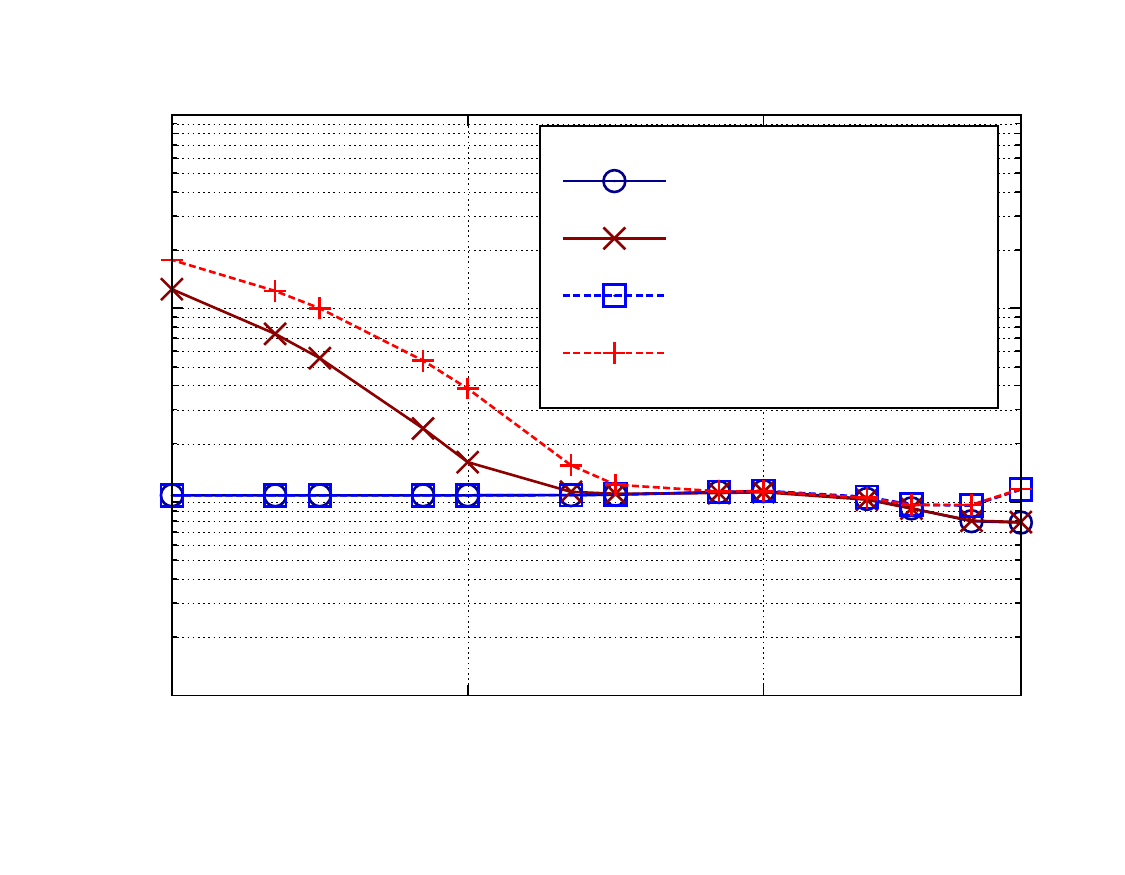
\end{minipage}
\vspace*{0.25cm}

\begin{minipage}[hbt]{0.32\textwidth}
\centering
\def\figscaling{0.5}
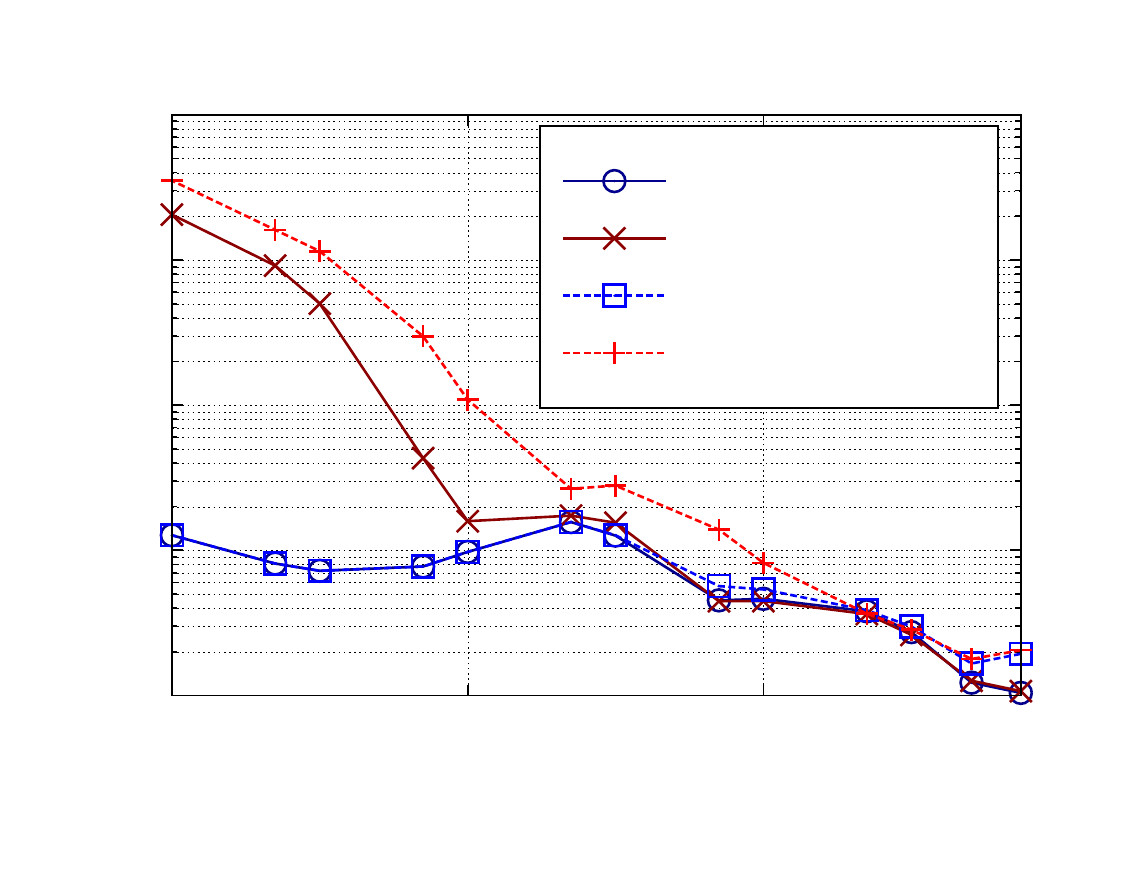
\end{minipage}
\begin{minipage}[hbt]{0.32\textwidth}
\centering
\def\figscaling{0.5}
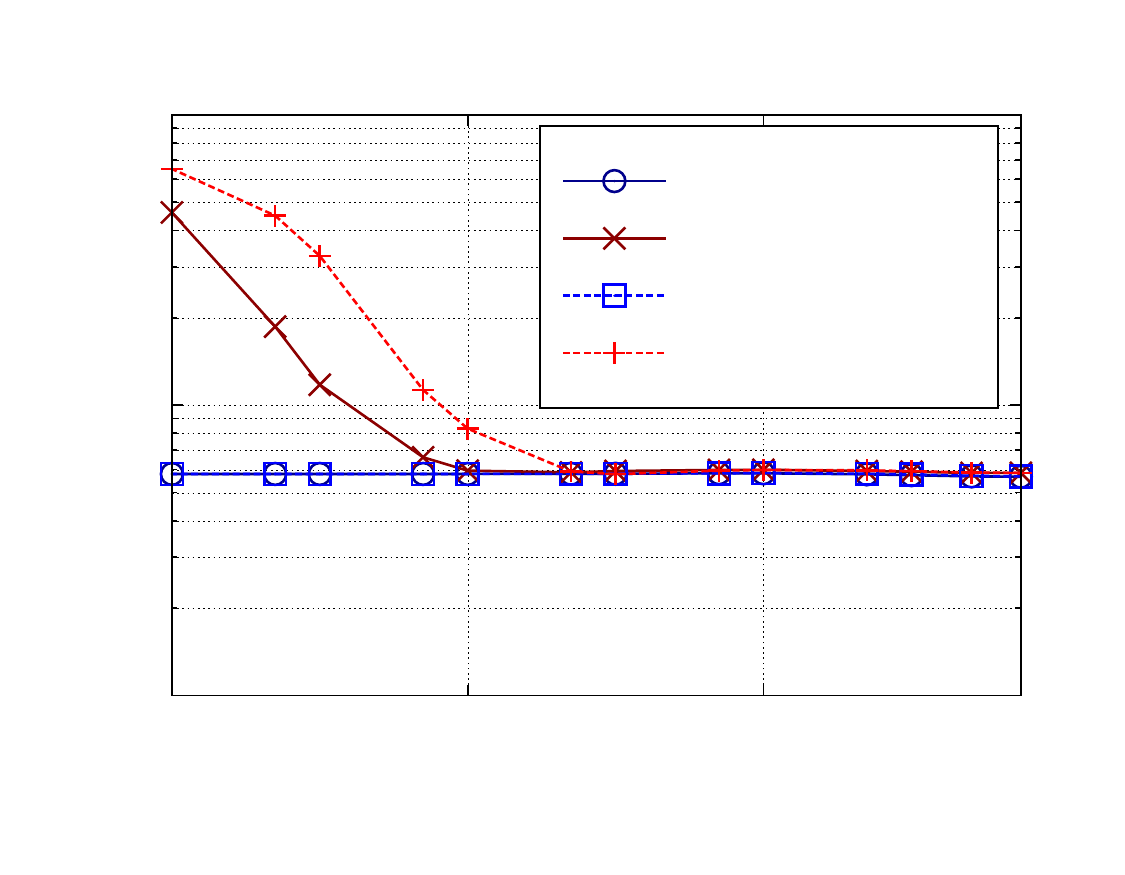
\end{minipage}
\begin{minipage}[hbt]{0.32\textwidth}
\centering
\def\figscaling{0.5}
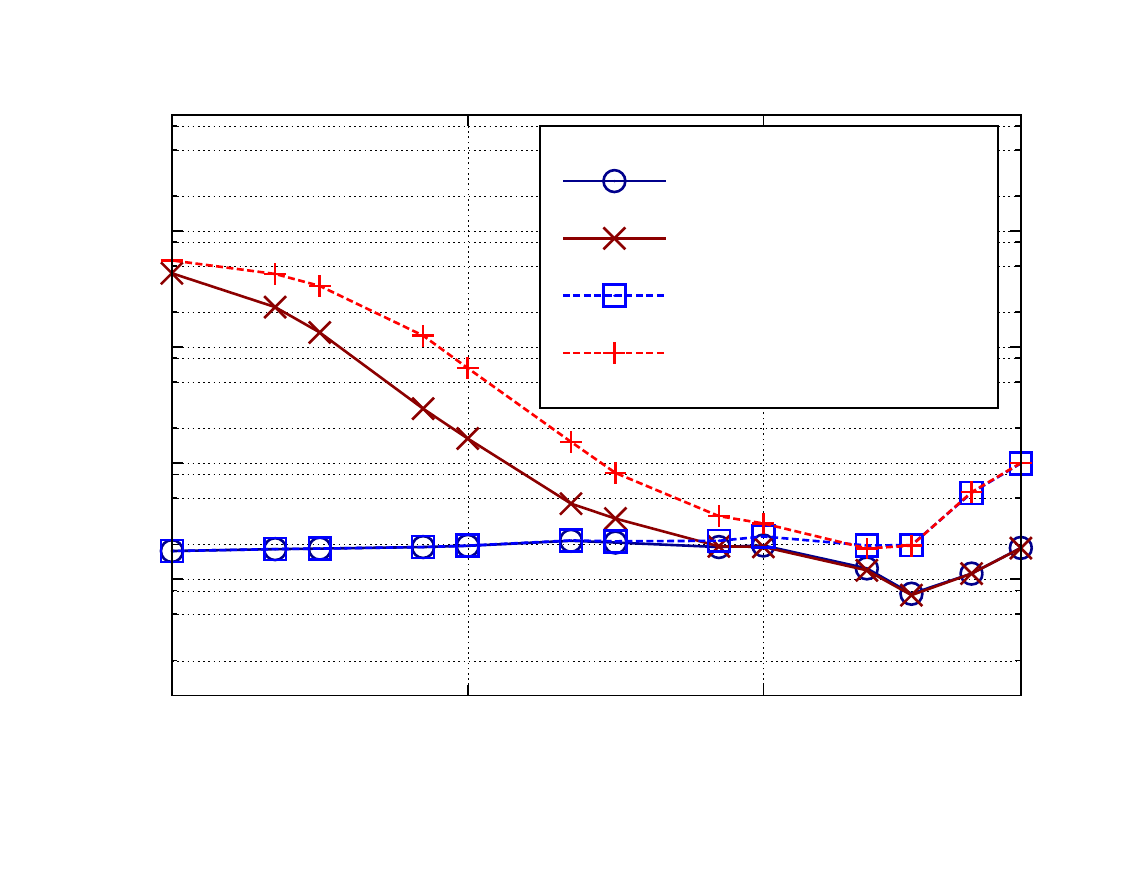
\end{minipage}
\caption{Computed interface error norms for varying porosity $\porosity$, varying material permeability $\matpermeabp$ according to the Kozeny-Carman formula \eqref{eq:kozcar}, varying permeability $\permeabp$, 
and varying sliplength $\sliplengh$ according to relation \eqref{eq: bjcoeff}.
Considered are the ``Substitution'' approach (``Sub'') with contribution \eqref{eq:w_fpi_t_sub} to the weak form and the Nitsche-based approach (``Nit'') with contribution \eqref{eq:w_fpi_t_nit} to the weak form.
The BJ condition is applied with coefficient $\bjcoeff =1.0$: (``$1.0$'') and $\bjcoeff =10.0$: (``$10.10$'').
Computed for mesh size $h=0.0078125$, the adjoint-inconsistent variant ($\xi=-1$) and the inverse penalty parameters $\gamma_n^{-1}=\gamma_t^{-1}=45$.
}
\label{fig:varperm}
\begin{minipage}[hbt]{0.32\textwidth}
\centering
\def\figscaling{0.5}
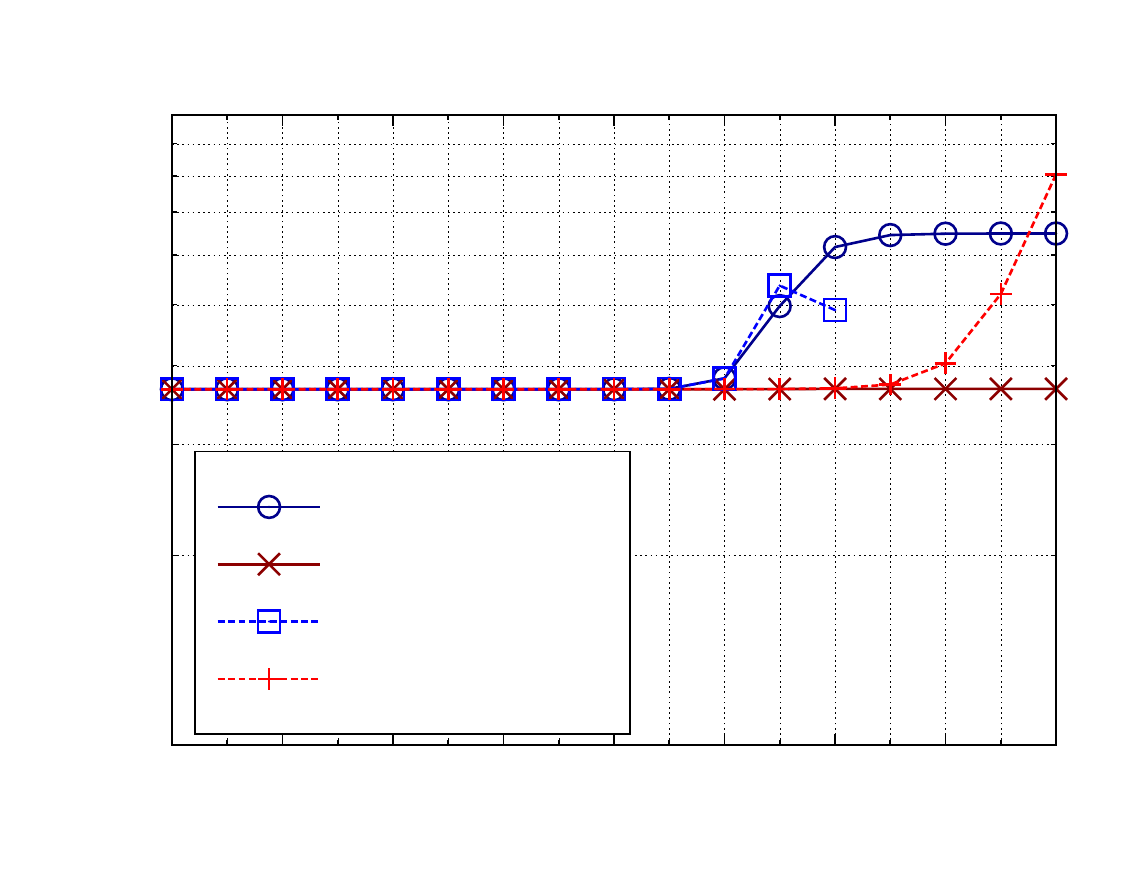
\end{minipage}
\begin{minipage}[hbt]{0.32\textwidth}
\centering
\def\figscaling{0.5}
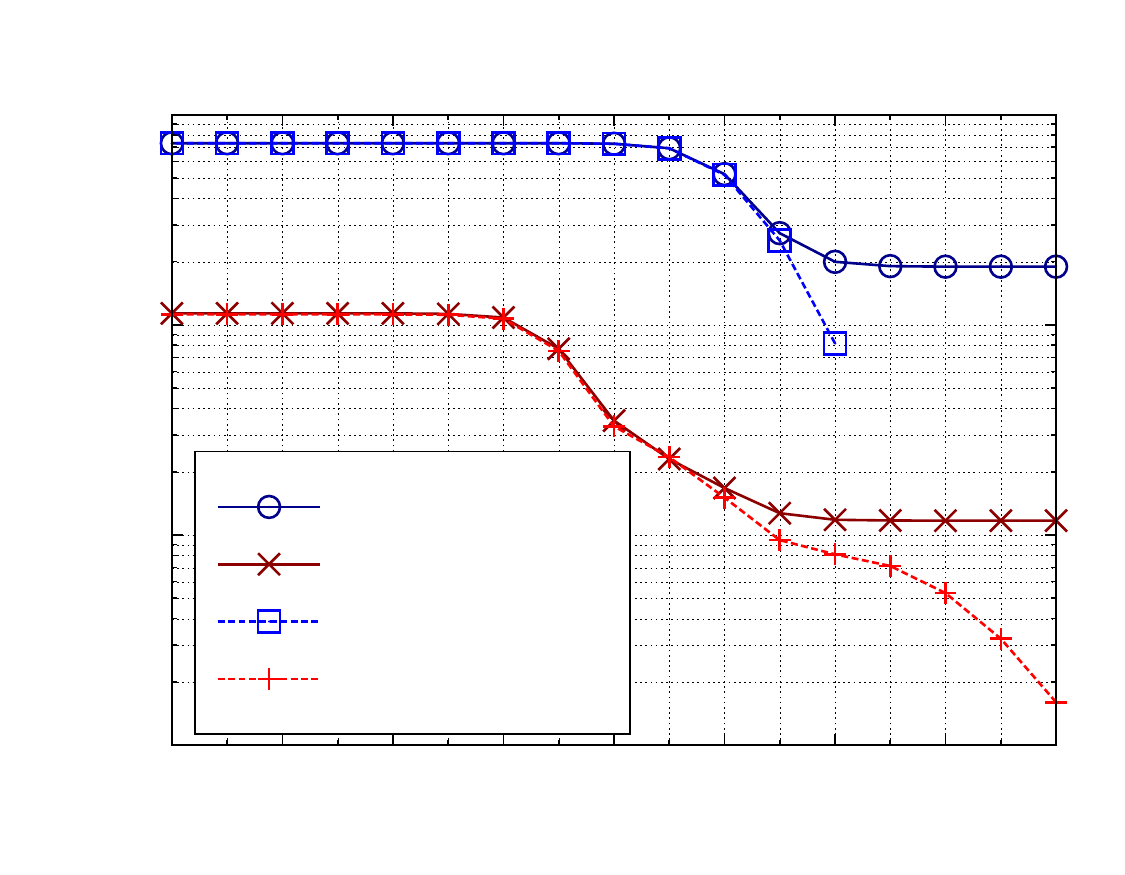
\end{minipage}
\begin{minipage}[hbt]{0.32\textwidth}
\centering
\def\figscaling{0.5}
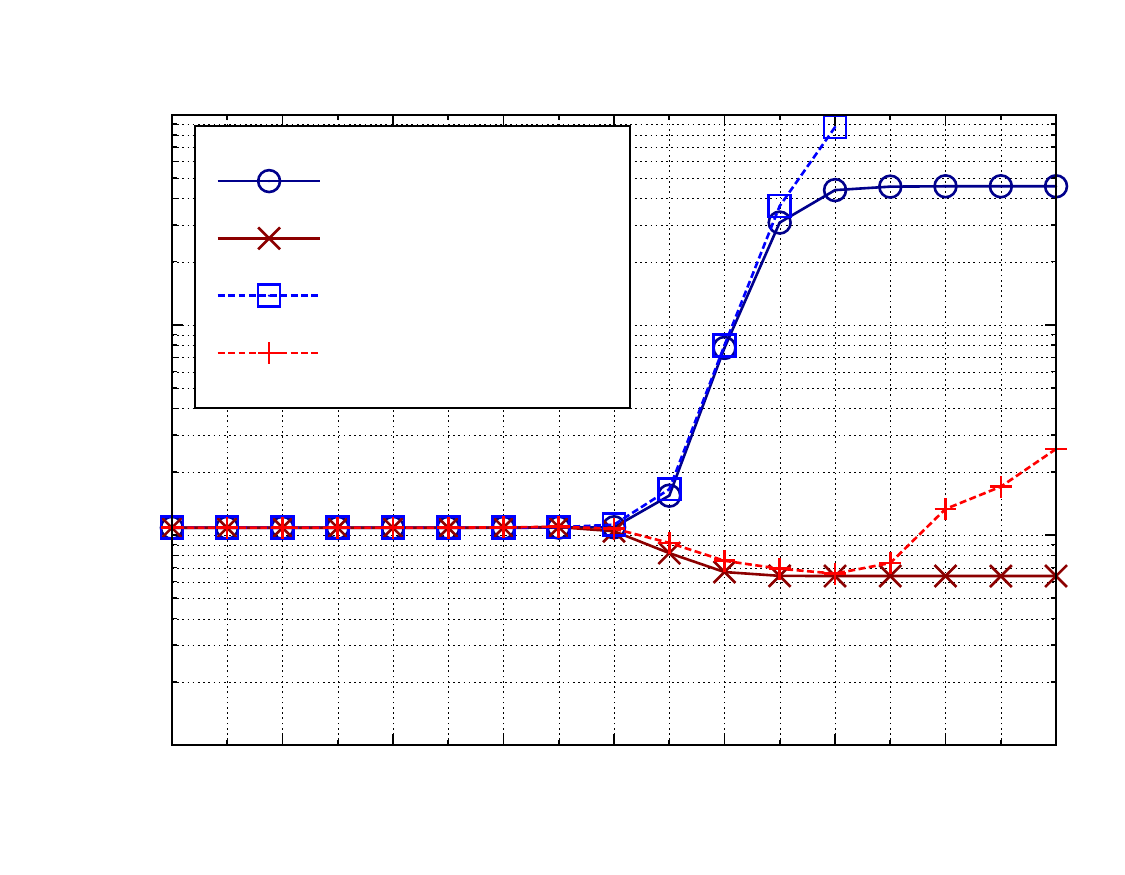
\end{minipage}

\begin{minipage}[hbt]{0.32\textwidth}
\centering
\def\figscaling{0.5}
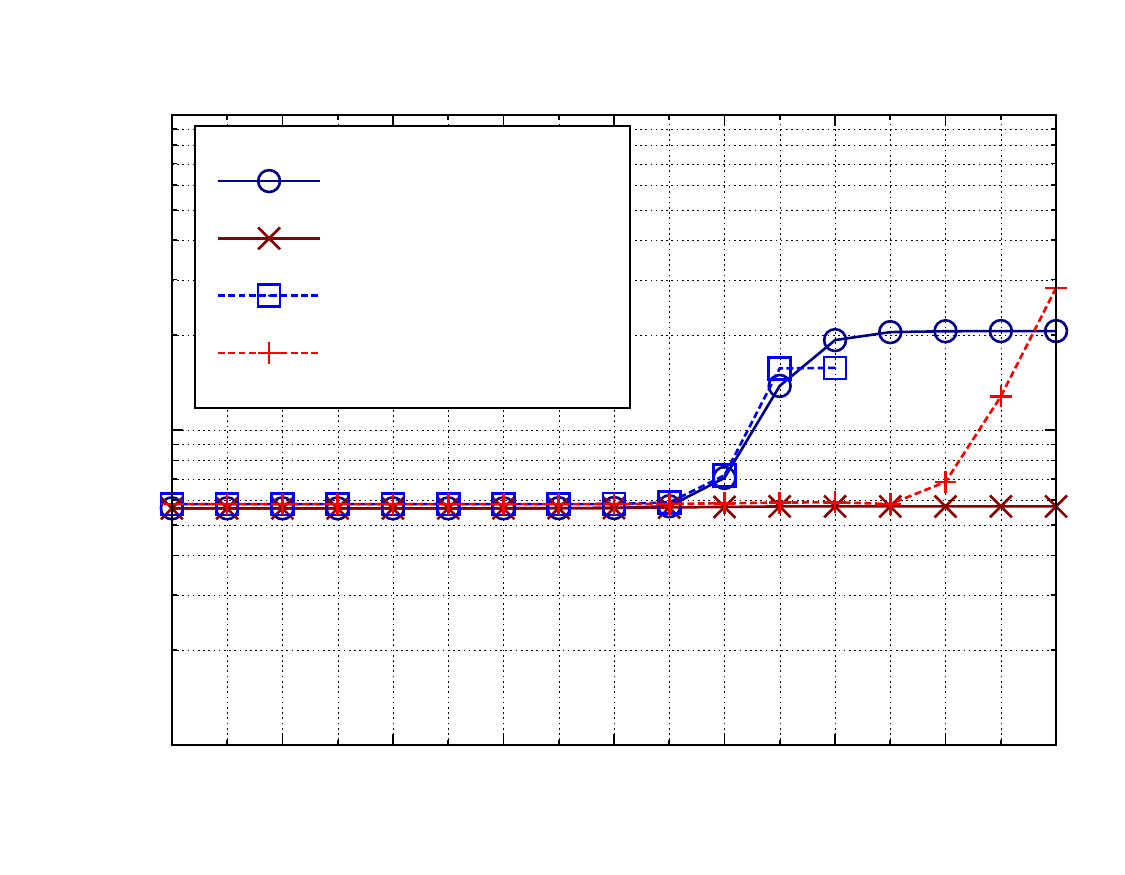
\end{minipage}
\begin{minipage}[hbt]{0.32\textwidth}
\centering
\def\figscaling{0.5}
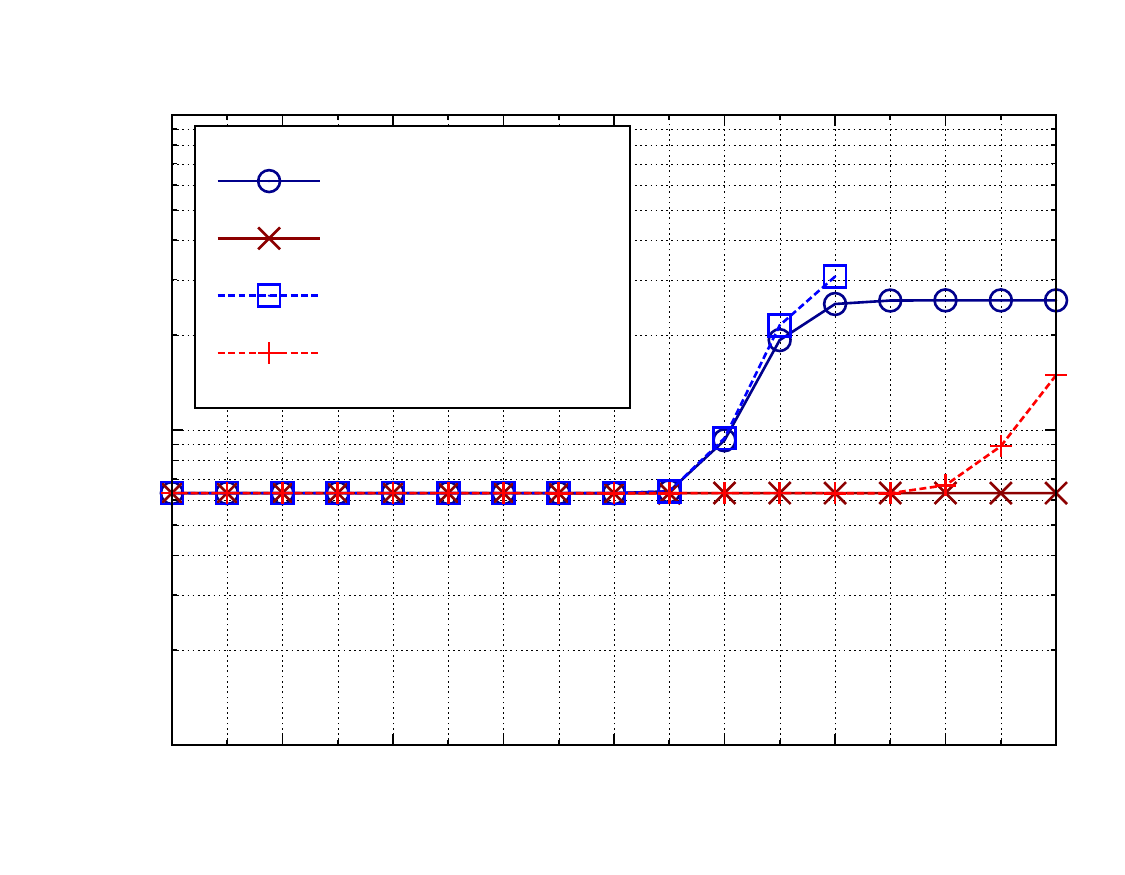
\end{minipage}
\begin{minipage}[hbt]{0.32\textwidth}
\centering
\def\figscaling{0.5}
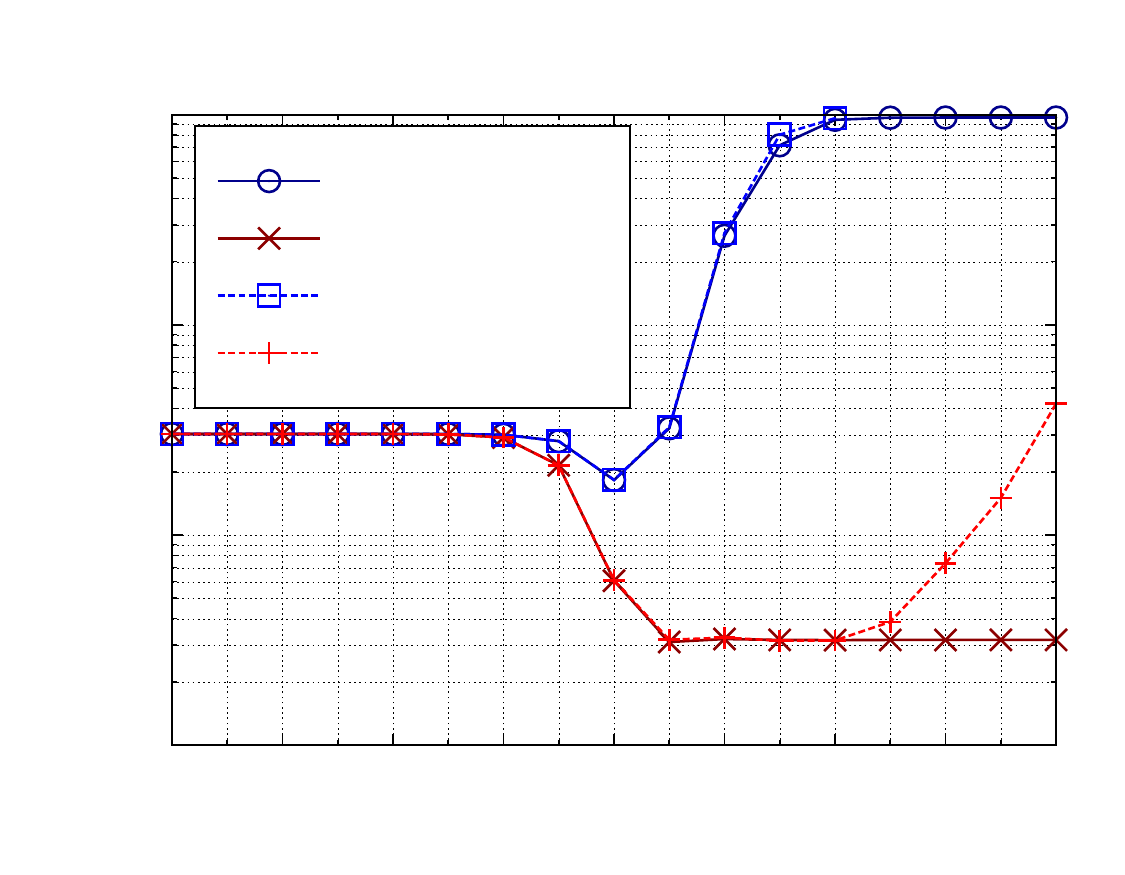
\end{minipage}
\caption{Computed domain and interface error norms for a varying coefficient $\bjcoeff$.
The computations are performed for the Beavers-Joseph (``BJ'', $\BJfac = 1$) and the Beavers-Joseph-Saffmann (``BJS'', $\BJfac = 0$) interface condition (see \eqref{eq:fpi_bj}).
Considered are the ``Substitution'' approach (``Sub'') with contribution \eqref{eq:w_fpi_t_sub} to the weak form and the Nitsche-based approach (``Nit'') with contribution \eqref{eq:w_fpi_t_nit} to the weak form.
Computed for mesh size $h=0.0078125$, the adjoint-inconsistent variant ($\xi=-1$) and the inverse penalty parameters $\gamma_n^{-1}=\gamma_t^{-1}=45$.
}
\label{fig:varbj}
\end{figure}
To analyze the applicability of the BJ or the simplified BJS for the coupled FPI problem,
a comparison of both approaches for a large range of $\bjcoeff = [10^{-8}, 10^8]$ is computed.
For the sake of completeness, both formulations to enforce the tangential constraint are considered.
In \cite{gartling1996,burman2007}, computations for the interface coupling of the Stokes and Darcy equation by the BJ condition with $\bjcoeff^{-1}=0$ lead to an oscillatory velocity solution
close to the interface. Mathematical analysis of the BJ condition of this problem setup can be found in \cite{cao2010,cao2010b}.

The computed results are presented in Figure \ref{fig:varbj}.
As expected for low coefficients $\bjcoeff$, similar results for all variants are computed.
Due to the vanishing kinematic contribution in \eqref{eq:fpi_bj}, when approaching the ``full slip'' limit $\sliplengh^{-1} = 0$, no difference between BJ and BJS occurs.
Only the tangential kinematic error $\mathcal{E}_t$ differs due to the varying definition for the BJ and BJS case.
For a growing coefficient, starting at around $\bjcoeff = 10.0$, a significant increase for almost all computed domain and interface errors can be observed for the BJ interface condition, 
which finally remains constant for large coefficients $\bjcoeff$. 
Solely the error $\mathcal{E}_t$ reduces towards the no-slip limit.
By contrast, the BJS variant does not lead to a drastic large change of the computed error norms in the whole range of $\bjcoeff$.
Only the error norm $\mathcal{E}_t$ behaves differently and shows similar behavior as for the BJ case. 
For the sake of completeness, the displacement domain error norm is shown in Figure \ref{fig:varbj}, which reduces for large coefficients $\bjcoeff$ and the BJS variant.
The difference between the results computed for the Nitsche-based and ``Substitution'' approach to enforce BJ or BJS conditions follows the same argumentation as in the previous section.
For small values of $\bjcoeff$, both methods perform similarly, 
while for large values of the coefficient $\bjcoeff$, closer to the no-slip limit, only the Nitsche-based variant retains a well-conditioned system of equations.
In contrary to the direct observation of oscillations of the velocity solution in the poroelastic domain for large coefficients $\bjcoeff$ in \cite{gartling1996,burman2007},
for the considered mesh size parameter $h=0.0078125$, these oscillations are not directly observable in a visualization such as Figure \ref{fig:ex1_solution}.

From these computed results, both interface conditions, the BJ and the BJS condition,
confirm their applicability for the coupled  FPI problem from a computational point of view for the whole relevant range of the coefficient $\bjcoeff$.
Therefore, both conditions can be considered for upcoming computations, whereas the BJ condition is preferred due to its experimental validation.
For small values of  $\bjcoeff < 10.0$, which includes the physical relevant range, no essential difference in the computed error norms can be observed.

\section{Numerical example: Fluid induced bending of a poroelastic beam}
\label{sec:ex2}
\FloatBarrier
In this section, the presented approach for solving the FPI is applied to the fluid induced dynamic bending process of a poroelastic beam.
As this includes large deformation and motion, the benefits of the presented poroelasticity formulation and CutFEM approach are validated.
\begin{figure}[htbp]
\centering
\input{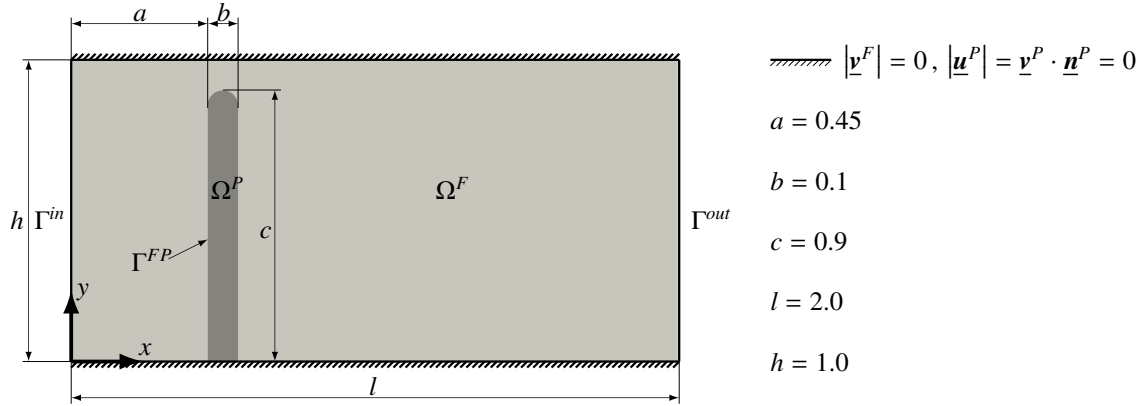}
\caption{Geometry and boundary conditions for the fluid induced bending process of a poroelastic beam.}
\label{fig:ex2_setup}
\end{figure}
\subsection{Problem description}
The principal problem setup including the geometry and basic boundary conditions is visualized in Figure \ref{fig:ex2_setup}.
It consists of a rectangular fluid domain and a poroelastic beam with a circular tip.
On the Dirichlet boundary $\Gamma^{in}$ a parabolic velocity inflow profile is prescribed: 
$\velfD = \left[0.2 \left(y-y^2\right)\left(2-2 \mycos{0.5 \pi t}\right) ,0\right]^T$ for $t \in \left[0.0,2.0\right]$ and
$\velfD = \left[0.8 \left(y-y^2\right),0\right]^T$ for $t \in \left[2.0,10.0\right]$.
On the boundary $\Gamma^{out}$ a zero traction Neumann boundary condition in $x$-direction is combined with a zero velocity Dirichlet boundary condition in $y$-direction.

\begin{figure}[t]
\centering
\includegraphics[width=1\textwidth]{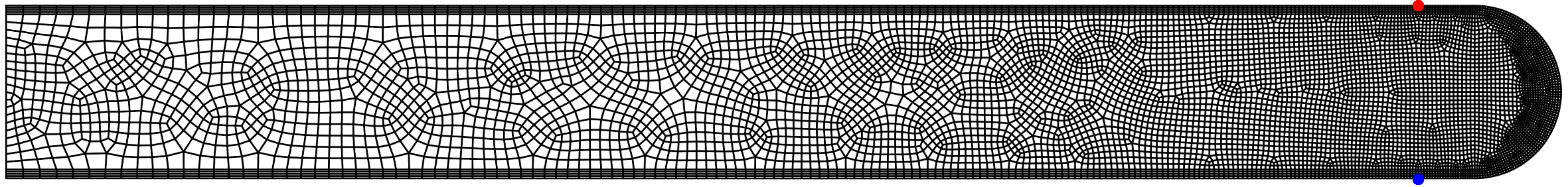}
\caption{Boundary and interface fitted computational mesh consisting of $7880$ bi-linear, quadrilateral elements for the discretization of the poroelastic equations in domain $\domainp$ (for visualization rotated by $-\pi/2$).
Computed solution is visualized in the position of the red marker (top) and in the position of the blue marker (bottom) in Figure \ref{fig:ex2_diag}.
}
\label{fig:ex2_poromesh}
\end{figure}
\begin{figure}[b!]
\centering
\begin{minipage}{0.065\textwidth}
\vspace*{-3.3cm}
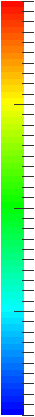
\end{minipage}
\includegraphics[width=0.46\textwidth]{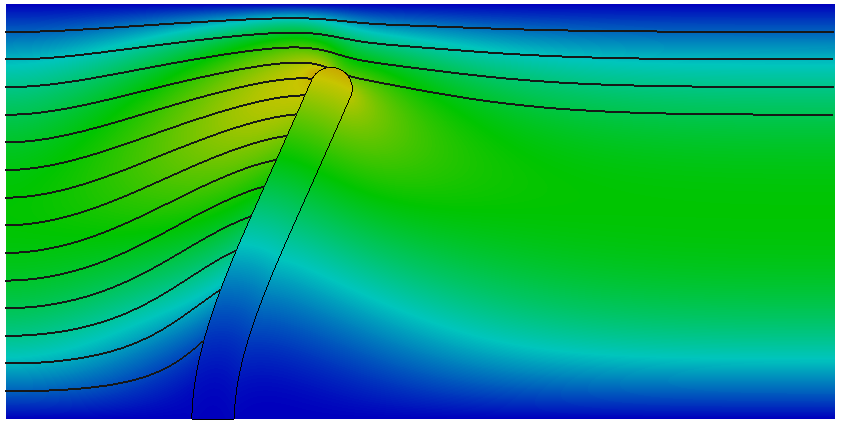}
\includegraphics[width=0.46\textwidth]{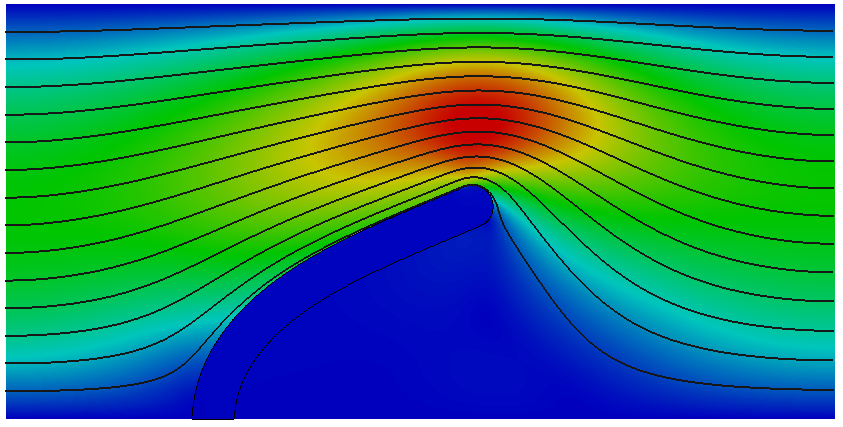}
\begin{minipage}{0.065\textwidth}
\vspace*{-3.3cm}
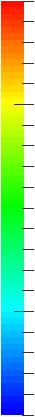
\end{minipage}
\includegraphics[width=0.46\textwidth]{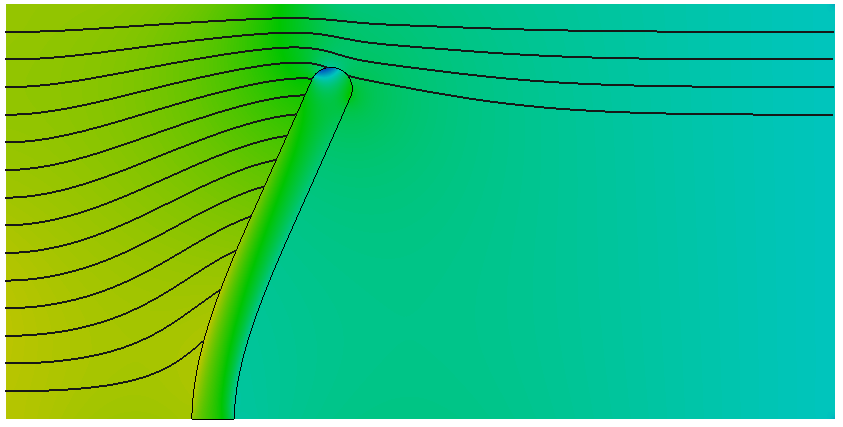}
\includegraphics[width=0.46\textwidth]{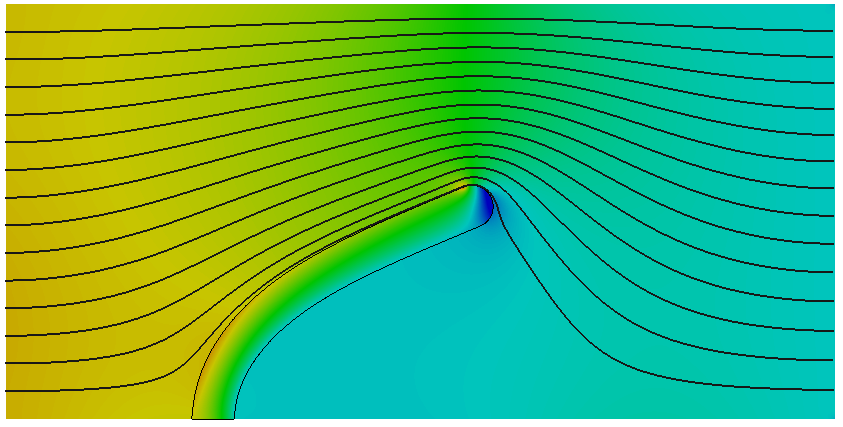}
\caption{Computed velocity magnitude ($\velf$ and $\velp$) in top row and pressure solution ($\pf$ and $\pp$) in bottom row at time $t=2$ (left) and $t=8$ (right). 
Black lines are the streamlines computed from the fluid velocity $\velf$.}
\label{fig:ex2_velpres}
\end{figure}

The fluid is characterized by a dynamic viscosity of $\viscf = 0.01$ and a density of $\densityf = 0.1$.
The initial porosity in the poroelastic beam is $\porosityB = 0.5$ 
and the initial isotropic material permeability is $\initmatpermeabp = \initmatpermeabpscalar \cdot \identity$ with $\initmatpermeabpscalar = 10^{-5}$.
To consider the dependence of the permeability on the porosity, the Kozeny-Carman formula is applied:
\begin{align}
\matpermeabp = \initmatpermeabp \frac{1-\porosityB^2}{\porosityB^3} \frac{\left(\Jp\porosity\right)^3}{1-\left(\Jp\porosity\right)^2}.
\label{eq:ex2_kozcar}
\end{align}
Equal to the first example in Section \ref{sec:ex1}, the macroscopic solid material behavior is given by a Neo-Hookean material model with the stain energy function $\strainenergy{P,skel}$ given in \eqref{ex1:strainenergysp} and 
the material parameters Young's modulus $E = 100$ and Poisson ration $\nu = 0.3$.
To consider a deformation and fluid pressure dependent varying porosity $\porosity$, the strain energy function $\strainenergy{P,vol}$ has to be considered as well. 
Here, the following formulation, with Bulk modulus $\kappa^P = 100$, is applied:
\begin{align}
\label{ex2:addstrainenergysp}
\strainenergy{P,vol} = \kappa^P \left[ \frac{(1-\porosity)\Jp}{1-\porosityB}-1-\text{ln}\left(\frac{(1-\porosity)\Jp}{1-\porosityB}\right)\right].
\end{align}
No contribution $\strainenergy{P,pen}$ is added to the overall strain energy function. The average initial density of the solid phase is set as $\refdensityps = 0.2$.
On the interface $\fpiinterface$ the BJ condition ($\BJfac = 1$) with a coefficient of $\bjcoeff = 1.0$ is weakly imposed by the Nitsche-based contribution \eqref{eq:w_fpi_t_nit}.

The computational mesh for the discretization of the fluid equations in domain $\domainf$, which is only fitted to the outer boundaries but not the interface $\fpiinterface$, covers the whole rectangular domain visualized in Figure \ref{fig:ex2_setup}.
It consists of $250\times120=30000$ structured, bi-linear, quadrilateral elements. Figure \ref{fig:ex2_poromesh} shows the boundary and interface fitted, unstructured computational mesh for domain $\domainp$.
Like for the computations in Section \ref{sec:ex1}, this is accomplished by a discretization with one layer of eight-noded, tri-linear hexahedreal elements. 

The discretization in time is performed by the backward Euler scheme $\theta = 1$ with a time step length of $\Delta t = 0.01$.
The initial state of the problem is the zero-state: $ \velfB = \disppB = \velpsB = \velpB = \zerovec$.

\subsection{Computed results and discussion}
In Figure \ref{fig:ex2_velpres}, the computed velocity and pressure solution for two instances in time is shown.
At $t=2$, the initiated bending motion of the poroelastic beam, due to the fluid inflow, which leads to a pressure difference between the two interface sides, can be seen.
The velocity in the fluid domain and the poroelastic domain is roughly continuous as the beam moves approximately with the same velocity as the fluid velocity.
For $t=8$ the poroelastic beam is already in a stationary position and therefore the small remaining relative velocity in the poroelastic domain cannot be seen by the color code.
The major part of the fluid mass flow does not pass the poroelastic beam, leading to the maximal velocity at the smallest constriction.
\begin{figure}[hbtp]
\hspace*{-1cm}
\begin{minipage}[bt]{0.33\textwidth}
\raggedleft
\def\figscaling{0.84}
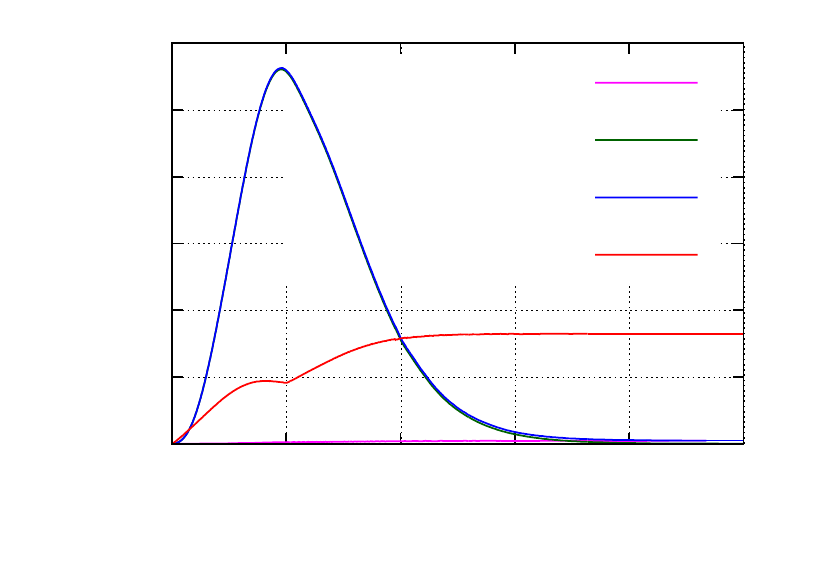
\end{minipage}
\begin{minipage}[hbtp]{0.33\textwidth}
\centering
\def\figscaling{0.84}
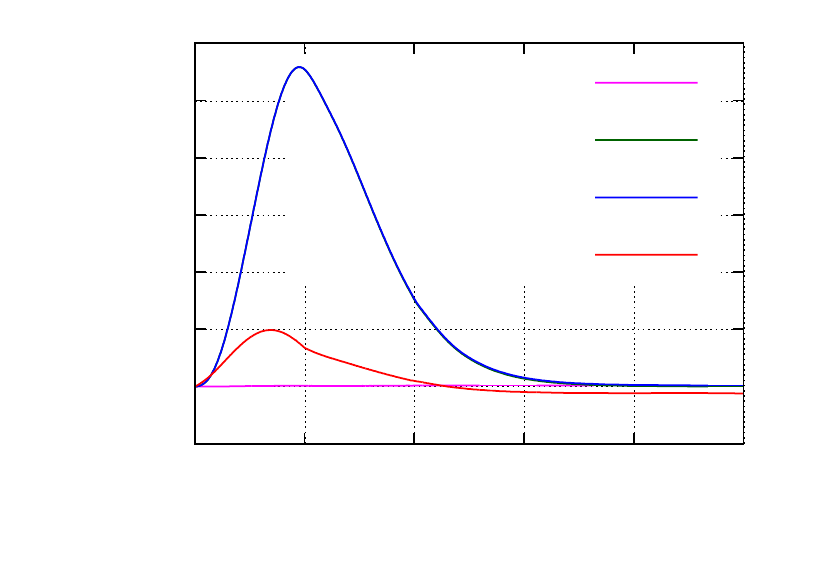
\end{minipage}
\begin{minipage}[hbt]{0.33\textwidth}
\centering
\def\figscaling{0.84}
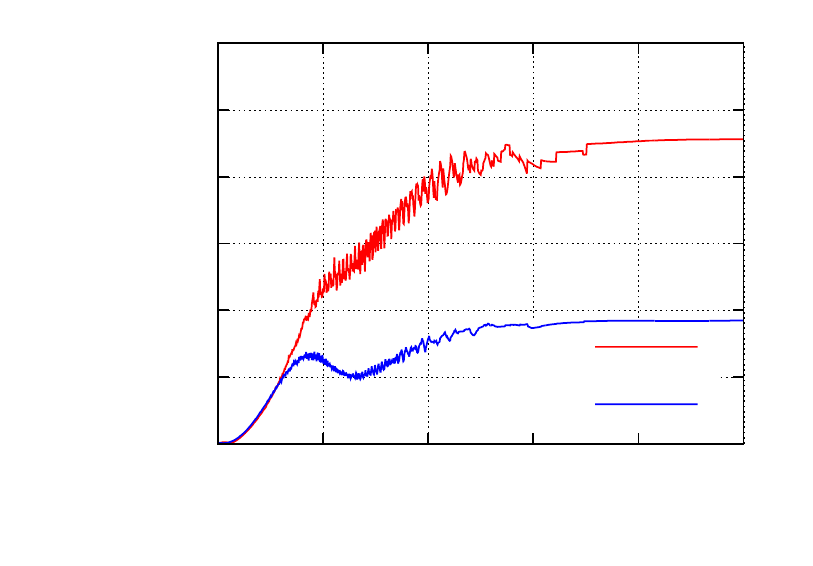
\end{minipage}
\caption{Computed pressure $\pp$, fluid velocity magnitude $|\velp|$, solid velocity magnitude $|\velpssmall|$, and relative velocity magnitude $|\velp - \velpssmall|$
for two selected points. Computed solution at position of red marker (left) and solution at position of blue marker (mid).
The relative velocity magnitude is visualized in detail for both markers in the last graph (right).
The markers can be found in Figure \ref{fig:ex2_poromesh}.}
\label{fig:ex2_diag}
\end{figure}
In Figure \ref{fig:ex2_diag}, the computed solution in the two selected points of the poroelastic domain is shown.
Whereas on the upper side (left) a continuous increase of the pressure $\pp$ during the bending motion can be observed, 
the lower side (mid) results first in an increase of the fluid pressure due to the increasing inflow velocity followed by a decrease of the pressure due the changing orientation 
caused by the bending motion of the beam.
At $t=2$, a kink in the pressure curve forms due to the change in the inflow function.
The fluid velocity $|\velp|$ and the solid velocity $|\velpssmall|$ are almost equal and represent the initial acceleration phase, 
followed by the slowdown phase due to the increasing elastic stress balancing the fluid stress.
A detailed view to the relative velocity $|\velp - \velpssmall|$ for both selected points is given in Figure \ref{fig:ex2_diag} (right).
In the time interval from approximately $t=2$ to $t=6$, where a high solid velocity $|\velpssmall|$ is prevalent, oscillations of the relative velocity can be observed.
Close inspection reveals these oscillations also exist for the evolution in time of the pressure $\pp$.
For a specific material point $\refcoordp$ on the interface $\fpiinterface$, this is due to the frequently changing neighboring 
discrete fluid solution space (of the fixed background computational mesh)
and the consequently varying computational error. 
A strong effect is expected to arise especially from the discontinuous viscous fluid stress on the interface between single fluid elements.
This issue can be resolved by an increase of the resolution of the computational fluid mesh close to the interface $\fpiinterface$ or a hybrid approach as presented in \cite{schott2016,schott2017b}.

\begin{figure}[hbtp]
\centering
\begin{minipage}{0.065\textwidth}
\vspace*{-3.3cm}
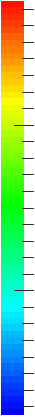
\end{minipage}
\includegraphics[width=0.3\textwidth]{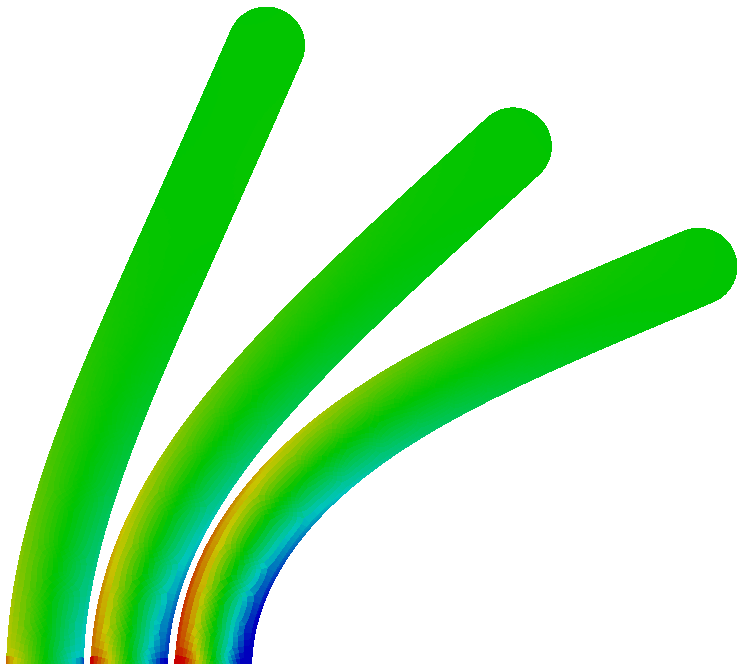}
\hspace*{0.1\textwidth}
\includegraphics[width=0.3\textwidth]{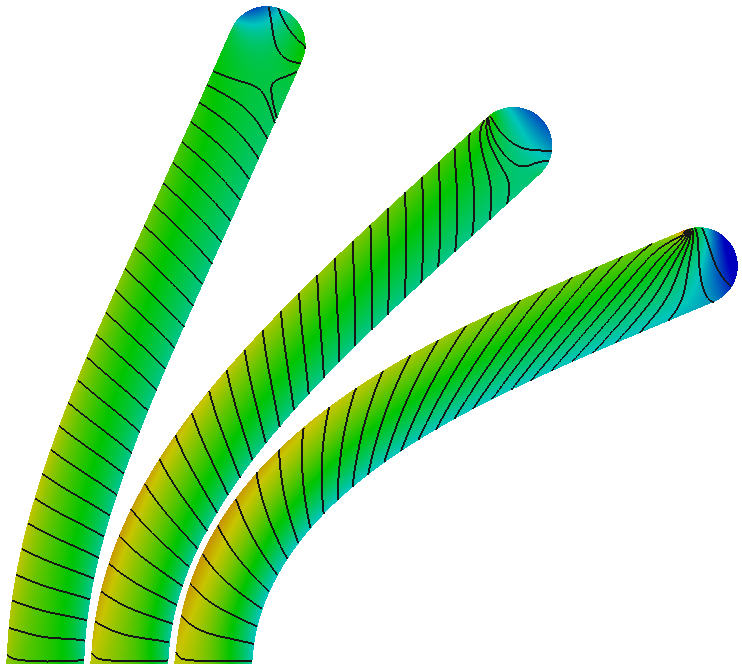}
\begin{minipage}{0.065\textwidth}
\vspace*{-3.3cm}
\input{pl.pdf_tex}
\end{minipage}
\caption{Computed porosity $\porosity$ (element-wise constant visualization) in the poroelastic domain $\domainp$ at three instances in time from left to right $t=2$, $t=3$, and $t=8$ (left).
Computed pressure $\pp$ and visualization of the  seepage velocity $\left(\velp-\velpssmall\right)$ by streamlines (black lines) in the poroelastic domain $\domainp$ at three instances in time from left to right $t=2$, $t=3$, and $t=8$ (right).}
\label{fig:ex2_poros}
\end{figure}
The computed porosity in the poroelastic domain is visualized for three instances in time in Figure \ref{fig:ex2_poros} (left).
Due to the expansion of the beam on the inflow side, an increase of the fluid fraction and therefore the porosity $\porosity$ can be observed.
On the other side, the compression leads to a reduction of the porosity.
At the tip of the beam, no significant deformation occurs and the porosity is almost equal to its initial value of $\porosity = \porosityB = 0.5$.
It can be seen that for this problem setup, no significant influence of the fluid pressure on the porosity can be observed.
In Figure \ref{fig:ex2_poros} (right), the streamline computed for the seepage velocity and the fluid pressure is visualized.
As one would expect, a flow from the high pressure inflow side to the low pressure side can be observed for a large part of the beam.
A fluid flow from the inside of the poroelastic domain leaving at the tip of the beam can be noticed solely at $t=2$.

\section{Conclusion}
\label{sec:conclusion}
In this contribution, we presented a finite element approach for solving the interface coupled problem of an incompressible, viscous fluid flow and poroelastic, fluid saturated structures.
The incorporation of a general formulation for the poroelastic domain, including varying porosity and the possibility to consider various different material behaviors, was one essential aspect.
Moreover, a CutFEM approach, which allows for the use of non-interface-fitted computational meshes for the discretization of the fluid domain, was applied.
All additional contributions to the discrete weak form arising from the discrete stabilization of the formulation were presented in detail.
In this coupled problem setup the CutFEM allows for the consideration of large interface motion and topological changes of the fluid domain.
Finally, a Nitsche-based formulation to incorporate the tangential Beavers-Joseph(-Saffmann) constraint was presented.
This formulation permits a good conditioning of the final system of equations to be solved even when approaching the ``no-slip'' limit case, 
which is the case for low permeabilities in the poroelastic structure.

To analyze the presented approach, a numerical example with a known solution for the coupled problem was considered and computed error norms were analyzed.
Therein, a spatial convergence study showed that the expected convergence rates could be reached.
A sensitivity study on the two included Nitsche penalty parameters allowed for a specification of a proper range for these parameters.
Furthermore, the comparison of the presented Nitsche-based formulation and the classic ``Substitution'' approach for incorporation of the tangential BJ interface condition was performed.
For low porosities and low permeabilities, the superiority over the classical approach was shown.
A comparison of the Beavers-Joseph and the simplified Beavers-Joseph-Saffmann interface condition confirmed the applicability of both approaches in the physical range of the model parameter.

Finally, a second numerical example analyzing the fluid induced bending of a poroelastic beam was presented. 
This includes fluid pressure and deformation dependent varying porosity and therefore provides an insight into the modeling potential of the poroelastic formulation.
The requirement for a sufficiently resolved fixed grid fluid domain close to the interface or a hybrid approach as presented in \cite{schott2016,schott2017b} is highlighted by oscillations of the relative velocity in the poroelastic domain.
Nevertheless, the large interface motion and deformation of the permeable structure for this problem configuration shows
the advantages of the applied poroelastic formulation and the CutFEM approach.

\section*{Acknowledgements}
\label{sec:acknowledgement}
The authors C.A.~and W.A.W.~gratefully acknowledge the support by the protject KonRAT: ``Komponenten von Raketentriebwerken f\"ur Anwendungen in
Transportsystemen der Luft- und Raumfahrt'', work package 2400 ``Fluid-structure-interaction (FSI)'' of the Ludwig B\"olkow Campus.
M.W.~is supported by the International Graduate School of Science and Engineering (IGSSE) of the Technical University of Munich, Germany, under project 6.02 and the Erasmus Mundus Joint Doctorate SEED project (European Commission with grant Ref. 2013-0043).

\bibliographystyle{elsarticle-num}
\bibliography{bib}
\end{document}